\RequirePackage{fix-cm}
\documentclass[smallextended]{svjour3}       
\smartqed  
\usepackage{graphicx}
\usepackage{mathptmx}      

\usepackage{amsfonts}
\usepackage{amsmath}
\usepackage{amssymb}
\usepackage{latexsym} 
\usepackage{commath}
\usepackage{calrsfs}
\usepackage{mathrsfs}
\usepackage{mathtools} 
\usepackage{bm}
\usepackage{xcolor}
\DeclareMathOperator*{\btimes}{\scalerel*{\times}{\sum}}

\usepackage{scalerel}

\newcommand{\I}{{\mathcal{I}}}
\newcommand{\T}{{\mathcal{T}}}
\newcommand{\cO}{{\mathcal{O}}}
\newcommand{\f}{\frac}

\renewcommand{\P}{{\mathcal{P}}}
\newcommand{\R}{{\mathcal{R}}}
\newcommand{\Rs}{{\mathbb{R}}}
\newcommand{\N}{{\mathcal{N}}}
\newcommand{\Ns}{{\mathbb{N}}}

\newcommand{\K}{{\mathcal{K}}}

\newcommand{\be}{\mathbf{e}}
\newcommand{\bh}{\mathbf{h}}
\newcommand{\bff}{\mathbf{f}}

\newcommand{\bk}{\mathbf{k}}
\newcommand{\bj}{\mathbf{j}}

\newcommand{\uk}{\underline{k}}

\newcommand{\uv}{\underline{v}}
\newcommand{\un}{\underline{n}}

\DeclareMathOperator*{\wlim}{w-lim}

\newcommand{\XXint}[3]{\setbox0=\hbox{$#1{#2#3}{\int}$ }
\vcenter{\hbox{$#2#3$ }}\kern-.6\wd0}

\newtheorem{thm}{Theorem}
\newtheorem{cor}[thm]{Corollary}
\newtheorem{lem}[thm]{Lemma}

\newtheorem{rem}[thm]{Remark}

\begin{document}

\title{Grand Canonical Evolution for the Kac Model
\thanks{\copyright~2021 by the authors. Reproduction of this article by any 
means permitted for non-commercial purposes.}
}

\author{Justin Beck \and Federico Bonetto}

\institute{J. Beck \at
              Department of Mathematics, University of Notre Dame,
			  255 Hurley Bldng, Notre Dame, IN 46556.
              \email{jbeck6@nd.edu} 
           \and
           F. Bonetto \at
              School of Mathematics, Georgia Institute of Technology,
              686 Cherry St., Atlanta, GA 30332. 
			  \email{bonetto@math.gatech.edu}
}

\date{Received: date / Accepted: date}

\maketitle

\begin{abstract}
We study a model of random colliding particles interacting with an infinite 
reservoir at fixed temperature and chemical potential. Interaction between the 
particles is modeled via a Kac master equation \cite{kac}. Moreover, particles 
can leave the system toward the reservoir or enter the system from the 
reservoir. The system admits a unique steady state given by the Grand Canonical 
Ensemble at temperature $T=\beta^{-1}$ and chemical potential $\chi$. We show 
that any initial state converges exponentially fast to equilibrium by computing 
the 
spectral gap of the generator in a suitable $L^2$ space and by showing 
exponential decrease of the relative entropy with respect to the steady state. 
We also show propagation of chaos and thus the validity of a Boltzmann-Kac type 
equation for the particle density in the infinite system limit. 
\end{abstract}

\section{Introduction}
\label{intro}
In 1955, Mark Kac \cite{kac} introduced a simple model to study the evolution of 
a dilute gas of $N$ particles with unit mass undergoing pairwise collisions. 
Instead of following the 
deterministic evolution of the particles until a collision takes place, he 
considered particles that collide at random times with every particle 
undergoing, on average, a given number of collisions per unit time. Moreover, 
when a collision takes place, the energy of the two particles is randomly 
redistributed between them. In such a situation, one can neglect the position of 
the particles and focus on their velocities. To obtain a model as 
simple as possible, he considered particles that move in one spatial dimension. 
This leads to an evolution governed by a master equation for the 
probability distribution $f(\uv_N)$, where $\uv_N\in \mathbb{R}^N$ are the 
velocities of the particles. Since collisions preserve the kinetic energy of 
the system, to obtain ergodicity one has to restrict the evolution to $\uv_N\in 
\mathbb{S}^{N-1}(\sqrt{2eN})$, that is on the surface of constant kinetic 
energy with $e$ the kinetic energy per particle. To further simplify the 
model, he neglected the dependence of a particle collision rate on its speed, 
a situation sometime referred as {\it Maxwellian particles}. In this  
setting, the dynamical properties of the evolution do not depend on $e$ and it 
is thus natural to set $e=1/2$, see \cite{kac,KacBook,McK} 
for more details. 

The study of the Kac master equation has been very useful to clarify and 
investigate notions and conjectures arising from the kinetic theory of diluted 
gases. We refer the reader to Kac's original works \cite{kac} and \cite{KacBook} 
for extensive discussion.

Kac's master equation also provides a natural setting to study approach to 
equilibrium. In the case of the standard Kac model \cite{kac}, equilibrium is 
represented by the uniform distribution on the surface of given kinetic energy. 
Uniform convergence in the sense of the $L^2$ gap was conjectured by Kac and it 
was established in \cite{Jeanvresse} while the gap was explicitly computed in 
\cite{CCL1}. 

A more natural way to define approach to equilibrium is via the relative 
entropy. This provides a better setting since the relative entropy, in general, 
grows only linearly with the number of particles. There is no result of 
exponential decay of relative entropy with a rate that is uniform in $N$ for the 
original Kac model. Moreover, estimates of the entropy production rate seem to 
point to a slow decay, at least for short times, see \cite{amit,villani}.

In \cite{BLV}, the authors studied the evolution of a dilute gas of $N$ 
particles brought to equilibrium via a Maxwellian thermostat, i.e. an infinite 
heat reservoir at fixed temperature $T=\beta^{-1}$. The velocities of the 
particles in the system evolve according to the standard Kac collision process 
described above. On top of this, particles in the system collide with particles 
in the thermostat at randomly distributed times. In this way, the system and the 
reservoir exchange energy, but there is no exchange of particles. In particular, 
the kinetic energy of the system is no more preserved. They proved that the 
system admits as a unique steady state the Canonical Ensemble, i.e. in the 
steady state the probability distribution $f(\uv_N)$ is the Maxwellian 
distribution at temperature $T$. Moreover, the steady state is approached 
exponentially fast and uniformly in $N$, both in the sense of the spectral gap, 
in a suitable $L^2$ space, and in the sense of the relative entropy. In both 
cases, the rate of approach is determined by the interaction with the thermostat 
while the rate of collision between particles in the system appears only in the 
second spectral gap. They also adapted McKean's proof \cite{McK} of propagation 
of chaos and obtained a Boltzmann-Kac type effective equation for the evolution 
of the one particle marginal in the limit $N\to\infty$.

In the present work, we study a different way to bring the system to  
equilibrium. As in \cite{BLV}, we study a system of $N$ particles evolving 
through pair collisions and interacting with an infinite reservoir at given 
temperature $T$; however, the system and the reservoir are allowed to exchange 
particles. The evolution of the the velocities of the particles in the system is 
again described by a standard Kac collision process. On top of these, at random 
times a particle in the system can leave it while, still at random times, a 
particle can enter the system from the reservoir with its velocity distributed 
according to the Maxwellian at temperature $T$. Since the reservoir is infinite, 
no particle can enter or leave the system more than once. Clearly, in this new 
setting, energy and number of particles are not preserved. We show that this new 
evolution admits as its unique steady state the Grand Canonical Ensemble. This 
means that, in the steady state, the probability that the system contains $N$ 
particles is given by a Poisson distribution while the probability distribution 
on the velocities, given the number of particles, is the Maxwellian at 
temperature $T$.
 
We also study the approach to equilibrium in a suitable $L^2$ space and in 
relative entropy. In both cases, we show that the rate of approach is uniform in 
the average number of particles. As in \cite{BLV}, the approach to equilibrium, 
both in $L^2$ and in relative entropy, is driven by the thermostat alone while 
the second spectral gap depends on the rate of binary particle collisions. 
Finally, we look at the emergence of an effective evolution for the particle 
density in the limit of a large system, that is when the average number of 
particles goes to infinity. This requires some adaptation of the concept of 
propagation of chaos since the number of particles in the system is not 
constant. Adapting the proof in \cite{McK}, we show that the relative particle 
density, defined in \eqref{marg} and \eqref{reld} below, satisfies a 
Boltzmann-Kac type of equation.

The rest of the paper is organized as follows. In section 2, we present the 
model and state our main results. Section 3 contains the proofs of our main 
results, while in 
section 4 we report some open problems and present possible areas of future 
work. Finally the appendix contains the proofs of some technical Lemmas used in 
section 3.

\section{Model and Results}
\label{sec:2}

Since we want to describe a dilute gas with uniform density exchanging particles 
with an infinite reservoir, it is natural to assume that, in a given time, each 
particle in the system has the same probability of leaving it independently from 
the total number $N$ of particles in the system. This implies that the flow of 
particles from the system to the reservoir is proportional to $N$. On the other 
hand, the probability of a particle to enter the system from the reservoir 
depends only on the characteristics of the reservoir, and not on $N$, so that 
the 
flow of particles in the system is independent from $N$. Finally, since the gas 
is dilute, given two particles in the system, their probability of colliding in 
a given time does not depend on the total number of particles in the system. 
Thus we expect the number of binary collisions in the system, in a given time, 
to be proportional to $\binom N 2$. These are the main heuristic considerations 
that lead to the formulation of our model to be introduced formally below. 

We consider a system of particles in one space dimension interacting with an 
infinite reservoir with which it exchanges particles. Since the number of 
particles in the system is not constant, the phase space is given by 
$\R=\bigcup_{N=0}^\infty \Rs^N$, where $\Rs^0=\{\emptyset\}$ represents the 
state where no particle is in the system.
 
The evolution of the system is governed by three separate random processes. 
First, at exponentially distributed times a particle is added to the system with 
a velocity randomly chosen from a Maxwellian distribution at temperature $T$. To 
simplify notation we chose $T^{-1}=2\pi$. Second, also at exponentially 
distributed times, a particle is chosen at random to exit the system and 
disappear forever with no chance of reentry. Finally, a pair of particles in 
the system is selected at random to undergo a standard Kac collision.

More precisely, let $L^1_s({\mathcal R})=\bigoplus_{N=0}^\infty L^1_s(\Rs^N)$ 
be the Banach space of all states $\bff=(f_N)_{N=0}^\infty$, with 
$f_N(\uv_N)$ symmetric under permutation of the $v_i$, defined by the 
norm $\|\bff\|_1:=\sum_N\|f_N\|_{1,N}$, where $\|f_N\|_{1,N}=\int 
d\uv_N|f_N(\uv_N)|$. We say that $\bff$ is positive if $f_N(\uv_N)\geq 0$ for 
every $N$ and almost every $\uv_N$. If $\bff$ is positive and $\|\bff\|_1=1$ 
then $\bff$ is a probability distribution on $\mathcal R$. In this case, for 
$N>0$, $f_N(\uv_N)$ represents the probability of finding $N$ particles in the 
system with velocities ${\uv}_N=(v_1,\dots,v_N)$ while $f_0\in \Rs$ is the 
probability that the system contains no particle. 

The master equation for the evolution is given by
\begin{equation}\label{master}
 \frac{d}{dt}\bff=\mathcal{L} [\bff]:=\mu (\I[\bff]-\bff) +\rho (\cO [\bff]- \N 
[\bff])+\tilde\lambda \K[\bff]
\end{equation}
where $\I$ is the {\it in} operator that represents the effect of introducing a 
particle into the system and, after symmetrization, is given by
\begin{equation}\label{eq:in}
(\I \bff)_{N}(\uv)=\frac{1}{N}\sum_{i=1}^N e^{-\pi
v_i^2}f_{N-1}(v_1,\dots,v_{i-1},v_{i+1},\ldots, v_N)
\end{equation}
while $\cO$ is the {\it out} operator that represents the effect of a random particle
leaving the system
\begin{equation}\label{eq:out}
(\cO \bff)_{N}(\uv)=\sum_{i=1}^{N+1}\int dw f_{N+1}(v_1,\ldots, 
v_{i-1},w,v_i,\ldots,v_N)
\end{equation}
and 
\[
 (\N \bff)_{N}(\uv)=Nf_N(v_1,\ldots, v_N)\, .
\]
Observe that, due to the symmetry of $f_{N+1}$, we can write
\[
 (\cO \bff)_{N}(\uv_N)=(N+1)\int dv_{N+1} f_{N+1}(\uv_{N+1})\, .
\]
We also define the {\it thermostat} operator $\T$ as
\begin{equation}\label{ther}
\T:=\mu(\I-{\rm Id}) +\rho (\cO - \N)\,.
\end{equation}
These definitions imply that, in every time interval $dt$, there is a 
probability $\mu dt$ of a particle being added to the system. This probability 
is independent of the number of particles already in the system. In the same 
time interval, every particle in the system has a probability $\rho dt$ of 
leaving the system, which is, again, independent of the number of particles in 
the system. Thus, as discussed at the beginning of this section, the outflow of 
particles is proportional to $N$ while the inflow does not depend on $N$.

Finally $\K$ represents the effect of the collisions among particles. It acts 
independently on each of the $N$ particles subspaces, that is  it is $(\K 
\bff)_N=K_N f_N$ with
\begin{equation}\label{Kcol}
K_N f_N:=\sum_{1\leq i<j\leq N}(R_{i,j}-{\rm 
Id})f_N:=Q_Nf_N-\binom{N}{2}f_N
\end{equation}
where $R_{i,j}$ represents the effect of a collision between particles $i$ and 
$j$:
\begin{equation}\label{rot}
(R_{i,j}f_N)(\uv_N)=\frac{1}{2\pi}\int
f_N(\dots,v_i\cos\theta-v_j\sin\theta,\dots,v_i\sin\theta+v_j\cos\theta,
\dots )d\theta\, ,
\end{equation}
that is, $R_{i,j}f_N$ is the average of $f_N$ over all rotations in the plane 
$(v_i,v_j)$.
In this way, the probability that two given particles suffer a collision in an 
interval $dt$ is proportional to $\tilde \lambda$ and does not depend on the 
number of particles in the system. 

Since $\mathcal L$ is a sum of unbounded operators that do not commute, we 
first need to show that \eqref{master} defines an evolution on $L^1_s(\R)$ 
and that such an evolution preserves probability distributions. Observe that, 
notwithstanding $\mathcal L$ is unbounded, the operator $\mathcal 
L_N\bff$, defined by $\mathcal L_N\bff:=(\mathcal L\bff)_N$, is bounded as an 
operator from $L_s^1(\R)$ to $L^1_s(\Rs^N)$ with $\|\mathcal L_N\|_{1,N}\leq 
2\mu+(2N+1)\rho+\tilde\lambda N^2$. Thus we will take 
$D^1=\{\bff \;|\,\sum_N N^2\|f_N\|_{1,N}<\infty\}$ as the domain 
of $\mathcal L$. It is easy to see that $D^1$ is dense in $L_s^1(\R)$.

In section \ref{sec:3.L1} we will build a semigroup of continuous operators 
$e^{t\mathcal{L}}$ that solves \eqref{master} for initial data $\bff\in 
D^1$ and show that $e^{t\mathcal{L}}$ preserves probability distributions. 

\begin{lem}\label{lem:L1}
There exists a semigroup of continuous operators $e^{t\mathcal{L}}$ such that
if $\bff\in D^1$ then $\bff(t)=e^{t\mathcal{L}}\bff$ solves \eqref{master}. For
every $\bff\in L^1_s(\R)$ we have
\[
 \|e^{t\mathcal{L}}\bff\|_1\leq \|\bff\|_1\, .
\] 
Moreover, if\, $\bff$ is positive so is $e^{t\mathcal{L}}\bff$ and 
$\|e^{t\mathcal{L}}\bff\|_1=\|\bff\|_1$. Thus \eqref{master} generates an 
evolution that preserves probability distributions.
\end{lem}
\noindent\emph{Proof.} See section \ref{sec:3.L1}.
\medskip

It is not hard to see that the evolution generated by \eqref{master} admits the 
steady state $\boldsymbol \Gamma$ given by
\begin{equation}\label{aNgN}
 (\boldsymbol \Gamma)_N(\uv_N)=\left(\frac{\mu}{\rho}\right)^N\frac{e^{-\frac{\mu}{\rho}}}{N!}e^{-\pi|\uv_N|^2}:=a_N\gamma_N(\uv_N)
\end{equation}
where $\gamma_N(\uv_N)=\prod_{i=1}^N\gamma(v_i)$, with $\gamma(v)=e^{-\pi 
v^2}$, is the Maxwellian distribution with $\beta=2\pi$ in dimension $N$ 
while 
$a_N=\left(\frac{\mu}{\rho}\right)^N\frac{e^{-\frac{\mu}{\rho}}}{N!}$ is a 
Poisson distribution on $\mathbb{N}$. We observe that $\boldsymbol 
\Gamma$ is a Grand Canonical Ensemble with temperature $T=\beta^{-1}=1/2\pi$, 
chemical potential $\chi=(2\pi)^{-1} \log(\rho/\mu)$, and average number of 
particles $\langle \N\boldsymbol\Gamma\rangle=\mu/\rho$ where
\[
 \langle \N\boldsymbol\bff\rangle:=\sum_{N=0}^\infty N\int f_N(\uv_N)d\uv_N\, .
\]
In section \ref{sec:3.L1} we show that $\boldsymbol \Gamma$ is the unique 
steady state of the evolution generated by \eqref{master}. Finally, from a 
physical point of view, it is natural to consider only initial states with 
finite average numebr of particle and average kinetic energy, that is 
probability distributions $\bff$ such that
\begin{equation}\label{finiteNE}
 \langle \N\bff\rangle<\infty,\quad\hbox{and}\qquad \langle 
\mathcal 
E\bff\rangle:=
 \sum_{N=0}^\infty \frac12\int\bigl(\sum_i v_i^2\bigr) 
f_N(\uv_N)d\uv_N<\infty\, .
\end{equation}
%
Since the Kac collision operator $\K$ preserves energy 
and number of particles we can derive 
autonomous equations for the evolutions 
of $N(t)=\langle \N \bff(t)\rangle$ and $E(t)=\langle \mathcal E 
\bff(t)\rangle$. 
Indeed, if $\bff$ is a probability distribution, we obtain
\begin{equation}\label{eveq}
 \begin{aligned}
  \frac{d}{dt}N(t)&=\mu-\rho N(t)\\
  \frac{d}{dt}E(t)&=\frac{\mu}{2\pi}-\rho E(t)
 \end{aligned}
\end{equation}
so that, if \eqref{finiteNE} holds at time $t=0$ it holds for every time $t>0$.
See Section \ref{sec:3.L1} for a derivation of these equations.
Letting $e(t)=E(t)/N(t)$, we get
\begin{equation}\label{Newton}
\frac{d}{dt}e(t)=\frac{\mu}{N(t)}\left(\frac{1}{2\pi}- e(t)\right)\,. 
\end{equation}
Eq. \eqref{Newton} looks like Newton law of cooling for a system like ours.
Notwithstanding this, $e(t)$ is not the natural definition of temperature since 
it is not the average kinetic energy per 
particle. A more interesting quantity is $\tilde e(t)=\langle 
v_1^2\bff\rangle$, but 
we were not able to obtain a closed form expression for its evolution. 

As discussed in the introduction, we are interested in properties that are 
uniform in the average number of particles in the steady state $\langle 
\N\boldsymbol\Gamma\rangle=\mu/\rho$ and eventually we want to consider the 
situation where the average number of particles goes to infinity, that is 
$\mu/\rho\to\infty$. A classical way to take such a limit is to require that the 
collision rate between particles decreases as the average number of 
particles increases in such a way that the average number of 
collisions a given particle suffers in a given time is independent from 
$\mu/\rho$, at least when $\mu/\rho$ is large. This is achieved by setting
\[
 \tilde\lambda=\lambda\frac{\rho}{\mu}\, .
\]
Observe that in this way, the scaling in $N$ of $K_N$ in \eqref{Kcol}  
differs from the scaling in the standard Kac model. Notwithstanding this, 
they can both be thought as implementations of the Grad-Boltzmann limit in the 
two different situations, see \cite{Grad}.

One way to study the approach of an initial state $\bff$ toward $\boldsymbol \Gamma$ is
by computing the spectral gap of $\mathcal{L}$. Since $\mathcal{L}$ is not self 
adjoint on
$L^2_s(\R)$ we perform a ground state transformation setting
\begin{equation}\label{ground} 
 f_N:=a_N\gamma_Nh_N\,. 
\end{equation}
We will express \eqref{ground} as $\bff=\boldsymbol \Gamma \bh$. Inserting the 
above definition in \eqref{master} we get
\begin{equation*}
\frac{d}{dt}\bh=\widetilde{\mathcal{L}} \mathbf h:=\rho(\P^+\bh-\N\bh)+\mu(\P^-\bh-\bh)+\tilde\lambda\K\bh
\end{equation*}
where we have set
\[
\begin{aligned}
&(\P^+\bh)_{N}=\sum_{i=1}^Nh_{N-1}(v_1,\dots,v_{i-1},v_{i+1},\dots,v_N)\\
&(\P^-\bh)_{N}=\frac{1}{N+1}\sum_{i=1}^{N+1}\int dwe^{-\pi 
w^2}h_{N+1}(v_1,\dots,v_{i-1},w,v_i,\dots,v_N)
\end{aligned}
\]
In this representation, the steady state is given by the vector $\be^0$ such 
that $(\be^0)_N\equiv 1$ for every $N$.
Thus $\widetilde{\mathcal{L}}$ is an unbounded operator on the Hilbert space
\[
L^2_s(\R,\boldsymbol \Gamma)=\bigoplus_{N=0}^\infty 
L^2_s(\Rs^N,a_N\gamma_N(\uv_N))
\]
of all states $\bh=(h_0,h_1,h_2,\ldots)$ with $h_N(\uv_N)$ symmetric under 
permutations of the $v_i$ and defined by the scalar product
\[
 (\bh_1,\bh_2):=\sum_{N=0}^\infty a_N (h_{1,N},h_{2,N})_N:=\sum_{N=0}^\infty 
a_N\int
h_{1,N}(\uv_N)h_{2,N}(\uv_N)\gamma_N(\uv_N)d\uv_N\, .
\]
As for $\mathcal L$, defining $\widetilde{\mathcal{L}}_M\bh=(\mathcal L\bh)_M$ 
we get a bounded operator from 
$L^2_s(\R,\boldsymbol \Gamma)$ to $L^2_s(\Rs^N,\gamma_N(\uv_N))$ so that, 
calling $\|h_N\|_{2,N}=(h_{N},h_{N})_N$, we can take 
\[
D^2=\bigl\{\bh\,\big|\,\sum_{N=0}^\infty 
a_N\|(\widetilde{\mathcal{L}}\bh)_N\|_{2,N}<\infty\bigr\}
\]
as the domain of 
$\widetilde{\mathcal{L}}$. The following Theorem shows that 
$\widetilde{\mathcal{L}}$ defines an evolution 
on $L^2_s(\R,\boldsymbol \Gamma)$.

\begin{thm}\label{thm:gap}
The generator $\widetilde{\mathcal{L}}$ is self adjoint and non-positive 
definite on
$L^2_s(\R,\boldsymbol \Gamma)$. Furthermore, if we define
\[
 \Delta=\sup \{(\bh,\widetilde {\mathcal{L}}\bh)\,|\, \bh\in D^2, \|\bh\|_2=1, 
\bh\perp\mathbf E_0\}
\]
where $\|\bh\|_2=(\bh,\bh)$ and $\mathbf E_0=\mathrm{span}\{\be^0\}$, we get
\[
 \Delta=-\rho\, .
\]
Moreover $\Delta$ is an eigenvalue and the associated eigenspace is $\mathbf 
E_1=\mathrm{span}\{\be_1,\be_{(0,0,1)}\}$ with 
$\be_1=\sqrt{\frac{\rho}{\mu}}\P^+\be^0-\sqrt{\frac{\mu}{\rho}}\be^0$ while
\[
(\be_{(0,0,1)})_N(\uv_N)= \sqrt{\frac{\rho}{2\mu}}\sum_{i=1}^N(2\pi v_i^2-1) \, 
.
\]
\end{thm}
\noindent\emph{Proof.} See section \ref{sec:3.1}.
\medskip

Due to the invariance of even, second degree polynomials under the Kac 
collision 
operator $\K$, Theorem \ref{thm:gap} shows that the spectral gap of the 
generator $\widetilde{\mathcal{L}}$ is completely determined by the presence of 
the reservoir. This is not surprising since all states $\bh$ such that $h_N$ is 
rotationally invariant for every $N$ are in the null space of $\K$.

As in \cite{BLV}, to see the effect of the Kac collision operator $\K$, we 
have to look at the second gap, defined as
\begin{equation}\label{Delta2}
 \Delta_2=\sup \{(\bh,\widetilde{\mathcal{L}}\bh)\,|\, \bh\in D^2, \|\bh\|_2=1, 
\bh\perp\mathbf E_0\oplus \mathbf E_1\}\, .
\end{equation}
 
\begin{thm}\label{thm:gap2}
If
\begin{equation}\label{cond}
\rho>\f{\lambda}4+2\lambda\sqrt{\f{\rho}{\mu}}\quad\mathrm{and}\quad\f{\mu}{\rho
}>256
\end{equation}
we have
 \begin{equation*}
-\rho-\frac{\lambda}{4}\leq\Delta_2<-\rho-\frac{\lambda}{4}+2\lambda\sqrt{\frac{
\rho}{\mu}}\, .
 \end{equation*}
Moreover $\Delta_2$ is an eigenvalue and the associated eigenspace is 
contained in 
the space of all states $\bh$ such that $h_N$ is an even, 
fourth degree polynomial.
\end{thm}
\noindent\emph{Proof.} See section \ref{sec:3.2}.
\medskip

Since $\mu/\rho$ is the average number of particles in the steady state, the 
conditions in \eqref{cond} are not too restrictive. 

It is possible to see that, as in the case of the standard Kac evolution, the 
$L^2$ norm discussed above does not scale well with the average number of 
particles in the system and thus it is not a good measure of distance from the 
steady state if $\mu/\rho$ is large. A better measure is the entropy of a 
probability distribution $\bff$ relative to the steady state 
$\boldsymbol\Gamma$ defined as
\[
 \mathcal S(\bff\,|\,\boldsymbol \Gamma)=\sum_Na_N\int dv_N h_N(\underline v_N)\log h_N(\underline 
v_N)\gamma_N(\underline v_N)
\]
where, as before, $\bff=\boldsymbol \Gamma\bh$ and $a_N$ and $\gamma_N$ are defined in \eqref{ground}.

As usual, it is easy to show using convexity that $\mathcal 
S(\bff\,|\,\boldsymbol \Gamma)\geq 0$, $\mathcal S(\bff\,|\,\boldsymbol 
\Gamma)=0$ if and only if $\bff=\boldsymbol \Gamma$. Moreover, from Lemma 
\ref{lem:L1} and convexity, it follows that $\mathcal S(\bff(t)\,|\,\boldsymbol 
\Gamma)\leq \mathcal S(\bff\,|\,\boldsymbol \Gamma)$ where 
$\bff(t)=e^{t\mathcal{L}}\bff$. In section \ref{sec:3.3}, we show that, thanks 
to the presence of the reservoir, the entropy production rate is strictly 
negative. More precisely, assuming that $\bff=\boldsymbol\Gamma \bh\in D^1$ and 
$\boldsymbol\Gamma \bh\log\bh\in D^1$
we essentially obtain that
\begin{equation}\label{decent}
 \frac{d}{dt}\mathcal S(\bff(t)\,|\,\boldsymbol \Gamma)\leq -\rho \mathcal
 S(\bff(t)\,|\,\boldsymbol \Gamma)\, .
\end{equation}
See Lemma \ref{lem:d+} and \ref{lem:ent} in section \ref{sec:3.3} below for a 
precise statement.
Form \eqref{decent} we obtain the following Theorem.

\begin{thm}\label{thm:entropy}
 If $\bff=\bh\boldsymbol\Gamma\in D^1$ is a probability distribution such that 
$\boldsymbol\Gamma\bh\log\bh\in D^1$
then
\begin{equation}\label{decay}
 \mathcal S(\bff(t)\,|\,\boldsymbol \Gamma)\leq e^{-\rho t}\mathcal S(\bff(0)\,|\,\boldsymbol \Gamma)\, .
\end{equation}
\end{thm}
\noindent\emph{Proof.} See section \ref{sec:3.3}. 
\medskip

As in the case of Theorem \ref{thm:gap}, convergence to equilibrium in entropy 
is completely dominated by the presence of the thermostat, that is, 
Theorem \ref{thm:entropy} remains valid in the case 
$\tilde\lambda=0$ where there is no 
collision among the particles.

We can now discuss the validity of a Boltzmann-Kac type equation when the 
average number of particles in the system goes to infinity. To follow the 
standard analysis in \cite{McK}, we have first to define what is a {\it chaotic 
sequence} in the present situation. It is natural to call 
$\bff=(f_0,f_1,f_2,\ldots)$ a {\it product state} if it has the form
\begin{equation}\label{prod}
 f_N(\uv_N)=e^{-\eta}\frac{\eta^N}{N!}\prod_{i=1}^Ng(v_i)
\end{equation}
where $g(v)$ is a probability density on $\mathbb{R}$ and $\eta>0$ is the 
average number of particles. 
We observe that for the state $\bff$ in \eqref{prod}, we have 
\begin{equation}\label{prodt}
\left( 
e^{t\T}\bff\right)_N=e^{-\eta(t)}\frac{\eta(t)^N}{N!}\prod_{i=1}^Ng(v_i,t)
\end{equation}
where $\T$ is defined in \eqref{ther} and, calling 
$l(v,t)=\frac{\rho}{\mu}\eta(t)g(v,t)$, we get
\begin{equation}\label{geta}
\begin{aligned}
 \eta(t)=&e^{-\rho t}\eta+(1-e^{-\rho t})\frac{\mu}{\rho}\\
l(v,t)=&e^{-\rho t}l(v)+(1-e^{-\rho t})\gamma(v)
\end{aligned}
\end{equation}
This implies that the thermostat preserves the product structure exactly. See 
Section \ref{app:bosons.eveq} for a derivation of \eqref{prodt} and 
\eqref{geta}.

Thus we call a sequence of states $\bff_n=(f_{n,0},f_{n,1},f_{n,2},\ldots)$ 
chaotic if 
it
approaches the structure \eqref{prod} while the average number of particles
$\langle \N\bff_n\rangle$ goes to infinity. More precisely, let $\mu_n$ be a
sequence such that $\lim_{n\to\infty}\mu_n=\infty$ and define
\begin{equation}\label{marg}
F_n^{(k)}(\uv_k)=\left(\frac{\rho}{\mu_n}\right)^{k}\sum_{N\geq 
k}\frac{N!}{(N-k) !}\int f_{n,N}(\uv_k,\uv_{N-k})d\uv_{N-k}
\end{equation}
where the factor $\frac{N!}{(N-k)!}$ accounts for the possible ways to choose 
the $k$ particles with velocities $\uv_k$. We also define 
\begin{equation}
 \|\bff\|_1^{(k)}=\sum_{N\geq k}\frac{N!}{(N-k)!}\|f_N\|_{1,N}
\end{equation}
so that $\|F_n^{(k)}\|_{1,k}\leq 
\left(\frac{\rho}{\mu_n}\right)^k\|\bff_n\|_1^{(k)}$. 

Observe that, if $\bff_n$ is a 
product state of the form \eqref{prod} with average number of particles 
$\eta_n$, that is if
\[
 f_{n,N}(\uv_N)=e^{-\eta_n}\frac{\eta_n^N}{N!}\prod_{i=1}^Ng(v_i)
\]
we get
\[
F_n^{(k)}(\uv_k)=\left(\frac{\eta_n\rho}{\mu_n}\right)^{k}
\prod_{i=1}^kg(v_i)\, .
\]
Thus the factor $\left(\frac{\rho}{\mu_n}\right)^{k}$ in \eqref{marg} assures 
that, at least in this case, if $\lim_{n\to\infty}\eta_n/\mu_n$ exists then 
also $\lim_{n\to\infty} F_n^{(k)}$ exists.

To generalize these observations, we say that $F_n^{(k)}$ converges weakly to 
$F^{(k)}$ if, for any continuous and bounded test function $\phi_k:\Rs^k\to\Rs$, 
we have 
\begin{equation*}
\lim_{n\to\infty}\int_{\Rs^k}F_n^{(k)}(\uv_k)\phi_k(\uv_k)d\uv_k=
\int_{\Rs^k}F^{(k)}(\uv_k)\phi_k(\uv_k)d\uv_k
\end{equation*}
and we write $\wlim_{n\to\infty}F_n^{(k)}=F^{(k)}$.
Given a sequence $\bff_n$ of probability distributions such 
that
\begin{equation}\label{bound1r}
 \|\bff_n\|_1^{(r)}\leq M^r\left(\frac{\mu_n}{\rho}\right)^r
\end{equation}
for some $M>0$ and every $n$ and $r$, we say that $\bff_n$ is {\it chaotic 
(w.r.t. $\mu_n$)} if, for some $F$ 
\begin{equation}\label{reld}
 \wlim_{n\to\infty}F^{(1)}_n=F
\end{equation}
while for every $k>1$ we have 
\begin{equation}\label{margprod}
\wlim_{n\to\infty}F_n^{(k)}=F^{\otimes k} 
\end{equation}
where $F^{\otimes k}(\uv_k)=\prod_{i=1}^k F(v_i)$.
Observe that 
\begin{equation}\label{relden}
 \int F(v)dv=\lim_{n\to\infty}\frac{\langle \N\bff_n\rangle\rho}{\mu_n}
\end{equation}
so that we can see  $F(v)$ as the {\it relative particle density}. 

In \cite{kac,McK} a sequence of probability distributions
$f_n:\mathbb{R}^n\to\mathbb{R}$ is said to be chaotic if, calling
\[
 \widetilde F_n^{(k)}(\uv_k)=\int f_{n}(\uv_k,\uv_{n-k})d\uv_{n-k}\, ,
\]
we have
\[
 \wlim_{n\to\infty}\widetilde F^{(1)}_n=\widetilde F\qquad\hbox{and}\qquad
 \wlim_{n\to\infty}\widetilde F_n^{(k)}=\widetilde F^{\otimes k}\, .
\]
If we consider the sequence of states $\bff_n$ defined as
\begin{equation*}
 (\bff_n)_N=\begin{cases}
             f_n & n=N\\
             0 & n\not = N
            \end{cases}
\end{equation*}
with the natural choice $\mu_n=n\rho$, since the number of particle in 
$\bff_n$ is exactly 
$n$, from \eqref{marg} we get $F=\widetilde F$ and thus $F^{(k)}=\widetilde 
F^{(k)}$. In this sense, \eqref{marg} and \eqref{margprod} can be 
considered as a 
generalization of the classical definition in \cite{kac}.

Let now 
\[
 \bff_n(t)=e^{\mathcal L_n t}\bff_n(0)
\]
where $\mathcal L_n$ is given by \eqref{master} with $\mu=\mu_n$ and
\begin{equation}\label{tl}
 \tilde\lambda=\tilde\lambda_n=\lambda\frac{\rho}{\mu_n}\, .
\end{equation}
In section
\ref{sec:3.4}, we prove that $e^{\mathcal L_n t}$ propagates chaos in the sense 
that, if $\bff_n(0)$ forms a chaotic sequence, then $\bff_n(t)$ also
forms a chaotic sequence for every $t$. This gives the following theorem.

\begin{thm}\label{thm:chaos}
 If\, $\bff_n(0)$ forms a chaotic sequence w.r.t. $\mu_n$, with 
$\lim_{n\to\infty}\mu_n=\infty$, then also $\bff_n(t)$ forms a chaotic 
sequence for every $t\geq 0$. Moreover the relative particle density 
 \[
F(v,t)=\wlim_{n\to\infty} \frac{\rho}{\mu_n}\sum_{N=1}^\infty N\int 
f_{n,N}(v,\uv_{N-1},t)d\uv_{N-1}
\]
satisfies the Boltzmann-Kac type equation
\begin{align}\label{BK}
\frac{d}{dt}&F(v,t)= -\rho(F(v,t)-\gamma(v))\\
&+\lambda\int_\Rs dw\int \frac{d\theta}{2\pi} 
[F(v\cos\theta+w\sin\theta,t)F(-v\sin\theta+w\cos\theta,t)-
F(w,t)F(v,t)]\, .\nonumber
\end{align}
\end{thm}
\noindent\emph{Proof.} See section \ref{sec:3.4}.
\medskip

\section{Proofs.}
\label{sec:3}

\subsection{Proof of Lemma \ref{lem:L1}.}
\label{sec:3.L1}

The results in this section are based on two observations. The first is 
that the collision operator $\K$ acts independently on each $L^1_s(\mathbb 
R^N)$ and thus preserves positivity and probability. The second is that, 
due to the different scaling in $N$ of the {\it in} and {\it out} operators, 
see \eqref{eq:in} and \eqref{eq:out}, for large $N$ the outflow of particles 
dominates the inflow. Thus even if the initial probability of having a number 
of particles much larger than the steady state average $\mu/\rho$ is high, this 
probability will rapidly decrease toward its steady state value, see 
\eqref{expa1} and \eqref{Nt1} below. In particular this prevents probability 
from ``leaking out at infinity''.

We will now construct a solution of \eqref{master} in three steps, starting 
from $\K$ alone, using a partial power series expansion, see \eqref{expaK} 
below, and then adding the {\it out} operator $\cO$ and finally the {\it in} 
operator $\I$, using a Duhamel style expansions, see \eqref{expa} and 
\eqref{expaI} below. These expansions are strongly inspired by the stochastic 
nature of the the evolution studied, see Remark \ref{rem:jumps} below for more 
details.

It is natural to define $\left(e^{t\tilde\lambda 
\K}\bff\right)_N=e^{t\tilde\lambda K_N}f_N$ where we can write
\begin{equation}\label{expaK}
 e^{t\tilde\lambda K_N}f_N=e^{-\tilde\lambda t\binom N2}\sum_{n=0}^\infty
\frac{\tilde\lambda^nt^n Q_N^n}{n!}f_N\, .
\end{equation}
Observing that
\begin{equation}\label{KN0}
\begin{aligned}
\| e^{t\tilde\lambda K_N}f_N-f_N\|_{1,N}\leq& \left(1-e^{-\tilde\lambda t\binom 
N2}\right)\|f_N\|_{1,N}+\Bigl\|e^{-\tilde\lambda t\binom N2}\sum_{n=1}^\infty
\frac{\tilde\lambda^nt^n Q_N^n}{n!}f_N\Bigr\|_{1,N}\\
\leq& 2\left(1-e^{-\tilde\lambda t\binom N2}\right)\|f_N\|_{1,N}
\end{aligned}
\end{equation}
and using that from Dominated Convergence we get
\[
\lim_{t\to 0^+}\sum_{N=0}^\infty\left(1-e^{-\tilde\lambda t\binom 
N2}\right)\|f_N\|_{1,N}=0
\]
we obtain that $\lim_{t\to0^+}e^{t\tilde\lambda \K}\bff=\bff$.
Similarly, we get
\[
\begin{aligned}
 &\frac 1t\| e^{t\tilde\lambda K_N}f_N-f_N-\tilde\lambda t K_Nf_N\|_{1,N}\\
 &\qquad\leq\frac 1t\left(e^{-\tilde\lambda t\binom 
N2}-1+\tilde\lambda t\binom N2\right)\|f_N\|_{1,N}+\left(1- e^{-\tilde\lambda 
t\binom N2}\right)\|\tilde\lambda Q_Nf_N\|_{1,N}\\
&\qquad\qquad\qquad\qquad\qquad\qquad\qquad\qquad\qquad
+\frac 1t\Bigl\|e^{-\tilde\lambda t\binom N2}\sum_{n=2}^\infty
\frac{\tilde\lambda^nt^n Q_N^n}{n!}f_N\Bigr\|_{1,N}
\\
&\qquad\leq \frac 2t\left(e^{-\tilde\lambda t\binom N2}-1+\tilde\lambda 
t\binom N2\right)\|f_N\|_{1,N}+\tilde\lambda 
\binom N2\left(1- 
e^{-\tilde\lambda t\binom N2}\right)\|f_N\|_{1,N}
\end{aligned}
\]
so that, if $\bff\in D_1$ then 
$\lim_{t\to0^+}\left(e^{t\tilde\lambda \K}\bff-\bff\right)/t=
\tilde\lambda\K\bff$.
Since $\|e^{t\tilde\lambda K_N}f_N\|_{1,N}\leq \|f_N\|_{1,N}$  we get 
$\|e^{t\tilde\lambda\K}\bff\|_1\leq \|\bff\|_1$. Moreover 
if $f_N$ is positive then also $e^{t\tilde\lambda K_N}f_N$ is 
positive and $\|e^{t\tilde\lambda K_N}f_N\|_{1,N}= \|f_N\|_{1,N}$. 
Thus if $\bff$ is positive then $e^{t\tilde\lambda\K}\bff$ is positive and 
$\|e^{t\tilde\lambda\K}\bff\|_1= \|\bff\|_1$. 

Let now $\bff(t)$ be a solution of
\begin{equation}\label{soloO}
\frac{d}{dt} \bff(t)=\tilde\lambda 
\K\bff(t)+\rho(\cO-\N)\bff(t)
\end{equation}
with $\bff(0)=\bff\in D^1$. If such a solution exists, it satisfies 
the Duhamel formula
\begin{equation}\label{Dua}
f_N(t)=e^{(\tilde\lambda 
K_N-\rho N)t}f_N+\rho\int_0^t e^{(\tilde\lambda 
K_N-\rho N)(t-s)}  \left(\cO \bff(s)\right)_N\,ds
\end{equation}
where the construction of $e^{(\tilde\lambda \K-\rho\N)t}$ is analogous to that 
of $e^{\tilde\lambda \K t}$.
From \eqref{Dua} we get
\begin{equation}\label{Dua1} 
\left\|f_N(t)\right\|_{1,N}\leq e^{-\rho Nt}\|f_N\|_{1,N}
+\int_0^t 
e^{-\rho N(t-s)} \rho (N+1)\left\|f_{N+1}(s)\right\|_{1,N+1}ds
\end{equation}
where we have used that
\begin{equation}\label{Opos}
\|(\mathcal O\bff)_N\|_{1,N}=(N+1)\int  \left |\int  
f_{N+1}(\uv_{N+1})dv_{N+1}\right|d\uv_N\leq (N+1)\|f_{N+1}\|_{1,N+1}\,.
\end{equation}
Observe that, in \eqref{Opos}, equality holds if and only if $f_{N+1}$ is 
everywhere positive or everywhere negative. To construct a solution of 
\eqref{soloO} we iterate \eqref{Dua} to define \begin{align}\label{expa}
\mathcal Q(t)\bff=&e^{(\tilde\lambda 
\K-\rho\N)t}\bff\\
+&\sum_{n=1}^\infty\int\limits_{0<t_1<\ldots<t_n<t} e^{(\tilde\lambda 
\K-\rho\N)(t-t_n)}  \rho\cO e^{(\tilde\lambda 
\K-\rho\N)(t_n-t_{n-1})} \nonumber\\
&\phantom{\sum_{n=1}^\infty\int\limits_{0<t_1<\ldots<t_n<t} 
e^{(\tilde\lambda 
\K-\rho\N)(t-t_n)}}\cdots
\rho\cO e^{(\tilde\lambda 
\K-\rho\N)t_1}\bff\,dt_1\cdots dt_n\nonumber
\end{align}
and then show that $\mathcal Q(t)$ is a semigroup of bounded operators and 
that $\bff(t)=\mathcal Q(t)\bff$ solves \eqref{soloO} if $\bff\in D^1$.
Using \eqref{Dua1} iteratively we get
\begin{align}\label{expa1}
\Bigl\|\left(\mathcal Q(t)\bff\right)_N\Bigr\|_{1,N} \leq&\sum_{n\geq 
0}e^{-\rho N 
t}\frac{(N+n)!}{N!}\int_{0<t_1<\cdots<t_n<t}\\
&\prod_{i=1}^{n}e^{\rho 
(N+n-i)t_{i}}\rho 
e^{-\rho (N+n-i+1)t_{i}}dt_1\cdots dt_n\|f_{N+n}\|_{1,N+n}\nonumber\\
 =&e^{-\rho N 
t}\sum_{n\geq 0}\binom{N+n}{N}\left(1-e^{-\rho 
t}\right)^{n}\|f_{N+n}\|_{1,N+n}\nonumber
\end{align}
where, in the last identity, we have used that
\begin{equation}\label{1me}
 \rho^n\int_{0\leq t_1\leq\cdots t_n\leq t} 
\prod_{i=1}^n e^{-\rho t_{i}}dt_1\cdots dt_n
=\frac{1}{n!}(1-e^{-\rho t})^n\, .
\end{equation}
After summing over $N$ we get
\begin{equation} 
\begin{aligned}\label{expa2}
 \|\mathcal Q(t)\bff\|_1\leq\sum_{N\geq 0} 
\sum_{n\geq 0} \binom{N+n}{N}e^{-\rho N 
t}\left(1-e^{-\rho t}\right)^{n}\|f_{N+n}\|_{1,N+n}=
\|\bff\|_1\, .
\end{aligned} 
\end{equation}
so that $\|\mathcal Q(t)\|_1\leq 1$. 
Observe also that, if $\bff$ is positive then $\mathcal Q(t)\bff$ is positive 
and $\|\mathcal Q(t)\bff\|_1=\|\bff\|_{1}$, see comment below \eqref{Opos}. 
Conversely, if for some $N$, $f_N$ takes both positive and 
negative values then $\|\mathcal Q(t)\bff\|_1<\|\bff\|_{1}$. 

From \eqref{expa}, we see that $\mathcal Q(t_1)\mathcal Q(t_2)=
\mathcal Q(t_1+t_2)$ while, using \eqref{expa1} and \eqref{expa2}, and the 
fact that
\[
 N(N-1)\binom{M}{N}=M(M-1)\binom{M-2}{N-2}
\]
we get 
\[
 \sum_{N=1}^\infty N^2 \left\|\left(\mathcal 
Q(t)\bff\right)_N\right\|_{1,N}\leq e^{-\rho t}\sum_{N=1}^\infty 
N^2\|f_N\|_{1,N}
\]
so that $\mathcal Q(t)\bff\in D^1$ if $\bff\in D^1$. Moreover observe that
\begin{align} \label{Qt0}
 \|\mathcal Q(t)\bff-\bff\|_1\leq&\sum_{N\geq 0} 
\sum_{n\geq 1} \binom{N+n}{N}e^{-\rho N 
t}\left(1-e^{-\rho t}\right)^{n}\|f_{N+n}\|_{1,N+n}\nonumber\\
+&\Bigl\|e^{(\tilde\lambda\K-\rho\N)t}\bff
-\bff\Bigr\|_1\\
=&\sum_{N\geq 0}\left(1-e^{-\rho N t}\right)\|f_N\|_{1,N}+
\Bigl\|e^{(\tilde\lambda\K-\rho\N)t}\bff
-\bff\Bigr\|_1\nonumber
\end{align}
so that $\lim_{t\to 0^+}\mathcal Q(t)\bff=\bff$.
Similarly we have
\begin{align} \label{dQt}
\frac{1}{t}\|\mathcal 
Q(t)\bff-\bff-t(\tilde\lambda\K-\rho(\cO-&\N))\bff\|_1\nonumber\\
\leq&\frac{1}{t}\sum_{N\geq 0} 
\sum_{n\geq 2} \binom{N+n}{N}e^{-\rho N 
t}\left(1-e^{-\rho t}\right)^{n}\|f_{N+n}\|_{1,N+n}\nonumber\\
+&\frac{1}{t}\left\|e^{(\tilde\lambda\K-\rho\N)t}-
\bff-t(\tilde\lambda\K-\rho\N)\bff\right\|_1\\
+&\frac{\rho}{t}\left\|\int_0^t e^{(\tilde\lambda\K-\rho\N)(t-s)}\cO
e^{(\tilde\lambda\K-\rho\N)s}\bff-t\cO\bff\right\|_1\nonumber\, .
\end{align}
If $\bff\in D^1$, proceeding as in \eqref{Qt0} we see that the second and third 
lines of the right hand side of \eqref{dQt} vanish as $t\to 0^+$ while writing
\begin{align}\label{diffdiff}
 \int_0^t e^{(\tilde\lambda\K-\rho\N)(t-s)}\cO
e^{(\tilde\lambda\K-\rho\N)s}\bff-t\cO\bff=&
\int_0^t e^{(\tilde\lambda\K-\rho\N)(t-s)}\cO\left(
e^{(\tilde\lambda\K-\rho\N)s}\bff-\bff\right)\nonumber\\
+&\int_0^t \left(e^{(\tilde\lambda\K-\rho\N)(t-s)}\cO -\cO\right)\bff
\end{align}
and using \eqref{KN0} we see that also the last line of \eqref{dQt} vanish as 
$t\to 0^+$. This implies that, for $\bff\in D^1$, we have $\lim_{t\to 0^+} 
\left(\mathcal Q(t)\bff-\bff\right)/t=\tilde\lambda\K\bff+\rho(\cO-\N)\bff$ 
and we can write
$\mathcal Q(t)=e^{t(\tilde\lambda \K+\rho(\cO-\N))}$.

We can now use a Duhamel style expansion once more to obtain
\begin{align}\label{expaI}
 e^{t{\mathcal L}}&\bff=e^{(\tilde\lambda 
\K+\rho(\cO-\N)-\mu\mathrm{Id})t}\bff\\
&+\sum_{n=1}^\infty\mu^n\int\limits_{0<t_1<\ldots<t_n<t} e^{(\tilde\lambda 
\K+\rho(\cO-\N)-\mu\mathrm{Id})(t-t_n)}  \I e^{(\tilde\lambda 
\K+\rho(\cO-\N)-\mu\mathrm{Id})(t_n-t_{n-1})} \nonumber\\
&\phantom{\sum_{n=1}^\infty\rho^n\int\limits_{0<t_1<\ldots<t_n<t} 
e^{(\tilde\lambda 
\K+\rho(\cO-\N))(t-t_n)}}\cdots
\I e^{(\tilde\lambda 
\K+\rho(\cO-\N)-\mu\mathrm{Id})t_1}\bff\,dt_1\cdots dt_n\nonumber
\end{align}
that, thanks to the fact that $\I$ is bounded, 
converges for every $\bff\in L^1(\R)$ to a solution of 
$\frac{d}{dt}\bff(t)=\mathcal L\bff(t)$. Lemma \ref{lem:L1} follows easily 
observing that $\|\I\bff\|_1= \|\bff\|_1$.\qed
\medskip

\begin{rem}\label{contra}\emph{The proof of Lemma \ref{lem:L1} above also shows 
that given  $\bff\in L^1(\R)$,
if for some $N$, $f_N$ takes both positive and negative values, then 
$\|e^{t\mathcal L}\bff\|_1<\|\bff\|_{1}$.
}\end{rem}

\begin{rem}\label{unique}\emph{From \eqref{Dua} it is not hard to see that, 
if $\bff_i(t)\in D^1$, $i=1,2$, are two solutions of \eqref{master} with 
$\bff_1(0)=\bff_2(0)$ then $\bff_1(t)=\bff_2(t)$.}
\end{rem}

\begin{rem}\label{rem:jumps}\emph{ Observe that \eqref{master} is the master 
equation of a 
jump process where jumps occur when two particles collide, a particle enters 
the system or a particle leaves it. Moreover, these jumps arrive according to a 
Poisson process. The expansions \eqref{expaK}, \eqref{expa} and \eqref{expa1} 
combined can be seen as a representation of the evolution of 
$\bff$ as an integral over all possible realizations of the 
jump process, sometime called {\it jump} or {\it collision histories}. A 
similar representation was used in \cite{BGLR} to study the 
interaction of a Kac system with a large reservoir. Clearly, such a 
representation is much more complex in the present situation then for the 
model studied in \cite{BGLR}. Here the arrival rate for the jumps depends on 
the state of the system via the number of particles $N$ and goes to infinity as 
$N$ increases.
} 
\end{rem}

Given a state $\bff=(f_0,f_1,\ldots)$ we set $\bar f_N=\int f_N(\uv_N)d\uv_N$. 
It is easy to see that
\begin{equation}\label{barf}
 \int (\cO \bff)_N(\uv_N)d\uv_N=(N+1)\bar f_{N+1}\, ,\qquad \int (\I 
\bff)_N(\uv_N)d\uv_N=\bar f_{N-1}
\end{equation}
while
\[
 \int (\K \bff)_N(\uv_N)d\uv_N=0\, ,
\]
so that we get 
\begin{align}\label{dotfN}
 \overline{({\mathcal L}\bff)_0}=&-\mu \bar f_0+\rho \bar 
f_1\\
 \overline{({\mathcal L}\bff)_N}=&-(N\rho+\mu)\bar f_N+\mu \bar 
f_{N-1}+\rho(N+1) \bar f_{N+1}& N>0\, .\nonumber
\end{align}
If $\boldsymbol \Gamma$ is a steady state, writing 
\[
\overline{\Gamma}_N=c_N \left(\frac\mu\rho\right)^N\frac{1}{N!}
\]
we see from  \eqref{dotfN} that $c_N=c_0$ for every $N$. 
Since $\sum_N\overline 
\Gamma_N=1$  we get 
$\overline{\Gamma}_N=a_N$, see \eqref{aNgN}. 
This implies that if $\boldsymbol \Gamma$ and $\boldsymbol \Gamma'$ are two 
steady states then
\[
 \int (\Gamma_N(\uv_N)-\Gamma'_N(\uv_N))d\uv_N=0
\]
for every $N$. From Remark \ref{contra} it follows that, if 
$\boldsymbol \Gamma\not=\boldsymbol 
\Gamma'$ then $\|e^{t\mathcal L}(\boldsymbol\Gamma-\boldsymbol\Gamma')\|_1<
\|\boldsymbol\Gamma-\boldsymbol\Gamma'\|_{1}$. Uniqueness of the steady state 
follows immediately.

We now prove a more general version of \eqref{eveq}. For 
$r\geq 0$ we define 
\begin{equation}\label{Nrdef}
 N_{r}(\bff)=\sum_{N=r}^\infty 
\frac{N!}{(N-r)!}\bar f_N\, .
\end{equation}
and, using \eqref{dotfN}, we get 
\begin{equation}\label{dotNr}
\begin{aligned}
 \frac{d}{dt} N_{r}(\bff)=&\sum_{N=r}^\infty 
\frac{N!}{(N-r)!}\left(-(N\rho+\mu)\bar f_N+ \mu \bar f_{N-1}+\rho (N+1)\bar 
f_{N+1}\right)=\\
&-\rho r N_r(\bff)+\mu rN_{r-1}(\bff)
\end{aligned}
\end{equation}
that, for $r=1$, would implies the first of \eqref{eveq} since for a 
probability distribution we have $N_0(\bff)=1$. 
This argument is suggestive but only formal since we need to show that we 
can exchange the sum with the derivative in the above derivation. 
Notwithstanding this, it shows that for $r=0$, if $\bff\in D^1$ then
\begin{equation}\label{sumdotf}
 \sum_{n=0}^\infty \overline{(\mathcal L)_N\bff}=0\, .
\end{equation}

To prove \eqref{eveq} we proceed more directly using the expansions derived 
previously. Indeed from \eqref{expa1} and \eqref{expa2} we get
\begin{equation}\label{Nt1}
\begin{aligned}
N_r  \left( e^{t(\tilde\lambda \K-\rho\N+\rho 
\cO)}\bff\right)=&\sum_{N\geq 0} 
\sum_{n\geq 0} \frac{N!}{(N-r)!}\binom{N+n}{N}e^{-\rho N 
t}\left(1-e^{-\rho t}\right)^{n}\bar f_{N+n}=\\
&e^{-\rho r t}N_r(\bff)\,\, . 
\end{aligned}
\end{equation}
Furthermore, using that  $N_r(\I\bff)=
 N_r(\bff)+r N_{r-1}(\bff)$, we get
\begin{equation}\label{Nt2}
 \begin{aligned}
  N_r\left(\bff(t)\right)=&N_r\left(e^{t(\tilde\lambda \K-\rho\N+\rho 
\cO-\mu\mathrm{Id})}\bff(0)+\mu\int_0^te^{(t-s)(\tilde\lambda 
\K-\rho\N+\rho\cO-\mu\mathrm{Id})}\I \bff(s)ds\right)\\
= & e^{-(\rho r+\mu)
 t}N_r(\bff(0))+\mu\int_0^t e^{-(\rho r+\mu)
 (t-s)}N_r(\bff(s))ds\\
 &\qquad +r\mu\int_0^t e^{-(\rho r+\mu)
 (t-s)}N_{r-1}(\bff(s))ds
 \end{aligned}
\end{equation}
that gives
\begin{equation}\label{Nr}
 N_r\left(\bff(t)\right)=e^{-\rho r
 t}N_r(\bff(0))+r\mu\int_0^t e^{-\rho r
 (t-s)}N_{r-1}(\bff(s)) ds\, .
\end{equation}
For $r=1$, if $\bff(0)$ is a probability distribution, we get
\[
 N(t)=e^{-\rho t}N(0)+(1-e^{-\rho t})\frac{\mu}{\rho}
\]
that proves the first of \eqref{eveq}.
We will need the following corollary in section \ref{sec:3.4} below.

\begin{cor}\label{cor:boundNr} Given a probability distribution 
$\bff$, assume that there exists 
$M$ such that $|N_r(\bff(0))|\leq M^r$ then we have
\begin{equation}\label{boundNr}
 |N_r(\bff(t))|\leq \max\left\{M,\frac{\mu}{\rho}\right\}^r
\end{equation}
for every $t\geq0$.
\end{cor}
\noindent\emph{Proof.} Clearly \eqref{boundNr} holds for $r=0$ since 
$N_0(\bff(t))=1$ for every $t\geq 0$. Calling 
$M_1=\max\left\{M,\frac{\mu}{\rho}\right\}$, assume that 
$|N_{r-1}(\bff(t))|\leq M_1^{r-1}$. Form \eqref{Nr} we get
\[
\begin{aligned}
  |N_r(\bff(t))|\leq& e^{-\rho r
 t}M^r+r\mu\int_0^t e^{-\rho r
 (t-s)}M_1^{r-1} ds\\
 =&e^{-\rho r
 t}M^r+\frac{\mu}{\rho}(1-e^{-\rho r
 t})M_1^{r-1}\leq \max\left\{M^r,\frac{\mu}{\rho}M_1^{r-1}\right\}\, .
\end{aligned}
\]
The corollary follows by induction on $r$.\qed

Let now
\[
 \tilde f_N=\sum_{i=1}^N\int v_i^2 f_N(\uv_N)d\uv_N
\]
so that $E(t)=\sum_{N=1}^\infty \tilde f_N$ and observe that
\begin{align*}
& \sum_{i=1}^N\int v_i^2(\cO \bff)_N(\uv_N)d\uv_N=N\tilde f_{N+1}\, .\\ 
& \sum_{i=1}^N\int v_i^2(\I \bff)_N(\uv_N)d\uv_N=\tilde 
f_{N-1}+\frac{1}{2\pi}\bar f_{N-1}
\end{align*}
while
\[
 \sum_{i=1}^N\int v_i^2(\K \bff)_N(\uv_N)d\uv_N=0\, .
\]
Again proceeding formally we get
\begin{align*}
 \frac{d}{dt}\sum_{N=1}^\infty \tilde f_N=&\sum_{N=1}^\infty 
\left(-(N\rho+\mu)\tilde f_N+ \mu \tilde f_{N-1}+\frac{\mu}{2\pi}\bar 
f_{N-1} + \rho N \tilde f_{N+1}\right)\\
=&\frac{\mu}{2\pi}\sum_{N=0}^\infty\bar f_N-\rho \sum_{N=1}^\infty \tilde f_N\, 
.
\end{align*}
It is not hard to adapt this argument, together with \eqref{Nt1} and 
\eqref{Nt2}, to prove the second of \eqref{eveq}.


\subsection{Proof of Theorem \ref{thm:gap}}
\label{sec:3.1}


To prove Theorems \ref{thm:gap} and \ref{thm:gap2}, we will construct a basis of
eigenvectors for the generator 
\begin{equation*}
\mathcal G=\rho (\P^+ - \N) + \mu (\P^--\mathrm{Id})
\end{equation*}
of the evolution due to the thermostat on $L^2_s(\R,\boldsymbol \Gamma)$.
We start by defining
\begin{equation}\label{P+P-}
\begin{aligned}
(\P^+(g)\bh)_{N}(\uv_N)=&\sum_{i=1}^N 
h_{N-1}(v_1,\ldots,v_{i-1},v_{i+1},\ldots, v_N)g(v_i)\\
(\P^-(g)\bh)_{N}(\uv_N)=&\frac{1}{N+1}\sum_{i=1}^{N+1} 
\int dw e^{-\pi w^2}g(w)h_{N+1}(\uv_{N,i}(w))
\end{aligned}
\end{equation}
with  $\uv_{N,i}(w)=(v_1,\ldots,v_{i-1},w,v_i,\ldots, 
v_N)$ and $g\in L^2(\Rs,\gamma)$. Moreover, we use the convention that the sum 
over an empty set is 0 so that $(\P^+(g)\bh)_{0}=0$ for every $\bh$. With this 
notation, $\P^+$ and $\P^-$ from the introduction are  $\P^+(1)$ and
$\P^-(1)$, respectively.

\begin{lem}\label{lem:adj} We have  
\begin{equation}\label{adj}
 \rho\P^+(g)^*=\mu \P^-(g)
\end{equation}
so that $\mathcal G$ is  self-adjoint.
\end{lem}
\noindent\emph{Proof.} 
Proceeding as in the definition of $D^2$, we take as domain of $\P^\pm(g)$ the 
subspaces 
\[
D^\pm=\bigl\{\bh\,\Big|\, 
\sum_{N=0}^\infty a_N\|(\P^\pm(g)\bh)_N\|^2_{2,N}<\infty\bigr\}\,.
\]
It is easy to see that $D^\pm$ are dense in $L^2(\R,\boldsymbol \Gamma)$.  

Calling
$\underline{v}_N^i=(v_1,\dots,v_{i-1},v_{i+1},\dots,v_N)$ 
 we get
\begin{equation}\label{adj1}
\begin{aligned}
(h_N,&(\mathcal{P}^+(g)\mathbf{j})_N)_N=
 \sum_{i=1}^N\int 
d\underline{v}_N\gamma_N(\underline{v}_N)h_N(\underline{v}_N)j_{N-1}(\underline{
v}_N^i)g(v_i)\\
&=\sum_{i=1}^N\int 
d\underline{v}_N^i\gamma_{N-1}(\underline{v}_N^i)\left(\int dv_ie^{-\pi 
v_i^2}g(v_i) h_N(\underline{v}_N)\right)j_{N-1}(\underline{v}_N^i)\\
&=N((\mathcal{P}^-(g)\mathbf{h})_{N-1},j_{N-1})_{N-1}\, .
\end{aligned}
\end{equation}
Assume now that $\bh$ is in the domain of $\P^+(g)^*$.
This means that for every $\bj$ in $D^+$ we have
\[
 (\P^+(g)^*\bh,\bj)=(\bh,\P^+(g)\bj)\, .
\]
Given $M$, choose $\bj$ such that $j_N\equiv 0$ if $N\not=M$. For such a $\bj$ 
we have $\bj\in D^+$ and
\[
\begin{aligned}
\rho a_M((\P^+(g)^*&\bh)_M,j_M)_M= 
\rho(\P^+(g)^*\bh,\bj)=\rho(\bh,\P^+(g)\bj)\\
&=\rho a_{M+1}(h_{M+1},(\P^+(g)\bj)_{M+ 1 })=
\mu a_M((\P^-(g)\bh)_M,j_M)_M
\end{aligned}
\]
where the last equality follows from \eqref{adj1} and the fact that
\begin{equation}\label{arec}
 \rho N a_N=\mu a_{N-1}\, .
\end{equation}

This implies that $\rho(\P^+(g)^*\bh)_M=\mu (\P^-(g)\bh)_M$ for every $M$ 
thus proving \eqref{adj}. This also implies that $\mathcal G$ is 
self adjoint.
\qed
\medskip

To obtain convergence toward $\be^0$, we first need to show that 
$\mathcal  G$ is non positive. This is the content of the following 
Lemma.

\begin{lem}\label{lem:G0}
${\mathcal{G}}$ is non positive and
${\mathcal{G}}\mathbf{h}=0$ if and only if 
$\mathbf{h}=c\mathbf{e}^0$, where
$\mathbf{e}^0$ is given by $e_N^0(\uv_N)=1$ for every $N$ and $\uv_N$.
\end{lem}
\noindent\emph{Proof.} 
From \eqref{adj}, 
we get $\rho(\mathbf{h},\mathcal{P}^+\mathbf{h})= 
\mu(\mathcal{P}^-\mathbf{h},\mathbf{h})$ so that
\begin{equation}\label{self}
 (\bh,\mathcal G \bh)=2\rho(\mathbf{h},\mathcal{P}^+\mathbf{h})-
 (\mathbf{h},(\rho\mathcal{N}+\mu)\mathbf{h})
\end{equation}
Moreover we have
\begin{align}\label{comp}
\rho(\mathbf{h},&\mathcal{P}^+\mathbf{h})=
\rho \sum_{N=1}^\infty a_N\int 
d\underline{v}_N\gamma_N(\underline{v}_N)h_N(\underline{v}_N)\left(\sum_{i=1}^
Nh_{N-1}(\underline{v}_N^i)\right)\nonumber\\
&=\sum_{N=1}^\infty\left[\sum_{i=1}^N\int 
d\underline{v}_N\gamma_N(\underline{v}_N)\left(\sqrt{\rho 
a_N}h_N(\underline{v}_N)\right)\left(\sqrt{\frac{\mu}{N}a_{N-1}}h_{N-1}(
\underline{v}_N^i)\right)\right]\\
&\leq\sum_{N=1}^\infty\sum_{i=1}^N\left[\frac{1}{2}\rho a_N\int 
d\underline{v}_N\gamma_N(\underline{v}_N)h_N(\uv_N)^2+\frac{1}{2}
\frac{\mu}{N}a_{N-1}\int d\underline{v}_N\gamma_N(\underline{v}_N)
h_{N-1}(\underline{v}_N^i) ^2\right]\nonumber\\
&=\sum_{N=0}^\infty\left[\frac{1}{2}N\rho a_N\int 
d\underline{v}_N\gamma_N(\underline{v}_N)h_N(\underline{v}_N)^2+\frac{1}{2}\mu 
a_N\int 
d\underline{v}_N\gamma_N(\underline{v}_N)h_N(\underline{v}_N)^2\right]\nonumber
\\
&=\frac{1}{2}(\mathbf{h},(\rho\mathcal{N}+\mu)\mathbf{h})\nonumber
\end{align}
where we have used \eqref{arec} to obtain the second line and that $ab\leq 
(a^2+b^2)/2$ in going 
from the second to the third line of \eqref{comp}. 
Non positivity follows immediately from \eqref{self} and \eqref{comp}.
Furthermore, we see that the inequality at the end of the second line of 
\eqref{comp} becomes an equality if and only if:
\begin{gather*}
\sqrt{\rho a_N}h_N(\underline{v}_N)=\sqrt{\frac{\mu}{N}a_{N-1}}h_{N-1}(\underline{v}_N^i)
\end{gather*}
or $h_N(\underline{v}_N)=h_{N-1}(\underline{v}_N^i)$ for every $i$ and $N$ which implies
that $h_N\equiv h_0$.\qed\medskip

Our construction of the eigenvalues and eigenvectors of $\mathcal G$ is 
inspired by the
construction of the Fock space for a bosonic quantum field theory, see for 
example Chapter 6 of \cite{Schweber}. The main observation is that the 
operators $\P^\pm(g)$ defined in \eqref{P+P-} have the form of the creation and 
annihilation operators. Since the ``ground state'' of 
$\mathcal G$ is $\be^0$, as opposed to the state with no particle $\mathbf n$, 
see \eqref{empty} below, we will introduce the operators $\R^\pm(g)$, see 
\eqref{RR} below, that can be thought as quasi particle operators, that is 
operators that create and destroy excitations above the ground state, see for 
example \cite{BenGal}. The proofs of the Lemmas in the remaining of this section 
should be familiar to readers with a background in QFT. 

We start with the commutation relations of the operators $\P^\pm(g)$ and $\N$. 
Setting
 $\{ \mathcal A,\mathcal B\}=\mathcal A\mathcal B-\mathcal B\mathcal A$,
we obtain the following Lemma.

\begin{lem}\label{lem:raise} We have
\begin{align*}
\{{\P}^+(g_1),{\P}^-(g_2)\}&=-(g_1,g_2)\mathrm{Id}\\
\{\mathcal{P}^+(g_1),\mathcal{P}^+(g_2)\}&=\{\mathcal{P}^-(g_1),\mathcal{P}
^-(g_2)\}=0\\
\{\mathcal{N},\mathcal{P}^\pm(g)\}&=\pm\mathcal{P}^\pm(g)
\end{align*}
where
\begin{equation*}
 (g_1,g_2)=\int_{\Rs}g_1(w)g_2(w)e^{-\pi w^2}dw\, .
\end{equation*}
\end{lem}
\noindent\emph{Proof}. We first observe that, due to the symmetry of $h_N$, we 
have
\[
 (\P^-(g)\bh)_N(\uv_{N})=\int 
\gamma(v_{N+1})g(v_{N+1})h_{N+1}(\uv_{N+1})dv_{N+1}:= (P^-_N(g)
h_{N+1})(\uv_{N})
\]
while
\[
 (\P^+(g)\bh)_N(\uv_{N})= \sum_{i=1}^{N}(P^+_{N,i}(g) 
h_{N-1})(\uv_{N})
\]
where
\[
 (P^+_{N,i} (g)
h_{N-1})(\uv_{N})=h_{N-1}(v_1,\ldots,v_{i-1},v_{i+1},\ldots,v_{N})g(v_i)\, .
\]
Thus we get
\begin{align*}
 (\P^-(g_1)\P^-(g_2)&\bh)_N(\uv_{N})=(P^-_N(g_1) P^-_{N+1}(g_2) 
h_{N+2})(\uv_{N})\\
&=\int \gamma(v_{N+1})\gamma(v_{N+2})
g_1(v_{N+1})g_2(v_{N+2})h_{N+2}(\uv_{N+2})dv_{N+1}dv_{N+2
}
\end{align*}
Using again that $h_N$ is symmetric we get $\{\P^-(g_1),\P^-(g_2)\}=0$.
Moreover, we have
\begin{align*}
& P^+_{N,i}(g_1) P^+_{N-1,j}(g_2) h_{N-2}(\uv_{N})\\
& \qquad\qquad=\begin{cases}
h_{N-2}(v_1,\ldots,v_{j-1},v_{j+1},\ldots,v_{i-1},v_{i+1},\ldots,v_{N})g_1(v_i) 
g_2(v_j) & i>j\\
h_{N-2}(v_1,\ldots,v_{i-1},v_{i+1},\ldots,v_{j},v_{j+2},\ldots,v_{N})g_1(v_i) 
g_2(v_{j+1}) & i\leq j
                                                    \end{cases}
\end{align*}
so that 
\[
 \begin{cases}
  P^+_{N,i}(g_1) P^+_{N-1,j}(g_2) h_{N-2}=P^+_{N,j}(g_2) 
P^+_{N-1,i-1}(g_1) h_{N-2} & i>j\\
P^+_{N,i}(g_1) P^+_{N-1,j}(g_2) 
h_{N-2}=P^+_{N,j+1}(g_2) P^+_{N-1,i}(g_1) h_{N-2} & i\leq j\, .
 \end{cases}
\]
Summing over $i$ and $j$ it follows that $\{\P^+(g_1),\P^+(g_2)\}=0$.

Similarly we have
\[
 (P^-_N(g_1)P^+_{{N+1},N+1}(g_2)h_N)(\uv_N)=h_N(\uv_N)\int 
g_1(v_{N+1})g_2(v_{N+1})\gamma(v_{N+1})dv_{N+1}
\]
while for $i\leq N$ we get
\begin{align*}
 (P^-_N(g_1)P^+_{{N+1},i}&(g_2)h_N)(\uv_N)\\
= &g_2(v_i)\int 
h_N(v_1,\ldots,v_{i-1},v_{i+1},\ldots,v_{N+1})g_1(v_{N+1})\gamma(v_{N+1})dv_{N+1
}\\
=& (P^+_{N,i}(g_2)P^-_{N-1}(g_1)h_N)(\uv_N)\, .
\end{align*}
Summing over $i$ we get $\{{\P}^+(g_1),{\P}^-(g_2)\}=-(g_1,g_2)\mathrm{Id}$.

Finally we observe that
\[
(\P^-(g)\N\bh)_N=P^-_{N}(g)(\N 
\bh)_{N+1}=(N+1)(\P^-(g)\bh)_N=((\N+\rm{Id})\P^+(g)\bh)_N
\]
so that $\{\mathcal{N},\mathcal{P}^-(g)\}=-\mathcal{P}^-(g)$. The commutation 
relation for $\P^+$ follows taking the adjoint. 
\qed\medskip

Observe that $\mathcal{P}^-(g)\mathbf{e}^0=(g,1)\mathbf{e}^0$ while from Lemma 
\ref{lem:raise} it follows that
\begin{equation}\label{comGP}
\begin{aligned}
 \{\mathcal G,\P^+(g)\}=&\{\P^+(1),\P^+(g)\}-
\rho\{\N,\P^+(g)\}+\mu\{\P^-(1),\P^+(g)\}\\
=&-\rho\P^+(q)+\mu (1,g)\mathrm{Id}
\end{aligned}
\end{equation}
that makes it natural to define the new creation and annihilation 
operators
\begin{equation}\label{RR}
\begin{aligned}
{\R}^+(g)=\sqrt{\frac{\rho}{\mu}}{\P}^+(g)-
\sqrt{\frac{\mu}{\rho}}(g,1)\,\mathrm{Id}\\
{\R}^-(g)=\sqrt{\frac{\mu}{\rho}}{\P}^-(g)-
\sqrt{\frac{\mu}{\rho}}(g,1)\,\mathrm{Id}\, .
\end{aligned}
\end{equation}
The following Corollary collects the relevant properties of ${\R}^\pm(g)$.
\begin{cor}\label{cor:crea}
 We have $\R^+(g)^*=\R^-(g)$, $\mathcal{R}^-(g)\mathbf{e}^0=0$, and
 \begin{align*}
\{{\R}^+(g_1),{\R}^-(g_2)\}&=-(g_1,g_2)\mathrm{Id}\\
\{\mathcal{R}^+(g_1),\mathcal{R}^+(g_2)\}&=\{\mathcal{R}^-(g_1),\mathcal{R}
^-(g_2)\}=0\\
\{\mathcal{N},\mathcal{R}^\pm(g)\}&=\pm\left(\mathcal{R}^\pm(g)+\sqrt{\frac{\mu}
{\rho }}(g,1)\mathrm{Id}\right)
\end{align*}
Moreover we also have
\begin{equation}\label{raise}
\{\mathcal G,\R^+(g)\}=-\rho\R^+(g)\,,\qquad\{\mathcal 
G,\R^-(g)\}=\rho\R^-(g)\,.
\end{equation} 

\end{cor}
\noindent\emph{Proof}. It is easy to verify that 
$\mathcal{R}^-(g)\mathbf{e}^0=0$. 
Moreover we only need to 
prove \eqref{raise} since the other relations are immediate consequences of 
Lemma $\ref{lem:raise}$. 
From \eqref{comGP} we get
\[
 \{\mathcal G,\R^+(g)\}=\sqrt{\frac{\rho}{\mu}}\{\mathcal G,\P^+(g)\}=
 -\rho\sqrt{\frac{\rho}{\mu}}\P^+(g)+\sqrt{\mu\rho}(g,1)\mathrm{Id}=
 -\rho \R^+(g)
\]
The second equation of \eqref{raise} follows by taking the 
adjoint of the first.\qed
\medskip

Since $K_N$ preserves the space of polynomials of a given degree, see 
\cite{BLV}, we choose as an 
orthonormal basis for $L^2(\Rs,\gamma(v))$ the 
polynomials
\begin{equation}\label{herm}
 L_n(v)=\frac{1}{\sqrt{n!}}H_n(\sqrt{2\pi}v)
\end{equation}
where
\begin{equation*}
 H_n(v)=(-1)^n e^{\frac{v^2}2}\frac {d^n}{dv^n}e^{\frac{-v^2}2}
\end{equation*}
are the standard Hermite polynomials.
For every sequence $\underline \alpha=(\alpha_0,\alpha_1,\alpha_2,\ldots)$ such 
that $\alpha_i\in\mathbb{N}$ and $\lambda(\underline
 \alpha):=\sum_{i=0}^\infty \alpha_i<\infty$, we 
define
\begin{equation}\label{ealpha}
 \be_{\underline \alpha}=\prod_{i=0}^\infty 
\frac{(\R^+_i)^{\alpha_i}}{\sqrt{\alpha_i!}}\be^0
\end{equation}
where $\R^\pm_n=\R^\pm(L_n)$.

\begin{lem}\label{lem:ortho}
 The vectors $\be_{\underline \alpha}$ form an orthonormal basis in
 $L^2_s(\R,\boldsymbol \Gamma)$. Moreover, we have
 \begin{equation}\label{spct}
  \mathcal G\be_{\underline \alpha}=-\rho\lambda(\underline \alpha)\be_{\underline \alpha}\, .
 \end{equation}
Finally we have $\|\K \be_{\underline \alpha}\|_2<\infty$, so that $ 
\be_{\underline \alpha}\in D^2$, for every 
$\underline\alpha$. 
\end{lem}
\noindent\emph{Proof}. If $n_1\not=n_2$ and $\alpha_1\alpha_2\not =0$, using 
Corollary \ref{cor:crea} we get
\begin{equation*}
((\R^+_{n_1})^{\alpha_1}\be^0,(\R^+_{n_2})^{\alpha_2}\be^0)=
(\be^0,(\R^+_{n_2})^{\alpha_2}(\R^-_{n_1})^{\alpha_1}\be^0)=0
\end{equation*}
 while
\begin{align*}
((\R^+_n)^{\alpha_1}\be^0,(\R^+_n)^{\alpha_2}\be^0)=&((\R^+_n)^{\alpha_1-
1}\be^0,\R^-_n(\R^+_n)^{\alpha_2}\be^0)\\
=&((\R^+_n)^{\alpha_1-
1}\be^0,\R^+_n\R^-_n(\R^+_n)^{\alpha_2-1}\be^0)+\\
&\qquad\qquad((\R^+_n)^{\alpha_1-
1}\be^0,(\R^+_n)^{\alpha_2-1}\be^0)\\
&\qquad\vdots\\
=&((\R^+_n)^{\alpha_1-
1}\be^0,(\R^+_n)^{\alpha_2}\R^-_n\be^0)+\\
&\qquad\qquad\alpha_2((\R^+_n)^{\alpha_1-
1}\be^0,(\R^+_n)^{\alpha_2-1}\be^0)\\
=&\alpha_2((\R^+_n)^{\alpha_1-
1}\be^0,(\R^+_n)^{\alpha_2-1}\be^0)\,.
\end{align*}
Assuming $\alpha_1\geq\alpha_2$ we get
\begin{equation}\label{a2}
((\R^+_n)^{\alpha_1}\be^0,(\R^+_n)^{\alpha_2}\be^0)=
\alpha_2!((\R^+_n)^{\alpha_1-\alpha_2}\be^0,\be^0)
\end{equation}
so that
\begin{equation*}
((\R^+_n)^{\alpha_1}\be^0,(\R^+_n)^{\alpha_2}\be^0)= 
\alpha_1!\delta_{\alpha_1,\alpha_2}
\end{equation*}
from which orthonormality follows easily. 
Observe now that
\begin{equation}\label{P0n}
 ((\P^+(1))^n\be^0)_N=\begin{cases} 0 & N<n\\
                       \frac{N!}{(N-n)!} & N\geq n
                      \end{cases}
\end{equation}
so that we can write
\begin{equation}\label{empty}
 \mathbf n=\sum_{n=0}^\infty \frac{(-1)^n}{n!}(\P^+(1))^n\be^0
\end{equation}
where $\mathbf n=(1,0,0,\ldots)$. Since 
$\P^+(1)=\sqrt{\frac{\rho}{\mu}}\R_0^++\sqrt{\frac{\mu}{\rho}}{\rm Id}$ we see 
that $\mathbf n$ is in the closure of the span of the $\be_{\underline \alpha}$. 
Calling $\P^+_i=\P^+(L_i)$, we observe that $(\P^+_i\mathbf n)_N=0$ for 
$N\not=1$ while $(\P^+_i\mathbf n)_1=L_i$. 
Since the $L_i$ form a basis for 
$L^2(\Rs,\gamma_1)$ we see that the closure of the span of $\{\mathbf n; 
\P^+_i\mathbf n, i\geq 0\}$ contains a basis for 
$L^2_s(\Rs^0,a_0)\oplus L^2_s(\Rs,a_1\gamma_1)$. Observe now that  
$\P^+_i=\sqrt{\frac{\mu}{\rho}}\R^+_i+\delta_{i,0}\frac{\mu}{\rho}\mathrm{Id}$ 
and that $\R^+_i\be_{\underline 
\alpha}=\sqrt{\alpha_i+1}\be_{\underline\alpha'}$, 
where $\alpha'_j=\alpha_j$ for $j\not=i$ while 
$\alpha'_i=\alpha_i+1$. Combining this with \eqref{empty} we get that closure 
of the span of the $\be_{\underline 
\alpha}$ contains $\P^+_i\mathbf n$ and thus it contains a basis for 
$L^2_s(\Rs^0)\oplus L^2_s(\Rs,a_1\gamma_1)$. Iterating this construction we 
obtain completeness. Equation \eqref{spct} follows easily from \eqref{raise}.

Finally, since $(h_N,R_{i,j}h_N)_N\leq \|h_n\|_{2,N}$, from \eqref{Kcol} we 
get 
\[
\|\K\bh\|_2^2\leq\sum_{N=0}^\infty a_N N^4\|\bh\|_{2,N}^2=\|\N^2\bh\|_2^2\,.
\]
Using the commutation relations in Corollary \ref{cor:crea} as in the 
derivation of \eqref{a2} we get 
\[
\N\left(\R^+_n\right)^\alpha=\left(\R^+_n\right)^\alpha\N+\alpha\left(\R^+_n
\right)^\alpha +\delta_{n,0}\alpha\sqrt{\frac{\mu}{\rho}}\left(\R^+_n
\right)^{\alpha-1}
\]
that, together with 
$\N\be^0=\sqrt{\frac{\mu}{\rho}}\R^+_0\be^0+\frac{\mu}{\rho}\be^0$, gives
\[
 \N\be_{\underline \alpha}=\left(\lambda(\underline 
\alpha)+\frac{\mu}{\rho}\right)\be_{\underline \alpha}+
\sqrt{\frac{\mu}{\rho}}(\sqrt{\alpha_0}\be_{\underline 
\alpha^-}+\sqrt{\alpha_0+1}\be_{\underline \alpha^+})
\]
where $\alpha^\pm_i=\alpha_i$, for $i>0$, while $\alpha_0^\pm=\alpha_0\pm 1$. 
Thus we have $\|\N^2\be_{\underline \alpha}\|_2<\infty$ and the proof is 
complete.
\qed
\medskip

In section \ref{sec:3.2} we will need a more explicit representation 
of the $\be_{\underline\alpha}$. To this end
observe that, if $n\not=0$, 
$(\R^+_n\be^0)_N(\uv_N)=\sqrt{\frac{\rho}{\mu}}\sum_{i=1}^N L_n(v_i)$ while 
for $n_1,n_2\not=0$ and $N\geq 2$ we can write
\[
(\R^+_{n_1}\R^+_{n_2}\be^0)_N(\uv_N)=\frac{\rho}{\mu}\sum_{i\not=j}
L_{n_1}(v_i)L_{n_2}(v_j)=\frac{1}{(N-2)!}\frac{\rho}{\mu}\sum_{\pi\in{\rm 
Sym}(N)}L_{n_1}(v_{\pi(1)})L_{n_2}(v_{\pi(2)})
\]
where ${\rm Sym}(N)$ is the group of permutations on $\{1,\ldots,N\}$.
More generally, given $n_i\not=0$, $i=1,\ldots,M$, we get, for $N\geq M$,
\begin{equation}\label{prodR}
\left(\prod_{i=1}^M\R^+_{n_i}\be^0\right)_N(\uv_N)=\frac{1}{(N-M)!}
\left(\frac{\rho}{\mu}\right)^{\frac M2}\sum_{\pi\in{\rm 
Sym}(N)}\prod_{i=1}^M L_{n_i}(v_{\pi(i)})\, .
\end{equation}
while $\left(\prod_{i=1}^M\R^+_{n_i}\be^0\right)_N\equiv 0$ for $N<M$.
Given $\underline \alpha$ with $\lambda(\underline \alpha)<\infty$, define
\[
 L_{\underline\alpha}=\bigotimes_{i=1}^\infty L_{i}^{\otimes\alpha_i}
\]
where $L_{i}^{\otimes0}=1$ and observe that $L_{\underline\alpha}$ is a 
polynomial 
in $\lambda_0(\underline\alpha):=\sum_{i=1}^\infty \alpha_i$ variables with 
degree $d(\underline\alpha):=\sum_{i=1}^\infty i\alpha_i$. Also for $\pi\in 
{\rm Sym}(N)$, define $\pi(\uv_N)=(v_{\pi(1)},v_{\pi(2)},\ldots 
v_{\pi(N)})$.  Using these definitions, together with \eqref{ealpha} and the 
fact that 
$\R_0^+=\sqrt{\frac{\rho}{\mu}}\P^+(1)+\sqrt{\frac{\mu}{\rho}}\mathrm{Id}$ we 
can write, for $N\geq \lambda_0(\underline\alpha)$,
\begin{equation}\label{beH}
 (\be_{\underline\alpha})_N(\uv_N)=c_{\underline\alpha,N}\sum_{\pi\in{\rm 
Sym}(N)} L_{\underline\alpha}(\pi(\uv_N))\, ,
\end{equation}
for suitable coefficients $c_{\underline\alpha,N}$, while 
$(\be_{\underline\alpha})_N(\uv_N)=0$ for $N<\lambda_0(\underline\alpha)$.

We now come back to the full operator $\widetilde{\mathcal L}$. 

\begin{cor} The operator $\widetilde{\mathcal L}$ is self-adjoint, non positive
and $\widetilde{\mathcal L}\bh=0$ if and only if $\bh=c\be^0$.
\end{cor}
\noindent\emph{Proof}. We can proceed exactly as in proof of Lemma 
\ref{lem:adj}. Assume that $\bh$ is in the domain of 
$\hbox{$\widetilde{\mathcal L}$ }^*$.
This means that for every $\bj$ in $D^2$ we have
\[
 (\hbox{$\widetilde{\mathcal L}$ }^*\bh,\bj)=(\bh,\widetilde{\mathcal L}\bj)\, .
\]
Given $M$, choose $\bj$ such that $j_N\equiv 0$ if $N\not=M$. Clearly 
$\bj\in D^2$ because $(\widetilde{\mathcal L}\bj)_N\not=0$ only for $N=M-1$, 
$M$, and $M+1$. Moreover $(\K\bh,\bj)=a_M(K_Mh_M,j_M)_M$ is well defined 
for every $\bh\in L_s^2(\R,\boldsymbol \Gamma)$. Finally we known that $K_M$ 
is non negative and self-adjoint for every $M$. Thus we get
\[
\begin{aligned}
a_M((\hbox{$\widetilde{\mathcal L}$ }^*\bh)_M,j_M)_M= &
((\hbox{$\widetilde{\mathcal L}$ }^*\bh,\bj)=
(\bh,\widetilde{\mathcal L}\bj)=
(\bh,\mathcal G\bj)+\tilde\lambda a_M(h_M,K_Mj_M)_M\\=&
a_M((\mathcal G\bh)_M,j_M)+\tilde\lambda a_M(K_Mh_M,j_M)_M
=a_M((\widetilde{\mathcal L}\bh)_M,j_M)_M\, .
\end{aligned}
\]
This implies that $(\hbox{$\widetilde{\mathcal L}$ }^*\bh)_M= 
(\widetilde{\mathcal L}\bh)_M$ for every $M$. This proves that 
$\widetilde{\mathcal L}$ is self-adjoint. Observe also that $\mathcal 
G\bh=0$ if and only if $\bh=c\be^0$, see Lemma \ref{lem:G0}, while 
$\K$ is positive and $\K\be^0=0$. This completes the proof.
\qed\medskip

 Let $\mathbf 
W_1=\mathrm{span}\{\be_{\underline \alpha}\,|\, \lambda(\underline 
\alpha)=1\}=\mathrm{span}\{\R^+_n\be^0\,|\,n\geq 0\}$. Observe that $\mathcal 
G\bh=-\rho \bh$ if $\bh\in\mathbf W_1$ 
while $(\bh,\mathcal G\bh)<-\rho (\bh,\bh)$ if $\bh\in D^2$, $\bh\perp 
\be^0$ but 
$\bh\not\in\mathbf W_1$. Thus we get
\[
 \Delta\leq -\rho+\sup \{(\bh,\mathcal{K}\bh)\,|\, \bh\in D^2, \|\bh\|_2=1, 
\bh\perp \bf E_0\}\leq-\rho
\]
From 
\cite{BLV} we know that $(f_N,K_Nf_N)\leq 0$ for every $f_N$ while 
$(f_N,K_Nf_N)= 0$ if and only if $f_N$ is rotationally invariant.
Since $(\R^+_n 
\be^0)_N=\sqrt{\rho/\mu}\sum_{i=1}^N 
L_{n}(v_i)$, for $n>0$, while $(\R^+_0 
\be^0)_N=\sqrt{\rho/\mu}N-\sqrt{\mu/\rho}$ we have that 
$\R^+_n\be^0$ is rotationally invariant if and only if $n=0$ or $n=2$. This 
implies that $(\bh,\widetilde{\mathcal L}\bh)=-\rho\|\bh\|_2$ if and only if 
$\bh\in\mathrm{span}\{\R_0^+\be^0,\R_2^+\be^0\}$. Since 
$\R_0^+\be^0=\be_{(1,0,\ldots)}$ and $\R_2^+\be^0=\be_{(0,0,1,0,\ldots)}$, this 
completes the proof of Theorem 
\ref{thm:gap}.\qed

\subsection{Proof of Theorem \ref{thm:gap2}}
\label{sec:3.2}

To prove Theorem \ref{thm:gap2}, we need more information on the action of $\K$ 
on the basis vectors $\be_{\underline \alpha}$.

As a basic step, we compute the action of $R_{1,2}$, see \eqref{rot}, on the 
product of two Hermite polynomials in $v_1$ and $v_2$. A simple calculation, 
see e.g. \cite{BLV}, shows that $(R_{1,2}F)(v_1,v_2)=0$ 
for every $F$ odd in $v_1$ or $v_2$. Thus, calling 
$H_{(m_1,m_2)}(v_1,v_2)=H_{m_1}(v_1)H_{m_2}(v_2)$, it follows that 
$R_{1,2}H_{(m_1,m_2)}\not=0$ if and only if $m_1$ and $m_2$ are both even 
while $R_{1,2}H_{(2n_1,2n_2)}$ is a rotationally 
invariant polynomial of degree $2(n_1+n_2)$ in $v_1$ and $v_2$. Moreover, if 
$m_1+m_2<2n_1+2n_2$, we get
\begin{align*}
&\int 
H_{(m_1,m_2)}(v_1,v_2)\bigl(R_{1,2}H_{(2n_1,2n_2)}\bigr)(v_1,v_2)\gamma(
v_1)\gamma(v_2)dv_1dv_2  \\
&\qquad=\int 
\bigl(R_{1,2}H_{(m_1,m_2)}\bigr)(v_1,v_2)H_{(2n_1,2n_2)}(v_1,v_2)\gamma(
v_1)\gamma(v_2)dv_1dv_2 =0
\end{align*}
where we have used that $H_{(2n_1,2n_2)}$ is orthogonal to any polynomial of 
degree less that $2(n_1+n_2)$. Thus we have 
$R_{1,2}H_{(2n_1,2n_2)}\in\mathrm{span}\{H_{(p_1,p_2)}\,|\, 
p_1+p_2=2n_1+2n_2\}$ and, since $H_n$ is a monic polynomial of degree $n$, we 
can write
\[
R_{1,2}H_{(2n_1,2n_2)}=\sum_{k=0}^{n_1+n_2}
a_{k,n_1,n_2} H_{(2k,2(n_1+n_2-k))}=\sum_{k=0}^{n_1+n_2}
a_{k,n_1,n_2}v_1^{2k}v_2^{2(n_1+n_2-k)}+Q
\]
for suitable coefficients $a_{k,n_1,n_2}$ and polynomial $Q(v_1,v_2)$ of degree 
strictly less then $2(n_1+n_2)$.
This, together with rotational invariance, implies that
\begin{equation}\label{poli} 
R_{1,2}H_{(2n_1,2n_2)}=\tilde\tau_{n_1,n_2}\sum_{k=0}^{n_1+n_2}
\binom{n_1+n_2}{k}H_{(2k,2(n_1+n_2-k))}
\end{equation} 
for suitable coefficients $\tilde \tau_{n,m}$. Using \eqref{poli}, together 
with \eqref{beH}, it is possible to give an explicit representation of $\K$ on 
the basis of the $\be_{\underline\alpha}$. For the purpose of this paper, we 
will only need some particular case discussed in details below.

Let now $\mathbf V_m=\mathrm{span}\{\be_{\underline\alpha}|\sum_{i=1}^\infty 
i\alpha_i=m\}=\mathrm{span}\{\prod_i (\R_i^+)^{\alpha_i}\be^0|\sum_{i=1}^\infty 
i\alpha_i=m\}$, that is $\mathbf V_m$ is the subspace of all states $\bh$ 
such that $h_N$ is a polynomial of degree $m$ orthogonal to all polynomials of 
degree less than $m$. From the above considerations and \eqref{beH} it follows 
that $\K \mathbf V_m\subset \mathbf V_m$ so that defining
\begin{equation}\label{deltam}
 \delta_{m}= \inf_{\substack{\bh\in \mathbf V_{m}\cap D^2\\ \|\bh\|_2=1,\, 
\bh\perp 
\mathbf{E}_1\oplus\mathbf{E}_0}}(\bh,-\widetilde{\mathcal{L}}\bh)\, .
\end{equation}
and observing that $L^2_s(\R,\boldsymbol \Gamma)=\bigoplus_{m=0}^{\infty} 
\mathbf V_m$, 
we get $\Delta_2=-\inf_m\delta_m$.

Since $\mathbf E_1=\mathrm{span}\{\R_0^+\be^0,\R_2^+\be^0\}$, we get 
\[
\begin{aligned}
\mathbf{V}_0\cap (\mathbf{E}_1\oplus\mathbf{E}_0)^\perp=&{\rm span}\{(\R^+_0)^n 
\be^0, n\geq 2\}\\
\mathbf{V}_2\cap 
(\mathbf{E}_1\oplus\mathbf{E}_0)^\perp=&{\rm 
span}\{(\R^+_0)^n \R_2^+ \be^0, n\geq 1; (\R^+_0)^m (\R^+_1)^2\be^0, m\geq 
0\}\, .
\end{aligned}
\]
Observing that  $\K(\R^+_0)^n\be^0=\K(\R^+_0)^n \R_2^+ \be^0=0$, due to 
rotational invariance, while $\K(\R^+_0)^m (\R^+_1)^2\be^0=0$, due to parity, 
we obtain 
$\delta_0=\delta_2=2\rho$. Moreover we have that, for $m\not=0,2$, $\mathbf 
V_m\perp\mathbf{E}_1\oplus\mathbf{E}_0$. Thus we need a 
lower bound on $\delta_m$ for $m$ odd and for $m$ even and greater than 2. 

Observe that $(\R^+_{m}\be^0,\mathcal G\R^+_{m}\be^0)=-\rho$ while 
$(\bh,\mathcal G\bh)\leq-2\rho(\bh,\bh)$ if $\bh\in\mathbf V_m$ and $\bh\perp 
\R^+_{m}\be^0$. Thus, if $\lambda$ is not too big, it is natural to search for 
the infimum of $(\bh,-\widetilde{\mathcal L}\bh)$ on $\mathbf V_m$ looking at 
states 
$\bh$ close to $\R^+_{m}\be^0$.
To do this, we need the 
representation of $\K\R^+_m\be^0$ on the basis formed by the $\be_{\underline 
\alpha}$. 
If $m=2n$, using \eqref{poli} for $n_2=0$ we get
\begin{equation}\label{poli1} 
R_{1,2}H_{(2n,0)}=\tau_{n}\sum_{k=0}^{n}
\binom{n}{k}H_{(2k,2(n-k))} 
\end{equation}
where $\tau_n=\tilde\tau_{n,0}$.
To compute $\tau_n$ we compare the coefficients of $v_1^{2n}$ on 
the left and right hand side of \eqref{poli1}. On the left hand side the only 
contribution comes from $R_{1,2}v_1^{2n}$ since $R_{1,2}$ preserve the degree. 
On the right hand side only the term with $k=n$ contains the monomial 
$v_1^{2n}$. Since the $H_n$ are monic and
\[
 R_{1,2}v_1^{2n}=\int_0^{2\pi}(v_1\cos \theta - v_2\sin\theta 
)^{2n}\frac{d\theta}{2\pi}=
 (v_1^2 + v_2^2 )^{n}\int_0^{2\pi}\cos^{2n}\theta\frac{d\theta}{2\pi}\, ,
\]
and we obtain
\[
\tau_n=\int_0^{2\pi} 
\cos^{2n}\theta\frac{d\theta}{2\pi}=\frac{1}{4^n}\binom{2n}{n}\, .
\]
Combining with \eqref{herm} we get 
\begin{equation*}
R_{i,j}L_{2n}(v_i)=\tau_n\sum_{k=0}^n\binom{n}{k}
\f{\sqrt{(2k)![2(n-k)]!}}{\sqrt{(2n)!}}L_{2k}(v_i)L_{2(n-k)}(v_j)\, .
\end{equation*}
Since for $n>0$ we have $(\R^+_{2n}\be^0)_N=\sqrt{\rho/\mu}\sum_{i=1}^N 
L_{2n}(v_i)$, a direct computation shows that
\begin{equation*}
\begin{aligned}
(\K\R^+_{2n}\be^0)_N=&
\sqrt{\frac{\rho}{\mu}}(N-1)(2\tau_n-1)
\sum_{i=1}^NL_{2n}(v_i)\\
+&\sqrt{\frac{\rho}{\mu}}\sum_{k=1}^{n-1}\sum_{i\neq j}
\sigma_{n,k}L_{2k}(v_i)L_{2(n-k)}(v_j)
\end{aligned}
\end{equation*} 
where
\begin{equation}\label{gammi2}
\sigma_{n,k}=\tau_n\f{\binom{n}{k}}{\sqrt{\binom{2n}{2k}}}=\sqrt{\tau_n\tau_k\tau_{n-k}}\,
.
\end{equation}
This gives us
\begin{align}\label{Rne}
\K\R^+_{2n}\be^0=&(2\tau_n-1)\R^+_{2n}\N\be^0
+\sqrt{\frac{\mu}{\rho}}\sum_{k=1}^{n-1}
\sigma_{n,k}\R^+_{2k}\R^+_{2(n-k)}\be^0\nonumber\\
=&\frac{\mu}{\rho}(2\tau_n-1)\R^+_{2n}\be^0+\sqrt{\frac{\mu}{\rho}}(2\tau_n-1)
\R^+_0\R^+_{2n}\be^0\nonumber\\
+&\sqrt{\frac{\mu}{\rho}}\sum_{k=1}^{n-1}
\sigma_{n,k}\R^+_{2k}\R^+_{2(n-k)}\be^0
\end{align}
where we have used that $\N
\be^0=\sqrt{\frac{\mu}{\rho}}\R^+_0\be^0+\frac{\mu}{\rho}\be^0$. 

If $m=2n+1$, $R_{1,2}H_{2n+1}(v_1)=0$ gives
\begin{equation}\label{Rno}
\K\R^+_{2n+1}\be^0=-\frac{\mu}{\rho}\R^+_{2n+1}\be^0-\sqrt{\frac{\mu}{\rho}}
\R^+_0 \R^+_{2n+1}\be^0\, .
\end{equation}
From \eqref{Rne} and \eqref{Rno} we get 
\[
\tilde\lambda(\R^+_{2n}\be^0,\K\R^+_{2n}\be^0)=-\lambda(1-2\tau_n)\, , \quad 
\tilde\lambda(\R^+_{2n+1}\be^0,\K \R^+_{2n+1}\be^0)=-\lambda
\]
so that $\delta_{2n}\leq\rho+\lambda(1-2\tau_n)$ and 
$\delta_{2n+1}\leq\rho+\lambda$. 

The following Lemma shows that, if the average number of particles in the 
steady state is large enough and $\lambda$ is not too large, one can find a
lower bound for $\delta_m$ close to the upper bound derived above.

\begin{lem} \label{lem:deltam} For $m=2n+1$ we have
\begin{equation}\label{oddm}
\delta_{2n+1}\geq \min\left\{\rho+\lambda -\lambda\sqrt{\f{\rho}{\mu}}\,,\; 
 2\rho-\lambda\sqrt{\f{\rho}{\mu}}\right\}
\end{equation}
while for $m=2n$, $n>1$, we have
\begin{equation}\label{evenm}
 \delta_{2n}\geq\min 
\left\{\rho+(1-2\tau_n)\lambda-2\lambda 
\sqrt{\f{\rho}{\mu}}\,,\;2\rho-2\lambda\sqrt{\f{\rho}{\mu}}\right\}\,.
\end{equation}
\end{lem}
\noindent\emph{Proof}. See Appendix \ref{app:bosons.deltam}.
\medskip

Since  $\tau_2=3/8$ and
$(\R_4\be^0,-\widetilde{\mathcal L}\R_4\be^0)=\rho+\lambda/4$, we get 
\[
\rho+\frac{\lambda}{4}-2\lambda 
\sqrt{\f{\rho}{\mu}}<\delta_4\leq \rho+\lambda/4. 
\]
Moreover, thanks to \eqref{cond},
\begin{equation*}
2\rho-\lambda\sqrt{\f{\rho}{\mu}}>\rho+\frac{\lambda}{4}\, 
,\qquad
\rho+\lambda -\lambda\sqrt{\f{\rho}{\mu}}>\rho+\frac{\lambda}{4}
\end{equation*}
so that $\delta_{2n+1}>\delta_4$ for every $n$. Finally we observe that 
$\tau_{n+1}<\tau_n$ 
and $\tau_3=5/16$. Using \eqref{cond} again it follows that, for $n\geq 3$,
\[
\delta_{2n}\geq\min 
\left\{2\rho-2\lambda\sqrt{\f{\rho}{\mu}},(1-2\tau_3)\lambda+\rho-2\lambda 
\sqrt{\f{\rho}{\mu}}\right\}>\rho+\f\lambda4\geq\delta_4
\]
so that $\Delta_2=-\delta_4$. 

To show that $\Delta_2$ is an eigenvalue, we need to construct an 
eigenstate, that is we need to find $\hat \bh\in \mathbf V_4$ such that 
$\widetilde{\mathcal L}\hat\bh=-\delta_4 \hat\bh$. To this end, it is enough to 
show that there exists $\hat \bh\in \mathbf V_4$ such that $(\hat 
\bh,\widetilde{\mathcal L}\hat\bh)=-\delta_4 (\hat\bh,\hat\bh)$.
Observe that if $\bh\in \mathbf V_4$ then $\K \bh$ is even. We 
thus restrict our search to $\hat\bh\in \mathbf 
V^e_4=\mathrm{span}\{(\R^+_0)^k\R^+_4\be^0,\, (\R^+_0)^k(\R^+_2)^2\be^0;\,k\geq 
0 
\}$.

Consider a sequence $\bh_n\in \mathbf V^e_4$ such that $\|\bh_n\|_2=1$
and $\lim_{n\to\infty}(\bh_n,-\widetilde{\mathcal L}\bh_n)=\delta_4$. Calling
$\mathbf V^e_{4,k}=\mathrm{span}\{(\R^+_0)^{k-1}\R^+_4\be^0, 
(\R^+_0)^{k-2}(\R^+_2)^2\be^0 \}$
for $k>2$, while $\mathbf V^e_{4,1}=\mathrm{span}\{\R^+_4\be^0\}$,
we can write $\bh_n=\sum_{k=0}^\infty \bh_{n,k}$ with $\bh_{n,k}\in \mathbf 
V^e_{4,k}$ and we 
can find a subsequence $\bh^0_n$ of $\bh_n$ such that 
$\lim_{n\to\infty}\bh_{n,0}=\hat\bh_0$. Similarly we can 
find a new subsequence $\bh^1_n$ of $\bh^0_n$ such that 
$\lim_{n\to\infty}\bh_{n,1}=\hat\bh_1$. Proceeding like 
this we find a sequence $\bh_n^\infty$ such that 
$\lim_{n\to\infty}\bh^\infty_{n,k}=\hat\bh_k$, for 
every $k$. Analogously, since $h_{n,N}$ is an even polynomial of degree 4 in 
$\uv_N$ we can assume, possibly at the cost of further extracting a 
subsequence, that $\lim_{n\to\infty} h^\infty_{n,N}=\hat h_N$ for every $N$.
From Fatou's Lemma 
we get  that $\lim_{n\to\infty}\bh^\infty_n=\hat\bh$ with $\|\hat\bh\|_2\leq 1$ 
while 
\[
(\hat \bh,-\mathcal G\hat\bh)=\rho\sum_{k=1}^\infty k\|\hat\bh_k\|^2
\leq\liminf_{n\to\infty}\rho\sum_{k=1}^\infty k\|\bh_{n,k}\|^2=
\liminf_{n\to\infty}(\bh^\infty_n,-\mathcal G \bh^\infty_n)
\]
and analogously, since $K_N$ is non positive,
\[
\begin{aligned}
(\hat\bh,-\K\hat\bh)=\sum_{N=0}^\infty(\hat h_N,-K_N \hat h_N)_N\leq&
\liminf_{n\to\infty}\sum_{N=0}^\infty(h^\infty_{n,N},-K_N h^\infty_{n,N})_N 
\\
\leq&\liminf_{n\to\infty}(\bh^\infty_n,-\K \bh^\infty_n)
\end{aligned}
\]
so that
\[ 
(\hat\bh,-\widetilde{\mathcal L}\hat\bh)\leq
\liminf_{n\to\infty}(\bh^\infty_n,-\widetilde{\mathcal L}
\bh^\infty_n)=\delta_4
\]
while $(\hat\bh,-\widetilde{\mathcal L}\hat\bh)\geq \delta_4 \|\hat\bh\|_2$ 
since $\hat\bh\in\mathbf V_4^e$. Thus we need to show that $\|\hat\bh\|_2=1$.

To this end observe that for every $M>0$ we have
\[
 \rho M\sum_{k=M+1}^\infty\|\bh_{n,k}\|_2^2\leq
\rho \sum_{k=1}^\infty k\|\bh_{n,k}\|_2^2=
(\bh_n,-\mathcal G\bh_n)\leq 
(\bh_n,-\widetilde{\mathcal L}\bh_n)\leq 2\delta_4
\]
definitively in $n$. Thus, for every $\epsilon$ there exists $M$ such that 
$\sum_{k=0}^M\|\bh_{n,k}\|_2^2\geq1-\epsilon$ definitively in $n$. Taking 
the limit this implies that for every $\epsilon$ there exists $M$ such that 
$\sum_{k=0}^M\|\hat \bh_{k}\|_2^2\geq1-\epsilon$ and thus we get 
$\|\hat\bh\|=1$. This concludes the proof of Theorem 
\ref{thm:gap2}.\qed

\subsection{Proof of Theorem \ref{thm:entropy}.}
\label{sec:3.3}

To simplify notation, given $\bff=\bh\boldsymbol \Gamma$, we set 
$S(\bh)={\mathcal 
S}(\bff\,|\,\boldsymbol \Gamma) $ and we define 
\begin{align*}
 \Psi(\bh)&=\sum_{N=0}^\infty a_{N}\int d\uv_{N+1}(h_{N+1}-h_{N})(\log 
h_{N+1}-\log h_{N})\gamma_{N+1}(\underline v_{N+1})\\
 E(\bh)&=\sum_{N=0}^\infty a_{N}\int d\uv_{N}h_N(\uv_{N})\gamma_N(\uv_{N}).
\end{align*}
Finally we observe that if $\bff\in L^1_s(\R)$ then 
$\bh\in L^1_s(\R,\boldsymbol\Gamma)$ and $e^{\mathcal 
Lt}\bff=(e^{\widetilde{\mathcal 
L}t}\bh)\boldsymbol \Gamma$ with 
$\widetilde{\mathcal L}=\mathcal G+\tilde\lambda\K$ defined in section 
\ref{sec:3.1} but now considered as an operators
on $L^1_s(\R,\boldsymbol{\Gamma})$.

To obtain an explicit expression for $\frac{d}{dt}S(\bh(t))$, where 
$\bh(t)=e^{\widetilde{\mathcal L}t}\bh$ we need to exchange the order of the 
derivative in $t$ with the sum over $N$ and the integral over $\uv_N$. To do 
this we will use the following two Lemmas that will allow us to use Fatou's 
Lemma to excahnge derivative and integrals. 

\begin{lem} \label{lem:limderi}
Given $\bff\in L^1(\R)$ we have
\begin{align*}
 &\lim_{t\to 0^+}\left(\left(e^{\mathcal L t
}\bff\right)_N(\uv_N)-f_N(\uv_N)\right)=0\\
  &\lim_{t\to 0^+} \frac{1}{t}\left(\left(e^{\mathcal L t
}\bff\right)_N(\uv_N)-f_N(\uv_N)\right)=\left(\mathcal L \bff\right)_N(\uv_N)
\end{align*}
for every $N$ and almost every $\uv_N$.
\end{lem}
\noindent\emph{Proof}. See appendix \ref{app:bosons.limderi}.
\medskip

\begin{lem}\label{lem:conve} If $\bh\boldsymbol\Gamma\in L^1_s(\R)$ then
 \[
 h_N(t)\log(h_N(t))\leq \left(e^{\widetilde{\mathcal L}t} 
 (\bh\log\bh)\right)_N\, .
\]
\end{lem}
\noindent\emph{Proof}. See appendix \ref{app:conve}.
\medskip

After setting
\[
 \frac{d_+}{dt}S(\bh(t)):=\limsup_{h\to 0^+} 
\frac{1}{h}(S(\bh(t+h))-S(\bh(t)))\,,
\]
we are ready to estimate of the variation in time of $S(\bh)$. 

\begin{lem} \label{lem:d+}
Let $\bh$ be such that $\bh\boldsymbol\Gamma\in D^1$ and 
$\bh\log\bh\boldsymbol\Gamma\in D^1$
then we have
\begin{gather*}
\frac{d_+}{dt}S(\bh(t))\leq-\mu\Psi(\bh(t))\, .
\end{gather*}
\end{lem}

\noindent\emph{Proof}. 
From Lemma \ref{lem:conve} we get
\begin{align*}
\frac{1}{t}\bigl( 
h_N(\uv_N,t)\log(h_N(\uv_N,t))-&h_N(\uv_N)\log(h_N(\uv_N))\bigr)-\\
&\frac{1}{t}\left(\left(e^{\widetilde{\mathcal 
L}t}(\bh\log\bh)\right)_N(\uv_N)-h_N(\uv_N)\log(h_N(\uv_N))\right)\leq 0\, .
\end{align*}
Since $\bh 
\log\bh\boldsymbol\Gamma\in L^1(\R)$, conservation of probability gives
\begin{align*}
 \sum_{N=0}^\infty a_N\int_{\Rs^N} \left(\left(e^{\widetilde{\mathcal 
L}t}(\bh\log\bh)\right)_N(\uv_N)\gamma_N(\uv_N)- 
h_N(\uv_N)\log(h_N(\uv_N))\gamma_N(\uv_N)\right)d\uv_N=0
\end{align*}
so that by Fatou's Lemma
\[
\begin{aligned}
 \limsup_{t\to 0^+} &\frac{1}{t}(S(\bh(t))-S(\bh))\\
 &\leq \sum_{N=0}^\infty 
a_N\int_{\Rs^N}\limsup_{t\to 0^+} \frac{1}{t}\bigl( 
h_N(\uv_N,t)\log(h_N(\uv_N,t))-h_N(\uv_N)\log(h_N(\uv_N))\bigr)\\
&-\sum_{N=0}^\infty 
a_N\int_{\Rs^N}\limsup_{t\to 0^+}\frac{1}{t}\left(\left(e^{\widetilde{\mathcal 
L}t}(\bh\log\bh)\right)_N(\uv_N)-h_N(\uv_N)\log(h_N(\uv_N))\right)
\end{aligned}
\]
and, using Lemma \ref{lem:limderi}, we get
\[
\begin{aligned}
 \limsup_{t\to 0^+} \frac{1}{t}(S(\bh(t))-S(\bh))&\leq \sum_{N=0}^\infty a_N
 \int_{\Rs^N} (\widetilde{\mathcal 
L}\bh)_N(\uv_N)(\log(h_N(\uv_N))+1)\gamma_N(\uv_N)d\uv_N\\
&-\sum_{N=0}^\infty a_N
 \int_{\Rs^N}\left(\widetilde {\mathcal 
L}(\bh\log\bh)\right)_N(\uv_N)\gamma_N(\uv_N)d\uv_N
\, .
\end{aligned}
\]
Since $\boldsymbol \Gamma \bh\in D^1$ and $\boldsymbol \Gamma \bh\log\bh\in 
D^1$, \eqref{sumdotf} gives
\begin{align*}
\frac{d_+}{dt}S(&\bh(t))\bigr|_{t=0}\leq
\sum_{N=0}^\infty a_N
 \int d\uv_N\gamma_N(\widetilde{\mathcal 
L}\bh)_N\log(h_N)
\\
=&\sum_{N=0}^\infty a_N\int d\uv_N\gamma_N\left(\rho 
(\P^+\bh)_{N}+\mu 
(\P^-\bh)_{N}-(\mu+\rho N)h_N+\tilde \lambda K_Nh_N\right)\log 
h_N
\\
\leq&\sum_{N=0}^\infty a_N\int d\uv_N\gamma_N\left(\rho 
(\P^+\bh)_{N}+\mu 
(\P^-\bh)_{N}-(\mu+\rho N)h_N\right)\log  
h_N
\end{align*}
where we have used that $\int d\uv_N\gamma_N (K_Nh_N)\log 
h_N\leq 0$. Observe finally that
\begin{align*}
\int d\uv_N\gamma_N\rho (\P^+\bh)_N\log h_N=&\int d\uv_N\gamma_NN\rho 
h_{N-1}\log h_N\\
\int d\uv_N\gamma_N\mu (\P^-\bh)_N\log h_N=&\int d\uv_{N+1}\gamma_{N+1}\mu h_{N+1}\log h_N
\end{align*}
from which we get
\begin{align*}
\frac{d_+}{dt}S(\bh(t))\bigr|_{t=0}\leq&\sum_{N=1}^\infty a_N\int 
d\uv_N\gamma_NN\rho 
(\tilde h_{N-1}(t)-\tilde h_N(t))\log \tilde h_N(t)\\
+&\mu\sum_{N=0}^\infty a_N\int d\uv_{N+1}\gamma_{N+1} (\tilde h_{N+1}(t)-\tilde 
h_N(t))\log \tilde h_N(t)
\end{align*}
The thesis follows by reindexing the first sum and using 
\eqref{arec}.\qed\medskip


Thus to show that $S(\bh(t))$ decays exponentially we need a lower bound for 
$\Psi(\bh)$ in terms of $S(\bh)$. This is the content of the following Lemma 
that 
is the main result of this section.

\begin{lem}\label{lem:ent} If $\boldsymbol \Gamma\bh\in L^1_s(\R)$ with 
$S(\bh)<\infty$, then 
\begin{equation}\label{Psi}
S(\bh)\leq E(\bh)\log E(\bh) +\f{\mu}{\rho}\Psi(\bh)\,.
\end{equation}
\end{lem}

\begin{rem}\emph{
The idea behind the proof of \eqref{Psi} is to think of the entry and exit 
processes 
defined by the thermostat as a continuous family of independent entry 
processes, 
one for each possible velocity $v$, with entry rates $\mu\gamma(v)dv$, while 
each particle in the system leaves with rate $\rho$ independent of its velocity.
Clearly such a description makes little mathematical sense and, as a first 
step, 
one may think of approximating the original process by restricting the velocity 
of each particle to assume only a finite number of values $\bar v_k$, 
$k=1,\ldots,K$, characterized by suitable entry rates $\omega_k$. After this, 
using convexity, we reduce the proof of \eqref{Psi} to the case with $K=1$, 
essentially equivalent to the case in which  all particles in the thermostat 
have the same velocity. In this situation, 
we further approximate the infinite reservoir by a large finite reservoir 
containing $M$ particles that enter and leave the system, independently from 
each other, at a suitable rate. Convexity will allow us to reduce this 
situation 
to that of a single particle jumping from the system to the reservoir and back. 
The final 
step is thus Lemma \ref{lem:bin} below that deals with this situation. This 
argument is inspired by the proof of the Logarithmic Sobolev Inequality in 
\cite{Gross1}.}\label{rem:strat}
\end{rem}

\begin{rem}\emph {In 
the proof of Lemma \ref{lem:d+} we required that $\bh\boldsymbol\Gamma\in D^1$ 
and $\bh\log\bh\boldsymbol\Gamma\in D^1$ only to differentiate 
$e^{t\widetilde{\mathcal L}}\bh$ and show that $\sum_{N=0}^\infty a_N\int 
(\widetilde{\mathcal L}\bh)_N \gamma_Nd\uv_N=0$ and similarly for $\bh\log\bh$. 
We 
believe it is possible to implement the strategy outlined in Remark 
\ref{rem:strat}, and developed in the proof below, directly to  
$S(\bh)$ thanks to the representation of the evolution described in Remark 
\ref{rem:jumps}. This would eliminate the need for conditions on $\bh$ but it 
would make the proof below unnecessarily involved. 
}\end{rem}

\noindent\emph{Proof of Lemma \ref{lem:ent}.}
A way to make the first step of the discussion in Remark \ref{rem:strat} 
rigorous is to {\it coarse 
grain}, that is to approximate each $h_N$ by a simple function obtained by 
averaging it over the element of a partition of $\Rs^N$ made by rectangles 
obtained as the Cartesian product of a finite number of measurable set of $\Rs$.

More precisely, we call  
$\mathcal{B}=\{B_k\}_{k=1}^K$ a {\it (measurable) partition} of $\Rs^N$ if  
$B_k\subset\Rs^N$ are measurable and
$\bigcup_k B_k=\mathbb{R}^N$ while $B_k\cap B_{k'}=\emptyset$ if $k\not=k'$.
Given a measurable partition $\mathcal{B}$ let $I_k(\uv_N,\underline w_N)$ be 
the indicator function of $B_k\times B_k\subset \Rs^{2N}$ and define the {\it 
coarse graining} kernel:
\[
 C_\mathcal{B}(\uv_N,\underline w_N)=\sum_{k=1}^K 
\frac{1}{\omega_{k}}I_k(\uv_N,\underline w_N)
\quad
\mathrm{with}
\quad
 \omega_{k}=\int_{B_k}\gamma_N(\uv_N)d\uv_N\,.
\]
Clearly, for every $\underline w_N$ we have
\[
 \int_{\Rs^N}C_\mathcal{B}(\uv_N,\underline w_N)\gamma(\underline v_N)d\uv_N=1
\]
while $C_\mathcal{B}(\uv_n,\underline w_N)=C_\mathcal{B}(\underline 
w_N,\uv_N)$.
Given a function $h_N$ is $L^1(\Rs^N)$ we can define its {\it coarse grained} 
version as
\[
 h_{N,\mathcal{B}}(\uv_N)=\int_{\Rs^N}C_\mathcal{B}(\uv_N,\underline 
w_N)h_N(\underline 
w_N)\gamma(\underline 
w_N)d\underline w_N\, .
\]
Observe that, if $\uv_N\in B_k$ then
\[
h_{N,\mathcal{B}}(\uv_N)=\frac{1}{\omega_{k}}\int_{B_k}\gamma(\underline 
w_N)h_N(\underline w_N)d\underline w_N\, .
\]
This means that $h_{N,\mathcal{B}}(\uv_N)$ is a simple function that assumes 
only $K$ possible values. Finally we have 
$\int_{\Rs^N}h_{N,\mathcal{B}}(\uv_N)\gamma(\uv_N)d\uv_N=\int_{\Rs^N}h_N(\uv_N)
\gamma(\uv_N) d\uv_N$.

Given measurable partitions $\mathcal B=\{B_k\}_{k=1}^K$ and 
$\mathcal B'=\{B'_j\}_{j=1}^J$of $\Rs^N$ and $\Rs^M$ respectively, we can define 
the {\it product} partition $\mathcal B\times \mathcal B'=\{B_k\times 
B'_j\,|k=1,\ldots,K\,\,\,
j=1,\ldots,J\}$ of $\Rs^{N+M}$. Observe that the coarse graining kernel of 
$\mathcal B\times \mathcal B'$ satisfies
\[
 C_{\mathcal B\times \mathcal B'}(\uv_N,\uv'_M,\underline w_N,\underline w'_M)=
C_{\mathcal B}(\uv_N,\underline w_N)C_{\mathcal B'}(\uv'_M,\underline w'_M)\, .
\]

Finally, given a partition $\mathcal B=\{B_k\}_{k=1}^K$ of $\Rs$, 
and $\uk=(k_1,\ldots,k_N)\in \{1,\ldots,K\}^N$
we consider the set $B_{\uk}=\btimes_i B_{k_i}\subset \mathbb{R}^N$. Clearly 
the $B_{\uk}$ form a 
measurable partition of $\Rs^N$ that we will denote as $\mathcal B^N$. As 
before, we can define the coarse graining kernel 
for $\mathcal B^N$ as
\[
 C_{\mathcal{B}^N}(\uv_N,\underline w_N)=\sum_{\uk\in\{1,\ldots, K\}^N} 
\frac{1}{\omega_{\uk}}I_{\uk}(\uv_N,\underline w_N)
\]
where $\omega_{\uk}=\prod_{i=1}^{N}\omega_{k_i}$ and 
$I_{\uk}(\uv_N,\underline w_N)$ is the characteristic function of 
$B_{\uk}\times B_{\uk}\in\Rs^{2N}$. Moreover the coarse grained version of 
$h_N\in L^1(\Rs^N,\gamma_N)$ is
\[
 h_{N,\mathcal{B}^N}(\uv_N)=\int_{\Rs^N}\gamma(\underline 
w_N)C_{\mathcal{B}^N}(\uv_N,\underline w_N)h_N(\underline w_N)d\underline w_N\, 
.
\]
Again, if $\uv_N\in B_{\uk}$ we have
\[
  h_{N,\mathcal 
B^N}(\uv_N)=\frac{1}{\omega_{\uk}}\int_{B_{\uk}}h_N(\uv_N)\gamma_{N}(\underline 
v_{N})d \uv_{N}:=\bar h_{N,\mathcal B^N}(\uk)
\]
and $h_{N,\mathcal{B}^N}(\uv_N)$ assumes only the $K^N$ possible values $\bar 
h_{N,\mathcal B^N}(\uk)$.
Observe finally that, since
\[
 C_{\mathcal{B}^N}(\uv_N,\underline w_N)=\prod_{i=1}^NC_{\mathcal B}(v_i,w_i)\, 
,
\]
we can write
\begin{equation}\label{same}
 h_{N-1,\mathcal{B}^{N-1}}(\uv_{N-1})=\int_{\Rs^N}\gamma(\underline 
w_N)C_{\mathcal{B}^N}(\uv_N,\underline w_N)h_{N-1}(\underline 
w_{N-1})d\underline w_N\, .
\end{equation}
Given a state $\bh$ and a partition $\mathcal B$ of 
$\Rs^N$, we define the coarse grained version $\bh_{\mathcal{B}}$ of $\bh$ over 
$\mathcal B$ 
by setting $h_{\mathcal{B},N}=h_{N,\mathcal B^N}$. 
Since $x\log(x)$ is convex in $x$ and 
$(x-y)(\log(x)-\log(y))$ is jointly convex in $x$ and $y$, for 
every partition $\mathcal B$ of $\Rs$, we get
\begin{equation}\label{relaSPE}
 S(\bh_{\mathcal B})\leq S(\bh),\qquad \Psi(\bh_{\mathcal{B}})\leq 
\Psi(\bh),\qquad E(\bh_{\mathcal{B}})=E(\bh)
\end{equation}
where in the inequality for $\Psi$ we used \eqref{same}.
On the other hand, we have the following Lemma. 

\begin{lem}\label{lem:coarse}
 Given $\bh$, for every $\epsilon$ we can find a finite measurable  
partition $\mathcal B$ of $\Rs$ such that
\[
S(\bh)-S(\bh_{\mathcal B})\leq\epsilon
\]
\end{lem}
\noindent\emph{Proof}. See Appendix \ref{app:bosons.coarse}.
\medskip

We thus claim that to prove Lemma \ref{lem:ent} we just need to show that, for 
every finite partition $\mathcal B$ of $\Rs$ and every state $\bh$ we have
\begin{equation}\label{entB}
 S(\bh_{\mathcal B})\leq E(\bh_{\mathcal B})\log E(\bh_{\mathcal 
B})+\frac{\mu}{\rho}\Psi(\bh_{\mathcal B})\, .
\end{equation}
To see this observe that Lemma \ref{lem:coarse}, together with \eqref{relaSPE} 
and \eqref{entB}, implies that for every 
$\epsilon$ we can find a partition $\mathcal B$ such that
\begin{align*}
 S(\bh)\leq S(\bh_{\mathcal B})+\epsilon\leq &E(\bh_{\mathcal B})\log 
E(\bh_{\mathcal 
B})+\frac{\mu}{\rho}\Psi(\bh_{\mathcal B})+\epsilon\\  
\leq&E(\bh)\log E(\bh)+\f{\mu}{\rho}\Psi(\bh)+\epsilon\, .
\end{align*}

Thus we consider a given finite partition 
$\mathcal B=\{B_k\}_{k=1}^K$ and a given state $\bh$. Since 
$h_{\mathcal B, N}$ takes only finitely many values, it should be possible to 
transform the integrals defining $E(\bh_{\mathcal B})$, $S(\bh_{\mathcal B})$ 
and $\Psi(\bh_{\mathcal B})$ into summations. To do this, given 
$\uk\in\{1,\ldots, K\}^N$, we define the {\it occupation numbers} $\underline 
n(\uk)=(n_1(\uk),\dots,n_K(\uk))\in \mathbb{N}^K$ as
\[
n_q(\uk)=\sum_i \delta_{q,k_i}\,.
\]
That is $n_q(\uk)$is the number of $i$ such that $k_i=q$.
In other words, if $\uv_N\in B_{\uk}$ then there are $n_q(\uk)$ particles with 
velocity in $B_q$.

The fact that $h_N$ is invariant under permutation 
of its arguments implies that $\bar h_{N,\mathcal B^N}(\uk)$ depends only on  
$\un(\uk)$ or, more precisely, if $\underline n(\uk)=\underline 
n(\uk')$ then $\bar h_{N,\mathcal B^N}(\uk)=\bar h_{N,\mathcal B^N}(\uk')$. 
This allow us to define
the function $F:\mathbb{N}^K\to \mathbb{R}$ given by
\[
 F(\un)=\bar h_N(\uk)\quad\mathrm{if}\quad \un=\un(\uk),\;\;\mathrm{ and }\;\; 
N=\sum_{k=1}^K n_k:=|\un|\, .
\]
Using this definition and the fact that $\sum_{k=1}^K\omega_k=1$, we can now 
write
\begin{equation}\label{tildeE}
\begin{aligned}
 E(\bh_{\mathcal 
B})=&\sum_N a_N\sum_{\uk\in\{1,\ldots,K\}^N}\bar 
h_{N,\mathcal B^N}(\uk)\omega_{\uk}\\
=&\sum_N \frac{e^{-\frac{\mu}{\rho}}}{N!}\left(\frac{\mu}{\rho}\right)^N
\sum_{|\underline n|=N}\binom{N}{n_1,\ldots,n_K}F(\un)\prod_{k=1}^K
\omega_k^{n_k}\\
=&\sum_{\un\in\Ns^K}F(\un)\prod_{k=1}^K \pi_{\alpha_k}(n_k):=\widetilde 
E_{\underline \alpha_K}(F)
\end{aligned}
\end{equation}
where $\underline 
\alpha_K=(\alpha_1,\ldots,\alpha_K)$ with $\alpha_k=\mu\omega_k/\rho$ and
\[
\pi_{\alpha}(n)=e^{-\alpha}\frac{\alpha^n}{n!}\,,
\]
that is $\pi_{\alpha_k}$ is the Poisson distribution with expected value
$\alpha_k$.
Similarly we have
\begin{equation}\label{tildeS}
\begin{aligned}
 S(\bh_{\mathcal B})=&\sum_N a_N\sum_{\uk\in\{1,\ldots,K\}^N}\bar 
h_{N,\mathcal B^N}(\uk)\log(\bar h_{N,\mathcal B^N}(\uk))\omega_{\uk}\\
 =&\sum_{\un\in\mathbb{N}^K} F(\un)\log(F(\un))\prod_{k=1}^K 
\pi_{\alpha_k}(n_k):=\widetilde S_{\underline \alpha_K}(F)
\end{aligned}
\end{equation}
Finally setting $\un^{q}=(n_1,\ldots,n_q+1,\ldots,n_K)$ we get
\begin{align}\label{tildePsi}
\Psi(\bh_{\mathcal B})=&\sum_N a_{N}\sum_{\uk\in\{1,\ldots,K
\}^{N}}\sum_{q=1}^K
(\bar h_{N+1, \mathcal 
B^{N+1}}(\uk,q)-\bar h_{N,\mathcal B^N}(\uk))\cdot\nonumber\\
&\qquad(\log\bar h_{N+1,\mathcal 
B^{N+1}}(\uk,q)-
\log\bar h_{N,\mathcal B^N}(\uk))\omega_{\uk}\omega_q\\
=&\frac{\rho}{\mu}\sum_{q=1}^K\alpha_q\sum_{\un\in\Ns^K} 
\left(F(\un^q)-F(\un)\right)\left(\log F(\un^q)-\log 
F(\un)\right)\prod_{k=1}^K \pi_{\alpha_k}(n_k)\nonumber\\
:=&\frac{\rho}{\mu}\widetilde 
\Psi_{\underline \alpha_K}(F)\, .\nonumber
\end{align}
so that, to prove \eqref{entB}, we need to show that, for every 
$F:\Ns^K\to\Rs_+$ and for every $K$ and $\underline \alpha_K\in \Rs_+^K$, if 
$\widetilde S_K(F)<\infty$ then
\begin{equation}\label{tildeentK}
\widetilde S_{\underline\alpha_K}(F)\leq \widetilde 
\Psi_{\underline\alpha_K}(F)+\widetilde E_{\underline\alpha_K}(F)\log 
\widetilde E_{\underline\alpha_K}(F)\, .
\end{equation}


We will prove \eqref{tildeentK} by induction over $K$. Assume that 
\eqref{tildeentK} is valid for every index less than $K$ for 
some $K>1$ and write
\[
\widetilde S_{\underline \alpha_{K-1}}(F(\cdot,n_K))= 
\sum_{\un'\in\Ns^{K-1}}F(\un',n_K)\log F(\un',n_K) 
\prod_{k=1}^{K-1} \pi_{\alpha_k}(n_k)
\]
and similar expression for $E_{\underline \alpha_{K-1}}(F(\cdot,n_K))$ and 
$\Psi_{\underline \alpha_{K-1}}(F(\cdot,n_K))$. 

Using the inductive hypothesis we obtain
\begin{align*}
 \widetilde S_{\underline \alpha_{K}}(F)=&\sum_{n_K=0}^\infty\widetilde 
S_{\underline \alpha_{K-1}}(F(\cdot,n_K))\pi_{\alpha_K}(n_K)
\leq\sum_{n_K=0}^\infty\widetilde\Psi_{\underline 
\alpha_{K-1}}(F(\cdot,n_K))\pi_{\alpha_K}(n_K)\\
+& \sum_{n_K=0}^\infty\widetilde E_{\underline \alpha_{K-1}}(F(\cdot,n_K))\log 
\widetilde 
E_{\underline \alpha_{K-1}}(F(\cdot,n_K))\pi_{\alpha_K}(n_K)\nonumber\, .
\end{align*}
Calling $F_1(n_K)=\widetilde E_{\underline \alpha_{K-1}}(F(\cdot,n_K))$ and 
using the inductive hypothesis again we get
\[
\begin{aligned}
\sum_{n_K=0}^\infty\widetilde E_{\underline \alpha_{K-1}}(F(\cdot,n_K))\log 
\widetilde E_{\underline\alpha_{K-1}}(F(\cdot,n_K))\pi_{\alpha_K}(n_K)=
&\widetilde S_{\alpha_K}(F_1)\\
\leq&\widetilde\Psi_{\alpha_K}(F_1) +
\widetilde E_{\alpha_K}(F_1)\log \widetilde E_{\alpha_K}(F_1) 
\end{aligned}
\]
so that
\begin{align}\label{induK}
 \widetilde S_{\underline \alpha_{K}}(F)
\leq\sum_{n_K=0}^\infty\widetilde\Psi_{{\underline 
\alpha_{K-1}}}(F(\cdot,n_K))\pi_{\alpha_K}(n_K)+
\widetilde\Psi_{\alpha_K}(F_1) +
\widetilde E_{\alpha_K}(F_1)\log \widetilde E_{\alpha_K}(F_1)\, .
\end{align}
Observing that $\widetilde 
E_{\alpha_K}(F_1)=\widetilde E_{\underline \alpha_{K}}(F)$ and that, by 
convexity,
\[
\begin{aligned}
\widetilde\Psi_{\alpha_K}(F_1)=&\alpha_K\sum_{n=0}^\infty 
(F_1(n+1)-F_1(n))(\log 
F_1(n+1)-\log F_1(n))\pi_{\alpha_K}(n)\\
\leq&\alpha_K\sum_{\un \in \Ns^K}
\left(F(\un^K)-F(\un)\right)\left(\log F(\un^K)-\log 
F(\un)\right)\prod_{k=1}^K \pi_{\alpha_k}(n_k)
\end{aligned}
\]
we get \eqref{tildeentK} for $K$. Thus, by induction, to prove 
\eqref{tildeentK} for every $K$ we just need to prove it for $K=1$.
This is the content of the following Lemma. 

\begin{lem}\label{lem:ent1} Let $\pi_\alpha$ be
the Poisson distribution on $\mathbb{N}$ with expected value $\alpha>0$ 
and 
$f:\mathbb{N}\to \mathbb{R}^+$ be such that 
\[
\sum_{n=0}^\infty f(n)\log f(n)\pi_\alpha(n)<\infty\,,
\] 
then we have
\begin{align}\label{Poisoonent}
 \sum_{n=0}^\infty f(n)\log f(n) \pi_\alpha(n)\leq &\left(\sum_{n=0}^\infty 
f(n) 
\pi_\alpha(n)\right)\log \left(\sum_{n=0}^\infty 
f(n) \pi_\alpha(n)\right)\\
+&\alpha\sum_{n=0}^\infty \left(f(n+1)-f(n)\right)\left(\log f(n+1)-\log 
f(n)\right)\pi_\alpha(n)\, .\nonumber
\end{align}
\end{lem}

\noindent\emph{Proof}. Observe first that since 
$\alpha\pi_\alpha(n)=(n+1)\pi_\alpha(n+1)$ we get
\[
\begin{aligned}
 \alpha\sum_{n=0}^\infty \left(f(n+1)-f(n)\right)&\left(\log f(n+1)-\log 
f(n)\right)\pi_\alpha(n)\\
&=\sum_{n=1}^\infty n\left(f(n)-f(n-1)\right)\left(\log f(n)-\log 
f(n-1)\right)\pi_\alpha(n)\, . 
\end{aligned}\,
\]
Let now $\pi_{\alpha,N}(n)$ be the binomial distribution with parameters $N$ 
and $\alpha/N$, that is
\[
\pi_{\alpha,N}(n)=\binom{N}{n}\left(\frac{\alpha}{N}\right)^n
\left(1-\frac{\alpha}{N}\right)^{N-n}\, .
\]
We will prove by induction that for every $N$ and every $\alpha\leq N$ we have
\begin{align}\label{ind}
 \sum_{n=0}^N f(n)\log f(n) \pi_{\alpha,N}(n)\leq &\left(\sum_{n=0}^N 
f(n) \pi_{\alpha,N}(n)\right)\log \left(\sum_{n=0}^N 
f(n) \pi_{\alpha,N}(n)\right)\\
+&\sum_{n=1}^N n\left(f(n)-f(n-1)\right)\left(\log f(n)-\log 
f(n-1)\right)\pi_{\alpha,N}(n)\,\nonumber
\end{align}
so that, taking the limit for $N\to\infty$, we will obtain \eqref{Poisoonent}.
The base case $N=1$ is covered by the following Lemma.

\begin{lem}\label{lem:bin}
 Let $\mu_x\geq 0$, $x\in\{0,1\}$, be such that 
$\mu_0+\mu_1=1$ then for every function $f:\{0,1\}\to\mathbb{R}^+$ we have
\begin{align}\label{bin}
 \sum_{x=0,1} f(x)\log f(x) \mu_x\leq& \left(\sum_{x=0,1} f(x) \mu_x\right)\log 
\left(\sum_ {x=0,1}f(x)\mu_x\right)\crcr
+&\mu_0\mu_1\left(f(1)-f(0)\right)\left(\log f(1)-\log f(0)\right)\, .
\end{align}
\end{lem}

\noindent\emph{Proof}. Calling 
$h(0)=f(0)/(\mu_0f(0)+\mu_1 f(1))$ and 
$h(1)=f(1)/(\mu_0f(0)+\mu_1 f(1))$, \eqref{bin} becomes
\[
 \sum_{x=0,1} h(x)\log h(x) \mu_x\leq 
\mu_0\mu_1\left(h(1)-h(0)\right)\left(\log h(1)-\log h(0)\right)\, .
\]
Since $\mu_0 h(0)+\mu_1 h(1)=1$ we can write $h(0)=1+\delta \mu_1$ and
$h(1)=1-\delta \mu_0$ and we get
\[
\begin{aligned}
\sum_{x=0,1}  h(x)&\log h(x) \mu_x\\
=&\mu_0\mu_1\delta(\log(1+\delta \mu_1)- 
\log(1-\delta \mu_0))+\mu_0\log(1+\delta\mu_1)+\mu_1\log(1-\delta\mu_0)\\
\leq&\mu_0\mu_1\delta(\log(1+\delta \mu_1)-\log(1-\delta 
\mu_0))\\
=&\mu_0\mu_1\left(h(1)-h(0)\right)\left(\log h(1)-\log h(0)\right)
\end{aligned}
\]
where we have used concavity of the logarithm.\qed
\medskip

Assume now that \eqref{ind} 
holds for every index less than $N$. Given $\alpha\leq N$ call 
$\beta=(N-1)\alpha/N$ so that $\beta\leq N-1$. Define also $\mu_0=1-\alpha/N$, 
$\mu_1=\alpha/N$, and observe that, for every $J:\mathbb{N}\to \mathbb{R}$,
\begin{equation}\label{Nm1}
 \sum_{n=0}^N J(n)\pi_{\alpha,N}(n)=\sum_{x=0,1}\sum_{n=0}^{N-1}
J(n+x)\pi_{\beta,N-1}(n)\mu_x\, .
\end{equation}
Calling
\[
 \bar f(x)=\sum_{n=0}^{N-1} f(n+x)\pi_{\beta,N-1}(n)
\]
and using \eqref{Nm1} and the inductive hypothesis for index $N-1$, we get
\begin{align}
\sum_{n=0}^N f(n)\log f(n) 
\pi_{\alpha,N}&(n)=\sum_{x=0,1}\sum_{n=0}^{N-1} f(n+x)\log 
f(n+x) \pi_{\beta,N-1}(n)\mu_x\nonumber\\ 
\leq&\sum_{x=0,1}\bar f(x)\log \bar f(x)\mu_x+
\sum_{x=0,1}\sum_{n=1}^{N-1} n\left(f(n+x)-f(n-1+x)\right)\cdot\nonumber\\
&\quad\qquad\qquad \left(\log f(n+x)-\log 
f(n-1+x)\right)\pi_{\beta,N-1}(n)\mu_x\nonumber
\end{align}
while using Lemma \ref{lem:bin} for the first term in the second line delivers
\begin{align}\label{toolong}
\sum_{n=0}^N f(n)\log f(n) 
\pi_{\alpha,N}(n)\leq&\left(\sum_{n=0}^N
f(n)\pi_{\alpha,N}(n)\right)\log\left(\sum_{n=0}^N
f(n)\pi_{\alpha,N}(n)\right)\nonumber\\
+&\mu_0\mu_1(\bar f(1)-\bar f(0))(\log \bar f(1)-\log\bar f(0))\\
+&\sum_{x=0,1}\sum_{n=1}^{N-1} n\left(f(n+x)-f(n-1+x)\right)\nonumber\\
&\qquad\cdot\left(\log f(n+x)-\log 
f(n-1+x)\right)\pi_{\beta,N-1}(n)\mu_x\nonumber
\end{align}
Finally using the joint convexity in $(x,y)$ of the function $(x-y)(\log x - 
\log y)$ and the fact that $\mu_0<1$ we can write
\[
\begin{aligned}
 \mu_0\mu_1&(\bar f(1)-\bar f(0))(\log \bar f(1)-\log\bar f(0))\\
\leq & \mu_1\sum_{n=0}^{N-1} (f(n+1)-f(n))(\log f(n+1)-\log 
f(n))\pi_{\beta,N-1}(n)
\end{aligned}
\]
that inserted in \eqref{toolong} gives
\begin{align*}
\sum_{n=0}^N f(n)\log f(n)  
\pi_{\alpha,N}(n)\leq&
\left(\sum_{n=0}^N
f(n)\pi_{\alpha,N}(n)\right)\log\left(\sum_{n=0}^N
f(n)\pi_{\alpha,N}(n)\right)\nonumber\\
+&\sum_{x=0,1}\sum_{n=1}^{N-1} 
(n+x)\left(f(n+x)-f(n-1+x)\right)\\
&\qquad\cdot\left(\log 
f(n+x)-\log f(n-1+x)\right)\pi_{\beta,N-1}(n)\mu_x\, .
\end{align*}
Changing summation variables from $(x,n)$ 
to $(x,n+x)$ and using \eqref{Nm1} we obtain \eqref{ind} for index $N$. Thus 
\eqref{ind} is valid for every $N\geq 1$ and every $\alpha\leq N$.

To complete the proof of Lemma \ref{lem:ent1} we need to show that we can take 
the limit for $N\to \infty$ in \eqref{ind}. To this end observe that given 
$\alpha$, for $N$ large enough we have
$0<\left(1-\frac{\alpha}{N}\right)^{N}\leq 2e^{-\alpha}$. Thus for large $N$ 
and $\alpha<n\leq N$ we get
\begin{align}\label{dom}
 \pi_{\alpha,N}(n)\leq &
2e^{-\alpha}\frac{(\alpha)^n}{n!}\left(1-\frac{\alpha}{N}\right)^{-n}
\prod_{i=1}^n\left(1-\frac{i}{N} \right)\nonumber\\
\leq&2e^{-\alpha}\frac{(\alpha)^n}{n!}\left(1-\frac{\alpha}{N}\right)^{-
\lfloor\alpha \rfloor}
\prod_{i=1}^{\lfloor\alpha
\rfloor}\left(1-\frac{i}{N} \right)\leq 4\pi_\alpha(n)\, .
\end{align}
Using Dominated Convergence, \eqref{dom} implies that, if $f(n)$ is bounded 
below and $\sum_{n=0}^\infty f(n)\pi_{\alpha}(n)\leq \infty$ then
\[
 \lim_{N\to\infty} \sum_{n=0}^N f(n)\pi_{\alpha,N}(n)=\sum_{n=0}^\infty 
f(n)\pi_{\alpha}(n)\, .
\]
We can now let $N\to\infty$ in \eqref{ind} to obtain  \eqref{Poisoonent}.
This concludes the proof of Lemma \ref{lem:ent1}.\qed
\medskip

To sum up, the validity of \eqref{Poisoonent} together with the inductive 
argument in \eqref{induK} shows that \eqref{tildeentK} is valid for 
every $K$ and $\alpha_k$, $k=1,\ldots, K$. This in turn, together with 
\eqref{tildeE}, \eqref{tildeS} and \eqref{tildePsi}, 
establishes the validity of \eqref{entB} for every state $\bh$ and every 
partition 
$\mathcal B$ of $\Rs$. This, together with Lemma \ref{lem:ent1} completes the 
proof of Lemma \ref{lem:ent}.\qed
\medskip

Observe now that if $\bff$ is a probability distribution $E(\bh)=1$ so that 
Lemma \ref{lem:ent}, together with Lemma 
\ref{lem:d+}, gives
\begin{equation}\label{precise}
 \frac{d_+}{dt} S(\bh(t))\leq -\rho S(\bh(t))
\end{equation}
To complete the proof of Theorem \ref{thm:entropy} we have to show that 
\eqref{precise} implies \eqref{decay}. To this end, take $\rho'<\rho$, assume 
that there exists $t$ such that $S(\bh(t))> e^{-\rho' t} S(\bh(0))$ and let
\[
T=\inf\{t\geq 0\,|\, S(\bh(t))> e^{-\rho' t} S(\bh(0))\}\, .
\]
By continuity we get $S(\bh(T))= e^{-\rho' T} S(\bh(0))$. From 
\eqref{precise}, for every $\epsilon$ we can find $\delta$ such that
\[
 S(\bh(T+h))\leq (1-\rho h)e^{-T\rho'}S(\bh(0))+h\epsilon
\]
for every $h\leq\delta$. Choosing $\epsilon=(\rho-\rho')e^{-T\rho'}S(\bh(0))$ 
we get 
\[
S(\bh(T+h))\leq e^{-(T+h)\rho'}S(\bh(0))
\]
which implies that $S(\bh(t))\leq e^{-t\rho'}S(\bh(0))$ for every $t\geq0$ and 
every $\rho'<\rho$.
\qed

\subsection{Derivation of \eqref{prodt}}
\label{app:bosons.eveq}

To prove \eqref{prodt}, we observe that  $\eta(t)$ and $g(v,t)$ in \eqref{geta} 
satisfy the equations 
\[
\begin{aligned}
 \dot \eta(t)=&\mu-\rho \eta(t)\\
 \dot g(v,t)=&\frac{\mu}{\eta(t)}(\gamma(v)-g(v,t))\, .
\end{aligned}
\]
Setting $\bff(t)=(f_0(t),f_1(t),f_2(t),\ldots)$ with
\[
f_N(\uv_N,t)=e^{-\eta(t)}\frac{\eta(t)^N}{N!}\prod_{i=1}^Ng(v_i,t)
\]
we get
\[
\begin{aligned}
\frac{d}{dt}f_N(\uv_N,t)=&(\mu-\rho\eta(t))
e^{-\eta(t)}\frac{\eta(t)^{N-1}}{(N-1)!}
\left(1-\frac{\eta(t)}{N}\right)\prod_{i=1}^Ng(v_i,t)\\
+&\mu e^{-\eta(t)}\frac{\eta(t)^{N-1}}{N!}
\sum_i\left((\gamma(v_i)-g(v_i,t))\prod_{j\not=i}g(v_j,t)\right)\\
=&\rho e^{-\eta(t)}\frac{\eta(t)^{N+1}}{N!}\prod_{i=1}^Ng(v_i,t) 
-\rho e^{-\eta(t)}\frac{\eta(t)^N}{(N-1)!}\prod_{i=1}^Ng(v_i,t)\\
+&\mu e^{-\eta(t)}\frac{\eta(t)^{N-1}}{N!}\sum_i
\gamma(v_i)\prod_{j\not=i}g(v_j,t)-
\mu e^{-\eta(t)}\frac{\eta(t)^N}{N!}\prod_{i=1}^Ng(v_i,t)\\
=&\rho((\cO \bff(t))_{N}(\uv)-Nf_N(\uv_N,t))+\mu((\I 
\bff(t))_{N}(\uv)-f_N(\uv_N,t))\,.
\end{aligned}
\]
Thus $\bff(t)$ solves \eqref{master} with $\tilde\lambda=0$. Clearly 
$\bff(t)\in D^1$ for every $t\geq 0$ so that, by Remark \ref{unique}, 
$\bff(t)=e^{t\T}\bff(0)$.

\subsection{Proof  of Theorem \ref{thm:chaos}}
\label{sec:3.4}

Given a continuous and bounded test function $\phi_k:\mathbb{R}^k\to 
\mathbb{R}$, symmetric with respect to the permutation of its variables, we 
define
\[
(\bff_n,\phi_k)_{k,n}=\left(\frac{\rho}{\mu_n}\right)^{k}\sum_{N\geq 
k}\frac{N!}{(N - k
)!}\int_{\mathbb{R}^N}f_{n,N}(\uv_N)\phi_k(\uv_k)d\uv_N\, .
\]
What we need to show is that, if $\bff_n$ forms a chaotic sequence and 
$\phi:\mathbb{R}\to \mathbb{R}$ is a test function then
\begin{equation*}
 \lim_{n\to\infty}(e^{\mathcal L_n t}\bff_n,\phi^{\otimes 
k})_{k,n}=\left(\lim_{n\to\infty}(e^{ \mathcal L_n
t}\bff_n,\phi)_{1,n}\right)^k
\end{equation*}
which implies propagation of chaos.

The argument to prove propagation of chaos introduced in \cite{McK} is 
based on the power series expansion of $e^{\lambda K_N t}$, which converges 
since $K_N$ is a bounded operator. After this, one can exploit a cancellation 
between $Q_N$ and $\binom{N}{2}\mathrm{Id}$, see \eqref{Kcol}, when they act 
on a function $\phi_k$ depending only on $k<N$ variables, see 
Section 3 of \cite{McK}. In the present case the analogue of 
such an argument formally works but it cannot be applied directly since, being 
$\K$ unbounded, the 
power series expansion of $e^{\tilde\lambda_n 
\K t}$ does not converge. To 
avoid this problem, one may try to use the convergent expansion \eqref{expaK} 
introduced in Section \ref{sec:3.L1}. But the different treatment of $Q_N$ and 
$\binom{N}{2}\mathrm{Id}$ in \eqref{expaK} would make it very hard to see the 
needed cancellation.

Thus we will introduce a partial expansion of 
$e^{\tilde\lambda_n \K t}$ and combine it with \eqref{expa} and \eqref{expaI}.
The idea is to expand this exponential in the {\it least 
possible way} to exploit the central cancellations of McKean's argument. We 
first decompose $K_N$ as 
\[
K_N=K_k+\widetilde K_{N-k}+(N-k)G_k
\]
with
\[
\begin{aligned}
 \widetilde K_{N-k}&=\sum_{k+1\leq i<j\leq N}(R_{i,j}-{\rm 
Id})\\
G_k&=\frac1{N-k}\sum_{i=1}^k\sum_{j=k+1}^N (R_{i,j}-{\rm 
Id})
\end{aligned}
\]
and obtain
\begin{equation}\label{deris}
 e^{\tilde\lambda_n K_N t}\phi_k=e^{\tilde\lambda_n K_k 
t}\phi_k+(N-k)\tilde\lambda_n\int_0^t e^{\tilde\lambda_n 
K_N(t-s)}G_ke^{\tilde\lambda_n K_k s}\phi_k ds
\end{equation}
where we used that $K_N$ is a bounded operator on $C^0(\Rs^N)$ 
and that $\widetilde K_{N-k}\phi_k=0$.
Since we are interested in integrating \eqref{deris} against a symmetric 
function $f_N$ we can write
\[
G_k[\phi_k](\uv_{k+1})=
\sum_{i=1}^k\int\frac{d\theta}{2\pi} [ \phi ( v_1 ,\ldots,v_{i-1},v_i\cos
\theta+v_{k+1}\sin\theta,v_{i+1},\ldots,v_k)-\phi(\uv_k)]
\]
To iterate we need to apply \eqref{deris} to the factor $e^{\tilde\lambda_n 
K_N(t-s)}$ inside the integral in \eqref{deris} itself. Since 
$G_ke^{\tilde\lambda_n 
K_k s}\phi_k$ is a function of $k+1$ variables we now have to write
\[
 K_N=K_{k+1}+\widetilde K_{N-k-1}+(N-k-1)G_{k+1}\, .
\]
Iterating this procedure we get
\[
\begin{aligned}
  e^{\tilde\lambda_n K_N t}\phi_k=&e^{\tilde\lambda_n K_{k} 
t}\phi_k\\
+&\sum_{p=1}^{N-k}\frac{\tilde\lambda_n^p(N-k)!}{(N-k-p)!}\int_{0<t_1<
\cdots < t_p < t } e ^ { \tilde \lambda_n K_{k+p} 
(t-t_{p})}G_{k+p-1}e^{\tilde\lambda_n K_{k+p-1} (t_p-t_{p-1})}\\
&\qquad\qquad\qquad \cdots e^{(t_2-t_1)\tilde\lambda_n 
K_{k+1}}G_ke^{\tilde\lambda_n K_k t_1}\phi_k\,dt_p\cdots dt_1 
\end{aligned}
\]
so that
\begin{align}\label{step2}
 (e^{\tilde\lambda_n\K t}\bff_n,\phi_k)_{k,n}=&(\bff_n,e^{\tilde\lambda_n K_{k}
t}\phi_k)_{k,n}\\
+&\sum_{p=1}^\infty\lambda^p\int_{0<t_1<
\cdots < t_p < t } \Bigl(\bff_n,e^{ \tilde \lambda_n K_{k+p} 
(t-t_{p})}G_{k+p-1}e^{\tilde\lambda_n K_{k+p-1} (t_p-t_{p-1})}
\nonumber\\
&\qquad\qquad\qquad\cdots e^{(t_2-t_1)\tilde\lambda_n 
K_{k+1}}G_ke^{\tilde\lambda_n K_k t}\phi_k\bigr)_{k+p,n}\,dt_p\cdots dt_1 
\nonumber
\end{align}
where the factor $\lambda^p$ in the second line of \eqref{step2}, 
comes from \eqref{tl} and \eqref{marg}.

Observe now that the $R_{i,j}$ are averaging operators so that 
$\|R_{i,j}\|_\infty\leq 1$ which gives 
\[ 
\bigl\|e^{t\tilde\lambda_n K_{N}}\bigr\|_\infty=
e^{-t\tilde\lambda \binom N2}\bigl\|e^{t\tilde\lambda\sum_{1\leq i<j\leq 
N}R_{i,j}}\bigr\|_\infty\leq 1\, .
\]
For the same reason we have
\[
 \|G_k\|_\infty\leq \frac1{N-k}\sum_{i=1}^k\sum_{j=k+1}^N 
(\|R_{i,j}\|_\infty+1)\leq 2k\, .
\]
Using \eqref{bound1r} we get
\[
\begin{aligned}
 \left|(e^{\tilde\lambda_n\K t}\bff_n,\phi_k)_{k,n}\right|\leq &
\left(\frac{\rho}{\mu_n}\right)^k\|\bff_n\|_1^{(k)}
\|\phi_k\|_{\infty}\\
+&\sum_{p=1}^\infty\frac{\lambda^pt^p}{p!}
\prod_{i=k}^{k+p-1}\|G_i\|_\infty\left(\frac{\rho}{\mu_n}\right)^{k+p}
\|\bff_n\|_1^{(k+p)}\|\phi_k\|_{\infty}\\
\leq& 
\|\phi_k\|_{\infty}K^k\sum_{p=0}^\infty2^p\lambda^pt^pK^p\binom{k+p-1}{p}\,.
\end{aligned}
\]
Observe that the series in the last line converges for $\lambda K t<1/2$.
On the other hand, since $\lim_{n\to\infty}\tilde\lambda_n=0$, for every $t$ we 
have
\[
 \lim_{n\to\infty}(\bff_n,e^{\tilde\lambda_n K_{k}
t}\phi_k)_{k,n}=\lim_{n\to\infty}(\bff_n,\phi_k)_{k,n}
\]
and similarly, calling $G_k^{*p}=\prod_{i=0}^p 
G_{k+i}$,
\[
\begin{aligned}
\lim_{n\to\infty} \int_{0<t_1<\cdots < t_p < t } &\Bigl(\bff_n,e^{ \tilde 
\lambda_n K_{k+p}(t-t_{p})}G_{k+p-1}e^{\tilde\lambda_n K_{k+p-1} 
(t_p-t_{p-1})}\\
\cdots &e^{(t_2-t_1)\tilde\lambda_n 
K_{k+1}}G_ke^{\tilde\lambda_n K_k t}\phi_k\bigr)_{k+p,n}\,dt_p\cdots dt_1 =
\lim_{n\to\infty}\left(\bff_n, 
G_{k}^{*p}\phi_k\right)_{k+p,n}
\end{aligned}
\]
so that we finally get
\begin{equation}\label{ntoinf}
 \lim_{n\to\infty}(e^{\tilde\lambda_n\K 
t}\bff_n,\phi_k)_{k,n}=\lim_{n\to\infty}\sum_{p=0}^\infty 
\frac{\lambda^pt^p}{p!}\left(\bff_n, 
G_{k}^{*p}\phi_k\right)_{k+p,n}\, .
\end{equation}
Observe now that 
$G_k$ acts as a 
derivation in the sense of \cite{McK}, that is, for every $\phi_{k_1}$ and 
$\psi_{k_2}$ with $k_1+k_2=k$, we have 
\[
 G_{k}(\phi_{k_1}\otimes 
\psi_{k_2})=(G_{k_1}\phi_{k_1})\otimes 
\psi_{k_2}+\phi_{k_1}\otimes(G_{k_2}\psi_{k_2})\, .
\]
This implies that
\begin{equation}\label{fact}
\frac{1}{p!}G_k^{*p}(\phi_{k_1}\otimes\psi_{k_2})=
\sum_{p_1+p_2=p}\frac{
1}{p_1!}\frac{1}{p_2!}(G^{*p_1}_
{k_1}\phi_{k_1})\otimes(G^{*p_1}_{k_2}\psi_{k_2})
\, .
\end{equation}
Observing that if $\bff_n$ forms a chaotic sequence then
\begin{equation}\label{distri}
 \lim_{n\to\infty} (\bff_n,\phi_{k_1}\otimes\psi_{k_2})_{k,n}=\lim_{n\to\infty} 
(\bff_n,\phi_{k_1})_{k_1,n}\lim_{n\to\infty} (\bff_n,\psi_{k_2})_{k_2,n}
\end{equation}
we get
\begin{align}\label{propa}
\lim_{n\to\infty}\sum_{p=0}^\infty &\frac{\lambda^pt^p}{p!}(\bff_n, 
G_{k}^{*p}\phi_{k_1}\otimes 
\psi_{k_2})_{k+p,n}\\
=&\lim_{n\to\infty}\sum_{p_1=0}^\infty 
\frac{\lambda^{p_1}t^{p_1}}{p_1!}\left(\bff_n, 
G_{k_1}^{*p_1}\phi_{k_1}\right)_{k_1+p_1,n}
\lim_{n\to\infty}\sum_{p_2=0}^\infty 
\frac{\lambda^{p_2}t^{p_2}}{p_2!}\left(\bff_n, 
G_{k_2}^{*p_2}\phi_{k_2}\right)_{k_2+p_2,n}\nonumber
\end{align}
which implies that $e^{\tilde\lambda_n \K t}$ propagates chaos, at least for 
$t\leq t_0=\frac{1}{2\lambda K}$. Finally we need to verify that 
\eqref{bound1r} 
still holds. Since $\bff_n$ are positive $\|\bff_n\|_1^{(r)}=N_r(\bff)$, see 
\eqref{Nrdef}. 
Thus Corollary \ref{cor:boundNr} implies that for every $t\geq 0$ we have 
$\|\bff_n(t)\|^{(r)}\leq 
K_1^r\left(\frac{\mu_n}{\rho}\right)^r$
with $K_1=\max\{K,1\}$. 
Thus $\bff_n(t_0)=e^{\tilde\lambda_n \K t_0}\bff_n$ forms a 
chaotic sequence that satisfies \eqref{bound1r} with $K_1$ in place of $K$. 
Using $\bff_n(t_0)$ as initial condition we get that propagation of chaos 
holds up to time $t_1=\frac{1}{2\lambda K}+\frac{1}{2\lambda K_1}$. Iterating 
this argument we see that $e^{\tilde\lambda_n \K t}$ propagates chaos for every 
$t\geq0$.

To add the {\it out} operator $\cO$, we observe that from \eqref{ntoinf} we get
\begin{align}\label{ntoinfO}
& \lim_{n\to\infty}(e^{(\tilde\lambda_n\K -\rho\N)
t}\bff_n,\phi_k)_{k,n}\\
&\;=\lim_{n\to\infty}\sum_{p=0}^\infty 
\frac{\lambda^pt^p}{p!}\left(\frac{\rho}{\mu}\right)^{k+p}\sum_{N\geq k+p} 
\frac{N!}{(N-k-p)!}e^{-\rho Nt}\int f_{n,N}(\uv_{N}) 
(G_{k}^{*p}\phi_k)(\uv_{k+p}) d\uv_{N}\, .\nonumber
\end{align}
Inserting \eqref{ntoinfO} into \eqref{expa}, after some long algebra that we 
report in Appendix \ref{app:bosons.propa}, we 
obtain
\begin{align}\label{densiO}
 \lim_{n\to\infty}\Bigl( e^{(\tilde\lambda_n 
\K+\rho(\cO-\N))t}&\bff_n,\phi_k\Bigr)_{k,n}=\lim_{n\to\infty}\sum_{p=0}^\infty 
\frac{t^p\lambda^p}{p!}
\left(\bff_n,e^{-\rho(k+p) t}G_k^{*p}\phi_k
\right )_{k+p,n}\, .
\end{align}
It is not hard to see that \eqref{densiO} implies that $e^{(\tilde\lambda_n 
\K+\rho(\cO-\N))t}$ propagates chaos.

Finally we consider the {\it in} operator $\I$. Observe that
\[
\begin{aligned}
\mu_n(\I\bff_n,\phi_k&)_{k,n}=\mu_n\left(\frac{\rho}{\mu_n}\right)^{k}
\sum_{N\geq k} \frac{(N-1)!}{(N-k)!}\int \sum_{i=1}^N 
f_{n,N-1}(\uv_{N-1}^i)\gamma(v_i)\phi_k(\uv_k)d\uv_N=\\
&\mu_n\left(\frac{\rho}{\mu_n}\right)^{k}\sum_{N\geq k} 
\frac{(N-1)!}{(N-k)!}\int \bigl((N-k)
f_{n,N-1}(\uv_{N-1})\gamma(v_N)\phi_k(\uv_k)d\uv_N+ \\
&\phantom{\mu_n\left(\frac{\rho}{\mu_n}\right)^{k}\sum_{N\geq k} 
\frac{(N-1)!}{(N-k)!}\int \bigl((}
kf_{n,N-1}(\uv_{N-1})\phi_k(\uv_{k-1},v_N)\gamma(v_N)d\uv_N\bigr)=\\
&\mu_n\left(\frac{\rho}{\mu_n}\right)^{k}\sum_{N> 
k}\frac{(N-1)!}{(N-1-k)!}\int 
f_{n,N-1}(\uv_{N-1})\phi_k(\uv_k)\gamma(v_N)d\uv_{N-1}+\\
&\quad k\rho\left(\frac{\rho}{\mu_n}\right)^{k-1}\sum_{N\geq 
k-1}\frac{N!}{(N-(k-1))!}\int 
f_{n,N}(\uv_{N})\phi_k(\uv_{k-1},w)\gamma(v_N)d\uv_Ndw
\end{aligned}
\]
so that
\begin{equation}\label{deriI} 
\mu_n (\I\bff_n-\bff_n,\phi_k)_{k,n}=
(\bff_n,I_k\phi_k)_{k-1,n}
\end{equation}
where
\[
I_k[\phi_k](\uv_{k-1}):=\rho k \int_{\mathbb{R}}\phi_k(\uv_{k-1},w)e^{-\pi 
w^2}dw\, .
\]
which clearly act as a derivative in the sense of \cite{McK}. We can now use an 
expansion similar to \eqref{expa1}
\begin{align}\label{expaIbis}
& e^{t{\mathcal L}_n}\bff_n=e^{(\tilde\lambda_n 
\K+\rho(\cO-\N))t}\bff_n\\
&\quad+\sum_{q=1}^\infty\mu_n^q\int\limits_{0<t_1<\ldots<t_q<t} 
e^{(\tilde\lambda_n 
\K+\rho(\cO-\N))(t-t_q)}  (\I-\mathrm{Id}) e^{(\tilde\lambda_n 
\K+\rho(\cO-\N))(t_q-t_{q-1})} \nonumber\\
&\quad\phantom{\sum_{n=1}^\infty\mu_n^q\int\limits_{0<t_1<\ldots<t_q<t} 
e^{(\tilde\lambda 
\K+\rho(\cO-\N))(t-t_n)}}\cdots
(\I-\mathrm{Id}) e^{(\tilde\lambda_n 
\K+\rho(\cO-\N))t_1}\bff_n\,dt_1\cdots dt_n\nonumber
\end{align}
that combined 
\eqref{deriI} with \eqref{densiO} gives 
\begin{align}\nonumber
\lim_{n\to\infty}(&e^{\mathcal L_n t}\bff_n,\phi_k)_{k,n}= 
\lim_{n\to\infty}\sum_{q\geq 0}\sum_{p_0, p_1 , \ldots,p_{q} \geq 0} 
\rho^q\lambda^{|p|}e^{-\rho k t}\, \cdot\\
&\int_{0\leq t_q\leq\cdots\leq t_1\leq t}\prod_{i=0}^q 
e^{-\rho 
(t_{i}-t_{i+1})(|p|_i-i)}\frac{(t_{i}-t_{i+1})^{p_i}}{p_i!}\, dt_1\cdots 
dt_q\cdot\label{densiOI}\\
&\qquad\left(\bff_n,G_{k+|p|_q-q}^{*p_{q}}I_{k+|p|_q-q+1}\cdots 
G_{k+p_0-1}^{*p_1}I_{k+p_0}G_k^{*p_0}\phi_k\right)_{k+|p|-q,n}\nonumber
\end{align}
where $|p|_{i}=\sum_{j=0}^{i-1} p_j$ and $t_0=t$, $t_{q+1}=0$ and the order 
of the $t_i$ in the integral is inverted due to the inversion of the order of 
the operators when taking the adjoint.
From \eqref{densiOI} it follows, after more long algebra 
reported in appendix \ref{app:bosons.propa}, we see that, if $k_1+k_2=k$, 
then
\begin{equation}\label{proptot}
 \lim_{n\to\infty}(e^{\mathcal L_n t}\bff_n,\phi_{k_1}\otimes\psi_{k_2})_{k,n}=
 \lim_{n\to\infty}(e^{\mathcal L_n 
t}\bff_n,\phi_{k_1})_{k_1,n}\lim_{n\to\infty}(e^{\mathcal L_n 
t}\bff_n,\psi_{k_2})_{k_2,n}
\end{equation}
that is, $e^{\mathcal L_n t}$ propagates chaos. 
The validity of the Boltzmann-Kac type equation \eqref{BK} follows exactly as 
in 
\cite{McK}.\qed

\section{Conclusions}

The central aim of this work is the extension of the analysis in \cite{BLV}, in 
which a thermostat idealizes the interaction with a large reservoir of particles 
kept at constant temperature and chemical potential.  While in \cite{BLV} the 
reservoir and the system could not exchange particles, here the main interaction 
is the continuous exchange of particles between the two.

However, it is in these same works which we hoped to extend that we also find 
points of possible extension to our current work. In the case of the 
standard Kac model, approach to equilibrium in the sense of the GTW metric 
$d_2$ was shown in \cite{Hagop} while for a Kac system interacting 
with one or more Maxwellian thermostats it was shown in \cite{Evans}. In the 
present situation though, it is not clear how to define an analogue of the GTW 
metric since the components $f_N$ of a state $\bff$ are not, in general,  
probability distributions on $\mathbb{R}^N$.

Furthermore, in \cite{BLTV} the authors show that, in a strong and uniform 
sense, the evolution of the Kac system with a Maxwellian thermostat can be 
thought of as an idealization of the interaction with a large heat reservoir, 
itself described as a Kac system. We 
think it is possible to replicate such an analysis in the present context and 
hope to come back to this issue in a forthcoming paper.

We based our proof of propagation of chaos on the work in \cite{McK}; therefore, 
as in \cite{McK}, it is not quantitative nor uniform in time. Recently, a 
quantitative and uniform in time result was obtained for the Kac system with a 
Maxwellian thermostat \cite{Hagop1}. It is unclear to us whether the methods in 
their work extend to the present model. 

Finally, the assumption that the rates $\rho$ and $\mu$ are independent of the 
number of particles is clearly unrealistic, allowing the possibility of an 
unbounded number of particles in the system. However, in the steady state (and 
in a chaotic state) the probability of having a number of particles in the 
system much larger then the average is extremely small, and so we do not 
consider this a serious problem. In any case, it would be interesting to 
investigate what happens if one assumes a maximum number of particles allowed 
inside the system.
\newpage

\appendix\normalsize

\section{Proofs of Technical Lemmas.}
\label{app:bosons}

\subsection{Proof of Lemma \ref{lem:deltam}} 
\label{app:bosons.deltam}


As already observed, we will search for 
the infimum of $(\bh,-\mathcal G\bh)$ on $\mathbf V_m$ looking at states 
$\bh$ close to $\R^+_{m}\be^0$. This is done using the representations 
\eqref{repho} and \eqref{rephe} below. Since $\mu/\rho$ is large, \eqref{Rne} 
and \eqref{Rno} suggest that the dominant term in $(\bh,\mathcal G\bh)$ for a 
state $\bh$ close to $\R^+_{m}\be^0$ is the ``diagonal term'', that is the 
first term on the right hand side of \eqref{Rne} or \eqref{Rno}. To prove 
Lemma \ref{lem:deltam} we thus need good bounds on the ``off diagonal'' terms. 
The proof in this section is thus loosely based on the proof of the Gershgorin 
circle theorem, see \cite{Golub}.

If $m=2n+1$, we can write any $\bh\in\mathbf V_m$ as
\begin{equation}\label{repho}
 \bh=a \R^+_{2n+1}\be^0+b\R^+_0\R^+_{2n+1}\be^0+\bj=a \R^+_{2n+1}\be^0+\bk
\end{equation}
with $\bj\perp\R^+_{2n+1}\be^0$ and $\bj\perp\R^+_0\R^+_{2n+1}\be^0$. 

From \eqref{Rno} we get $(\R^+_{2n+1}\be^0,\K\R^+_{2n+1}\be^0)=-\mu/\rho$ so 
that
\begin{equation*}
 \begin{aligned}
 (\bh,-\widetilde{\mathcal L}\bh)=& (\bh, (-\mathcal 
G-\lambda\frac{\rho}{\mu}\K)\bh)\\
 =&a^2(\lambda+\rho)+(\bk,(-\mathcal 
G-\lambda\frac{\rho}{\mu}\K)\bk)+2a(\R^+_{2n+1}\be^0,(-\mathcal 
G-\lambda\frac{\rho}{\mu}\K)\bk)
\end{aligned}
\end{equation*}
By construction $\bk$ is in the span of the $\be_{\underline \alpha}$ with 
$\lambda(\underline\alpha)\geq 2$ so that $(\R^+_{2n+1}\be^0, \mathcal G\bk)=0$ 
and $(\bk,-\mathcal G\bk)\geq2\rho\|\bk\|_2=2\rho(b^2+\|\bj\|_2^2)$ while from 
\eqref{Rno} we get
\begin{equation}\label{contoo}
(\R^+_{2n+1}\be^0,\K\bk)=b(\K\R^+_{2n+1}\be^0,\R^+_{2n+1}\be^0)+(\K\R^+_{2n+1}
\be^0,\bj)=-\sqrt {\frac{\mu}{\rho}} b
\end{equation}
This gives

\begin{equation*}
 \begin{aligned}
 (\bh,-\widetilde{\mathcal L}\bh)\geq
 &a^2(\lambda+\rho)+2\rho(b^2+\|\bj\|_2^2)-2\lambda 
|ab|\sqrt{\f{\rho}{\mu}}\\
\geq &a^2\left(\rho+\lambda 
-\lambda\sqrt{\f{\rho}{\mu}}\right)+b^2\left(2\rho-\lambda\sqrt{\f{\rho}{\mu}}
\right)+2\rho\|\bj\|_2^2\, .
\end{aligned}
\end{equation*}
Since $\|\bh\|^2=a^2+b^2+ \|\bj\|^2$ we get \eqref{oddm}.

Similarly, every $\bh\in \mathbf V_{2n}$ with $n\geq 2$ can be written as
\begin{equation}\label{rephe}
 \bh=a \R^+_{2n}\be^0+\sum_{k=0}^{n/2}b_k\R^+_{2k}\R^+_{2(n-k)}\be^0+\bj=a 
\R^+_{2n}\be^0+ \bk
\end{equation}
where $\bj\perp \R^+_{2n}\be^0$ and $\bj\perp \R^+_{2k}\R^+_{2(n-k)}\be^0$. 
Observe that
\[
\biggl\|\sum_{k=0}^{n/2}b_k\R^+_{2k}\R^+_{2(n-k)}\be^0\biggr\|_2=\sum_{k=0}^{n/2
}
\epsilon_{n,k}b_k^2
\]
where, due to \eqref{ealpha}, $\epsilon_{n,k}=2$ if $k=n-k$ and $1$ otherwise. 
Analogously to \eqref{contoo}, using \eqref{Rne}, we get
\[
\begin{aligned}
 (\R^+_{2n}\be^0,\K\R^+_{2n}\be^0)=&\f{\mu}{\rho}(2\tau_n-1)\\
 (\R^+_{2n}\be^0,\K\bk)=&2 
\sqrt{\f{\mu}{\rho}}\sum_{k=1}^{n/2}b_k\sigma_{n,k}
+ b_0\sqrt{\f{\mu}{\rho}}(1-2\tau_n)\, .
\end{aligned}
\]
Proceeding as before we obtain
\begin{equation*}
 \begin{aligned}
 (\bh,&-\widetilde{\mathcal L}\bh)= (\bh, (-\mathcal 
G-\lambda\frac{\rho}{\mu}\K)\bh)\\
 &=a^2((1-2\tau_n)\lambda+\rho)+(\bk,(-\mathcal 
G-\lambda\frac{\rho}{\mu}\K)\bk)+
2a(\R^+_{2n}\be^0,(-\mathcal G-\lambda\frac{\rho}{\mu}\K)\bk)\\
 &\geq a^2((1-2\tau_n)\lambda+\rho)+2\rho 
\left(\sum_{k=0}^{n/2}\epsilon_{n,k}b_k^2+\|\bj\|_2^2\right)\\
 &\qquad\qquad-4\lambda \sqrt{\f{\rho}{\mu}}\sum_{k=1}^{n/2}|ab_k|\sigma_{n,k}
-2\lambda |ab_0|\sqrt{\f{\rho}{\mu}}(1-2\tau_n)\, .
 \end{aligned}
\end{equation*}
which gives
\begin{equation*}
 \begin{aligned}
 (\bh,-&\widetilde{\mathcal L}\bh)\geq a^2((1-2\tau_n)\lambda+\rho)+ 2\rho 
\left(\sum_{k=0}^{n/2}\epsilon_{n,k}b_k^2+\|\bj\|^2\right)-\\
&\qquad\qquad\qquad\lambda\sqrt{\f{\rho}{\mu}}
\left[2\sum_{k=1}^{n/2}a^2\sigma_{n,k}^2+a^2(1-2\tau_n)^2+2\sum_{k=0}^{n/2}b_k^2
\right]\geq\\
&\left(2\rho-2\lambda\sqrt{\f{\rho}{\mu}}\right) 
\left[\sum_{k=0}^{n/2}\epsilon_{n,k}b_k^2\right]+ 
a^2\left((1-2\tau_n)\lambda+\rho-\lambda\sqrt{\f{\rho}{\mu}}A_{2n}\right) 
+2\rho\|\bj\|_2^2
\end{aligned}
\end{equation*}
where 
\[
A_{2n}=(1-2\tau_n)^2+2\sum_{k=1}^{n/2}\sigma_{n,k}^2\, .
\]
We thus need an upper bound on $A_n$. To this end, observe that
\[
\log\tau_n = \log\prod_{i=1}^n \left(1- \f{1}{2i}\right)\leq -\sum_{i=1}^n 
\f{1}{2i}
\leq -\f{1}{2}\log n\quad
\Rightarrow \quad\tau_{n}\leq \f{1}{\sqrt{n}}
\]
while, from \eqref{gammi2}, we have
\[
\sigma_{n,k}^2\leq \tau_n\f{1}{n}\f{1}{\sqrt{\f{k}{n}}\sqrt{1-\f{k}{n}}}
\]
so that
\begin{gather*}
\sum_{k=1}^{n/2}\sigma_{n,k}^2\leq
\tau_n\int_{0}^{\frac12}\f{1}{\sqrt{x(1-x)}}dx=\frac{\pi}{2}\tau_n\, .
\end{gather*}
Finally we get $A_{2n}\leq 2$ which implies \eqref{evenm}.\qed

\subsection{Proof of Lemma \ref{lem:limderi}} 
\label{app:bosons.limderi}
 
Proceeding as in \eqref{KN0} we can write
\begin{align*}
 e^{\tilde\lambda K_Nt}f_N=&e^{-\tilde\lambda \binom 
N 2 t}f_N+t e^{-\tilde\lambda \binom 
N 2 t}\sum_{k=1}^\infty \frac{\tilde\lambda_n^kt^{k-1}}{k!} Q_N^kf_N\\
 \frac{1}{t}\left(e^{\tilde\lambda K_Nt}f_N-f_N\right)=&
e^{-\tilde\lambda \binom N 2 t}\tilde\lambda 
Q_Nf_N+\frac{1}{t}\left(e^{-\tilde\lambda \binom 
N 2 t}-1\right)f_N+\\
&t e^{-\tilde\lambda \binom 
N 2 t}\sum_{k=2}^\infty \frac{\tilde\lambda_n^k t^{k-2}}{k!} Q_N^kf_N
\end{align*}
Since $R_N(\uv_N,t)=\sum_{k=1}^\infty \frac{\tilde\lambda_n^kt^{k-1}}{k!} 
(Q_N^kf_N)(\uv_N)$ 
is a sum of positive increasing terms and $\|R_N(t)\|_{1}<\infty$, we see 
that $t e^{-\tilde\lambda \binom 2 Nt}R_N(\uv_N,t)$ converges to 0 as $t\to 
0$ for almost every $\uv_N$. A similar argument implies that 
$\left(e^{\tilde\lambda 
K_Nt}f_N-f_N\right)/t$ converges almost everywhere to $\tilde\lambda K_N f_N$.

Using the Duhamel formula we can write
\begin{align}\label{1stor}
 \left(e^{(\tilde\lambda 
\K+\rho(\cO-\N))t}\bff\right)_N=&
e^{\left(\tilde\lambda K_N - \rho N\right)t}f_N\\
+&\rho e^{\left(\tilde\lambda K_N - \rho N\right)t}\int_0^t 
e^{-\left(\tilde\lambda K_N - \rho N\right)s}\left(\cO 
e^{\left(\tilde\lambda \K + \rho(\cO-\N)\right)s}\bff\right)_{N}ds\nonumber\\
:=&e^{\left(\tilde\lambda K_N - \rho N\right)t}(f_N+\overline R_N(t))\, 
.\nonumber
\end{align}
Since $\overline R_N(t,\uv_N)$ is increasing in $t$ and $\|\overline 
R_N(t)\|_{1,N}\to 0$ as $t\to 0^+$ we see that 
\[
\lim_{t\to 0^+}\left(e^{(\tilde\lambda 
\K+\rho(\cO-\N))t}\bff\right)_N(\uv_N)= f_N(\uv_N)
\]
for almost every $\uv_N$. Similarly using the Duhamel formula once more 
we get
\[
\overline R_N(t)=\rho\int_0^t 
e^{-\left(\tilde\lambda K_N - \rho N\right)s}\left(\cO 
e^{\left(\tilde\lambda \K - \rho\N\right)s}\bff\right)_{N}ds+\overline 
R_{1,N}(t)
\]
where 
\[
\overline R_{1,N}(t)=\rho^2\int_0^t \int_0^s
e^{-\left(\tilde\lambda K_N - \rho N\right)s}\left(\cO  
e^{\left(\tilde\lambda \K - \rho \N\right)(s-s_1)}\cO
e^{\left(\tilde\lambda \K + \rho(\cO-\N)\right)s_1}\bff\right)_{N}ds_1ds
\]
Reasoning as in \eqref{1stor} we get $\overline R_{1,N}(t,\uv_N)/t\to 0$ as 
$t\to 0^+$ for almost every $\uv_N$ while proceeding as in \eqref{diffdiff} we 
get
\[
\lim_{t\to 0^+}\frac1t\int_0^t 
e^{-\left(\tilde\lambda K_N - \rho N\right)s}\left(\cO 
e^{\left(\tilde\lambda \K - 
\rho\N\right)s}\bff\right)_{N}ds=\left(\cO\bff\right)_N\, .
\]
Finally a similar argument using \eqref{expaI} concludes 
the proof.\qed


\subsection{Proof of Lemma \ref{lem:conve}} 
\label{app:conve}

Since $R_{i,j}$ is an average, we have $R_{i,j}h_N\log(R_{i,j}h_N)\leq 
R_{i,j}(h_n\log h_N)$ ,
from which, calling $\overline Q_N={\binom N 2}^{-1} Q_N$, see \eqref{Kcol}, it 
follows that $\overline Q_N h_N\log(\overline Q_N h_N)\leq \overline 
Q_N(h_n\log h_N)$. Finally writing
\[
 e^{\tilde\lambda K_N}h_N=e^{-\tilde\lambda \binom N 2 t}\sum_{n=0}^\infty 
{\binom N 2}^{n} \frac{\tilde\lambda^n t^n}{n!}\overline Q_N^n h_N
\]
we get $e^{\tilde\lambda K_N}h_N\log(e^{\tilde\lambda K_N}h_N)\leq 
e^{\tilde\lambda K_N}(h_n\log h_N)$.

Proceeding as in section \ref{sec:3.L1} we can write
\begin{equation}\label{duaPP}
e^{\widetilde{\mathcal L}t}\bh=e^{(\tilde\lambda 
\K-\rho\N-\mu\mathrm{Id})t}\bh+\int_0^t e^{(\tilde\lambda 
\K-\rho\N-\mu\mathrm{Id})(t-s)} (\rho \P^+ + \mu \P^-)e^{\widetilde{\mathcal 
L}s}\bh\,ds\, .
\end{equation}
so that, writing 
$\bh(t)=(h_{0}(t),h_{1}(v_1,t),\ldots)$ and using the notation introduced in 
the proof of Lemma \ref{lem:raise}, we get 
\[
\begin{aligned}
 h_{N,t}=&e^{-(\rho N+\mu\mathrm{Id})t}\left(e^{\tilde\lambda 
K_N t}h_N(0)\right)+\\
&\int_0^t e^{-(\rho 
N+\mu\mathrm{Id})(t-s)}\rho\sum_{i=1}^Ne^{\tilde\lambda 
K_N (t-s)}P^+_{N,i}h_{N-1}(s)ds+\\
&\int_0^t e^{-(\rho 
N+\mu\mathrm{Id})(t-s)}\mu e^{\tilde\lambda 
K_N (t-s)}P^-_{N}h_{N+1}(s)ds
\end{aligned}
\]
Observing that
\[
 e^{-(\rho N+\mu\mathrm{Id})t}+\int_0^t e^{-(\rho 
N+\mu\mathrm{Id})(t-s)}(\rho N+\mu)ds=1
\]
while 
\[
\begin{aligned}
P^+_{N,i}h_{N-1}\log( P^+_{N,i}h_{N-1})=&P^+_{N,i}(h_{N-1}\log h_{N-1})\\
P^-_{N}h_{N+1}\log( P^-_{N}h_{N+1})\leq&P^-_{N}(h_{N+1}\log h_{N+1})
\end{aligned}
\]
we get
\[
\begin{aligned}
 h_N(t)\log h_N(t)\leq &e^{(\tilde\lambda 
K_N-\rho N-\mu\mathrm{Id})t}(h_N(0)\log h_N(0))+\\
&\int_0^t e^{(\tilde\lambda 
K_N-\rho N-\mu\mathrm{Id})(t-s)}\rho\sum_{i=1}^NP^+_{N,i}(h_{N-1}(s)\log 
h_{N-1}(s))ds+\\
&\int_0^t e^{(\tilde\lambda 
K_N-\rho N-\mu\mathrm{Id})(t-s)}\mu P^-_{N}(h_{N+1}(s)\log h_{N+1}(s))ds=\\
&\left(e^{(\tilde\lambda 
\K-\rho\N-\mu\mathrm{Id})t}\bh\log\bh\right)_N+\\
&\qquad\left(\int_0^t e^{(\tilde\lambda 
\K-\rho\N-\mu\mathrm{Id})(t-s)} (\rho \P^+ + \mu \P^-)e^{\widetilde{\mathcal 
L}s}(\bh\log \bh)\,ds\right)_N\, .
\end{aligned}
\]
This, together with \eqref{duaPP}, completes the proof.\qed

\subsection{Proof of Lemma \ref{lem:coarse}} 
\label{app:bosons.coarse}

Given $h_N\in L^1(\Rs^N,\gamma_N)$ and a measurable set $A\in\Rs^N$
\[
 s(h_N):=\int_{\Rs^N}\gamma_N(\uv_N)h_N(\uv_N)\log h_N(\uv_N)d\uv_N\, .
\]
and
\begin{align}
m(A)&:=\int_A\gamma_N(\uv_N)d\uv_N\nonumber\\
e(h_N,A)&:=\frac{1}{m(A)}\int_{A}h_N(\uv_N)\gamma_N(\uv_N)d \uv_N\nonumber\\
s(h_N,A)&:=\frac{1}{m(A)}\int_{A}h_N(\uv_N)\log(h_N(\uv_N))
\gamma_N(\uv_N)d \uv_N\nonumber\\
d(h_N,A)&:=s(h_N,A)-e(h_N,A)\log(e(h_N,A))\, .\nonumber
\end{align}
Observe that $m$ defines a probability measure on $\mathbb{R}^N$ while 
$d(h_N,A)\geq 0$ for every $A$.

\begin{lem}\label{1pe} Let $h_N>0$ be such that 
$s(h_N)<\infty$. Then for every $\epsilon>0$ there exists $\delta>0$ such that, 
if $A$ is a measurable set 
with $m(A)\leq\delta$ then $m(A)d(h_N,A)\leq\epsilon$.
\end{lem}
\noindent\emph{Proof}. 
Observe that
\begin{align}\label{md}
 m(A)d(h_N,A)=&\int_A h_N(\uv_N)\log( h_N(\uv_N))\gamma_N(\uv_N) 
d\uv_N-\nonumber\\
 &\int_A h_N(\uv_N)\gamma_N(\uv_N)d\uv_N\log\left( \int_A  
h_N(\uv_N)\gamma_N(\uv_N)d\uv_N\right)+\\
&\log(m(A))\int_A h_N(\uv_N)\gamma_N(\uv_N)d\uv_N\nonumber
\end{align}
The last term on the right hand side of \eqref{md} is negative 
while continuity of the Lebesgue integral implies that, given $\epsilon>0$ we 
can find $\delta>0$ such that, if $A$ is a measurable set with $m(A)\leq\delta$ 
then the first and second terms in the right hand side of \eqref{md} are 
less then $\epsilon/2$.
\qed
\medskip

Given two partitions $\mathcal B_0$ and $\mathcal B_1$ of $\Rs$ we say that 
$\mathcal B_1$ {\it refines} $\mathcal B_0$ if every 
element of $\mathcal B_0$ can be written as a union of elements of $\mathcal 
B_1$. 
By convexity, if $\mathcal B_1$ refines $\mathcal B_0$ then
$s(h_{N,\mathcal B_0^N})\leq s(h_{N,\mathcal B_1^N})$.
It is also easy to see that given two 
partition $\mathcal B_0$ and  $\mathcal B_1$ there always exists a partition 
$\mathcal B_2$ that refines both $\mathcal B_0$ and  $\mathcal B_1$.

The following Lemma is the main result of this Appendix.

\begin{lem}\label{coarseN} Let $h_N>0$ be such that 
 $s(h_N)<\infty$. Then for every $\epsilon$ there exists a finite
partition $\mathcal B$ of $\Rs$ such that
\[
 s(h_N) - s(h_{N,\mathcal B^N})\leq\epsilon\, .
\]
\end{lem}
\noindent\emph{Proof}. Given a partition $\mathcal B=\{B_k\}_{k=1}^K$ of 
$\mathbb R$, we have
\[
\begin{aligned}
s(h_{N})=&\sum_{\uk\in \{1,\ldots,K\}^N} m(B_{\underline k})s(h_N,B_{\uk}) \\
s(h_{N,\mathcal B^N})=&\sum_{\uk\in \{1,\ldots,K\}^N} 
m(B_{\uk})e(h_N,B_{\uk})\log( e(h_N,B_{\uk}))
\end{aligned}
\]
so that we need to find $\mathcal B$ such that
\[
\sum_{\uk\in \{1,\ldots,K\}^N} m(B_{\underline k})d(h_N,B_{\uk})\leq \epsilon\, 
.
\]
To simplify notation, in what follows, 
we will write $d(A)$ for $d(h_N,A)$. 
Thanks to Lemma \ref{1pe}, given $\epsilon$ we can find $\delta$ such that 
for every $A$ with $m(A)<\delta$ we have 
$m(A)d(A)<\epsilon/2$. Moreover there exists $L$ such that calling 
$Q_L=(-L,L)^N$ we have $m(R)\leq \delta/2$ for every $R\subset\Rs^N\backslash 
Q_{L}$.\footnote{Observe 
that 
for our purpose it is enough to work with a partition $mod\, 0$, that is a 
family of set $B_k$ such that $m(\Rs\backslash\bigcup_k B_k)=0$ and $m(B_k\cap 
B_{k'})=0$ for $k\not= k'$. For this reason, the boundaries of the rectangles 
defined in this proof are irrelevant.}

Let now 
$q_l(\uv_N)=\{\underline w_N\,|\,  
|w_i-v_i|< l\}$, 
that is $q_l(\uv_N)$ is the cube of side $2l$ centered at $\uv_N$. By 
Lebesgue Differentiation Theorem, see e.g. Chapter 3 of \cite{Folland}, we get 
that $\lim_{l\to 0} d(q_l(\uv_N))=0$ for $m$ almost every $\uv_N$. 
Thus, given 
$\epsilon_0$ to be fixed later, there exists $\bar l$ such that $m(\{\uv_N\,|\, 
d(q_{\bar l}(\uv_N))>\epsilon_0\})\leq \delta/2$. Let $Q_0=\{\uv_N\,|\, 
d(q_{\bar 
l}(\uv_N))\leq\epsilon_0\}\cap Q_L$.

Let $K$ be the smallest integer such that $l_0=L/K< \bar l$ and consider the 
partition $\mathcal A$ of $[-L,L)$ formed by the sets $A_k=[kl_0,(k+1)l_0)$ with 
$-K\leq k< K$. For every $A_{\underline k} \in \mathcal A^N$ let 
$C_{\underline 
k}=\emptyset$ if $A_{\underline k}\cap Q_0=\emptyset$. Otherwise select a 
point $\uv_N\in A_{\underline k}\cap Q_0$ and set $C_{\underline 
k}=q_{\bar l}(\uv_N)$. Observe that, for every $\underline k$, $A_{\underline 
k}\cap Q_0\subset C_{\underline k}$ so that $Q_0\subset \bigcup_{\underline k} 
C_{\underline k}:=Q_1$. This means that the $C_{\underline k}$ form a covering 
of $Q_1$ but not necessarily a partition. Let $\mathcal 
D=\{D_j\}_{j=1}^J$ be the minimal partition of $Q_1$ such that, for every $j$, 
$D_j\subset C_{\underline k}$ for some ${\underline k}$.\footnote{The 
partition $\mathcal D$ can be constructed by taking intersections of the 
$C_{\underline k}$ and their complements.} We claim that 
\begin{equation}\label{5N}
 \sum_{j=1}^J m(D_j)d(D_j)\leq 5^N \epsilon_0 m(Q_1)\, .
\end{equation}
To see this, let $n_j$ be the number of $\underline k$ such that $D_j\subset 
C_{\underline k}$. By construction we have $n_j\geq 1$. On the other hand, for
$\uv_N\in A_{\underline k}$ and $\underline w_N\in A_{\underline k'}$, 
since $l_0\geq\bar l/2$, we have $q_{\bar l}(\uv_N)\cap  q_{\bar 
l}(\underline w_N)=\emptyset$ if $\sum_{i=1}^N |k_i-k'_i|>2$. This implies that 
$n_j\leq 5^N$. Calling $J_{\underline k}=\{j\,|\, D_j\subset C_{\underline 
k}\}$, 
by convexity, we have
\[
  \sum_{j\in J_{\underline k}} m(D_j)d(D_j)\leq 
m(C_{\underline k})d(C_{\underline k})\leq \epsilon_0 m(C_{\underline k})
\]
because, by construction, the center of $C_{\underline k}$ is in $Q_0$.
It follows that
\[
\begin{aligned}
 \sum_{j=1}^J m(D_j)d(D_j)\leq& \sum_{j=1}^J n_j m(D_j)d(D_j)=
 \sum_{\underline k} \sum_{j\in J_{\underline k}} m(D_j)d(D_j)\leq 
\epsilon_0\sum_{\underline k}m(C_{\underline k})\\
=&\epsilon_0\sum_{j=1}^J n_j m(D_j)\leq \epsilon_0 5^N m(Q_1)
\end{aligned}
\]
where we used the bound on $n_j$ and the fact that $\mathcal D$ is a partition.

We can now extend $\mathcal D$ to a partition $\widetilde{\mathcal 
D}=\{D_j\}_{j=1}^{\widetilde J}$ of $\Rs^N$ by adding finitely many rectangles. 
By construction, $m(\bigcup_{j=J+1}^{\widetilde J} D_j)=m(\Rs^N\backslash 
Q_1)\leq \delta$ so that, choosing $\epsilon_0=5^{-N}\epsilon/2$, we 
get $s(h_N)-s(h_{N,\mathcal D})\leq \epsilon_0 5^N m(Q_1)+m(\Rs^N\backslash 
Q_1)d(\Rs^N\backslash Q_1)\leq\epsilon$. Finally, since every $D_j\in 
\widetilde {\mathcal D}$ is a 
rectangle, we can find a finite partition $\mathcal B$ of $\Rs$ such that 
$\mathcal B^N$ refines $\widetilde{\mathcal D}$. This concludes the proof of 
Lemma \ref{coarseN}. 
\qed
\medskip 


We are now ready to prove Lemma \ref{lem:coarse}. Consider $\bh$ such that 
$S(\bh)<\infty$ and call
\begin{align*}
 E_M(\bh)&=\sum_{N>M}a_N\int_{\Rs^N}\gamma(\uv_N)h_N(\uv_N)d \uv_N\\
S_M(\bh)&=\sum_{N>M}a_N\int_{\Rs^N}\gamma(\uv_N)h_N(\uv_N)
\log\left(h_N(\uv_N)\right)d \uv_N\, .
\end{align*}
By convexity, for every $M$ and every partition $\mathcal B$, we get
\[
E_M(\bh)\log(E_M(\bh))
\leq S_M(\bh_{\mathcal B})\leq 
S_M(\bh)\,.
\]
Since $E(\bh),S(\bh)<\infty$ for every $\epsilon$ there exists $M$ such that 
$|E_M(\bh)\log(E_M(\bh))|\leq\epsilon/4$ and $|S_M(\bh)|\leq\epsilon/4$. This 
implies that, for every partition $\mathcal B$, we have 
$S_M(\bh)-S_M(\bh_{\mathcal B})<\epsilon/2$. Moreover, from Lemma 
\ref{coarseN}, for every $N\leq M$ we can find a partition $\mathcal B_N$ of 
$\Rs$ such that
\[
  s(h_N) - s(h_{N,\mathcal B_N^N})\leq\frac{\epsilon}{2M}\,.
\]
Finally let $\mathcal B$ be a partition of $\Rs$ that refines every $\mathcal 
B_N$ for $N\leq M$. Since $s(h_{N,\mathcal B_N^N})\leq s(h_{N,\mathcal 
B^N})\leq s(h_N)$, we get
\[
 S(\bh)=\sum_{N\leq M}a_N s(h_N)+S_M(\bh)\leq \sum_{N\leq M}a_N s(h_{N,\mathcal 
B^N})+S_M(\bh_{\mathcal B})+\epsilon=S(\bh_{\mathcal B})+\epsilon
\]
\qed

\begin{rem}
\emph{
An alternative approach to coarse graining is as follows. Let $I_{s}$ be the 
normalized characteristic function of the segment $(-s,s)$ and let $(\bh_s)_N= 
\int I_{s}^{\otimes N}(\uv_N-\underline w_N)h_N(\underline w_N)d\underline 
w_N$. Clearly $(\bh_s)_N$ is continuous for every $N$. Moreover we have 
$S(\bh_s)\leq S(\bh)$ and $\Psi(\bh_s)\leq \Psi(\bh)$, see the argument around 
\eqref{same}. Finally, 
$S(\bh_s)\to_{s\to 0} S(\bh)$. Thus, reasoning like in \eqref{entB}, we can 
restrict our attention to continuous 
states $\bh$. In this case, the analogous of Lemma \ref{coarseN} is simpler to 
proof. Indeed if $h_{N}$ is continuous,  $\lim_{l\to 0} d(q_l(\uv_N),h_{N})=0$ 
for every $\uv_N$ and, since $Q_L$ is compact, we can find 
$\bar l$ such that $d(q_{\bar l}(\uv_N),h_{N})\leq \epsilon$ for $\uv_N\in 
Q_L$. In such a 
situation, the regular partition $\mathcal A$ built before \eqref{5N} already 
provides the solution. }
\end{rem}

\subsection{Derivation of \eqref{densiO} and
\eqref{proptot}.}
\label{app:bosons.propa}

We start with \eqref{densiO}. Expanding the terms in \eqref{expa} the form 
$(e^{(\tilde\lambda\K-\rho \N)t}\bff_n,\phi_k)_{k,n}$ using 
\eqref{ntoinfO} recursively starting from the most external one gives 
\begin{align}\label{Oprop}
&\lim_{n\to\infty}\left( e^{(\tilde\lambda_n 
\K+\rho(\cO-\N))t}\bff_n,\phi_k\right)_{k,n}\\
&\qquad=\lim_{n\to\infty}\sum_{q\geq 0}\sum_{p_0,
p_1 , \ldots,p_{q} \geq 
0}\rho^{q}\lambda^{|p|}\left(\frac{\rho}{\mu_n}\right)^{k+|p|}\sum_{N\geq 
k+|p|}\frac{N!}{(N-k-|p|)!}\frac{(N+q)!}{N!} \\
&\qquad\qquad\cdot\int_{0\leq 
t_q\leq\cdots t_1\leq t}\prod_{i=0}^q\left(e^{-\rho(N+q-i)(t_{i}-t_{i+1})}
\frac{(t_{i}-t_{i+1})^{p_i}}{p_i!}\right)dt_1\cdots dt_q\\
&\qquad\qquad\qquad\cdot\int 
f_{N+q}(\uv_{N+q})(G_k^{*|p|}\phi_k)(\uv_{k+|p|})d\uv_{N+q} 
\end{align}
where $t_{q+1}=0$, $t_0=t$ and $|p|=\sum_{i=0}^{q} p_i$. 
Call now 
\begin{align*}
 b_{q,P}=&\sum_{\genfrac{}{}{0pt}{}{p_0,
p_1 , \ldots,p_{q} \geq 
0}{|p|=P}}
&\int_{0\leq 
t_q\leq\cdots t_1\leq t}\prod_{i=0}^q\left(e^{-\rho(N+q-i)(t_{i}-t_{i+1})}
\frac{(t_{i}-t_{i+1})^{p_i}}{p_i!}\right)dt_1\cdots dt_q\, .
\end{align*}
We first sum over 
the $p_i$ using that
\[
 \sum_{\genfrac{}{}{0pt}{}{p_0,p_1 , \ldots,p_{q} \geq 
0}{|p|=P}}\prod_{i=0}^q\frac{(t_{i}-t_{i+1})^{p_i}}{p_i!}=\frac{t^p}{p!}\,,
\]
then we integrate over the $t_i$ using \eqref{1me} and we get
\[
 b_{q,P}=\frac{t^p}{p!}\frac{1}{q!}e^{-\rho N t}(1-e^{-\rho t})^q
\]
Inserting in \eqref{Oprop} gives
\[
\begin{aligned}
&\lim_{n\to\infty}\left( e^{(\tilde\lambda_n 
\K+\rho(\cO-\N))t}\bff_n,\phi_k\right)_{k,n} 
=\lim_{n\to\infty}\sum_{q\geq 0}\sum_{p\geq 
0}\frac{t^p\lambda^{p}}{p!}\left(\frac{\rho}{\mu_n}\right)^{k+p}\\
&\qquad\cdot\sum_{N\geq 
k+p}\frac{(N+q)!}{(N-k-p)!\,q!}
e^{-\rho N t}(1-e^{-\rho t})^q
\int 
f_{N+q}(\uv_{N+q})(G_k^{*|p|}\phi_k)(\uv_{k+|p|})d\uv_{N+q}\, .
\end{aligned}
\]
Finally we write
\[
 \frac{(N+q)!}{(N-k-p)!\,q!}=\frac{(N+q)!}{(N+q-k-p)!}\binom{N+q-k-p}{q}
\]
so that, setting $M=N+q$ and summing over $q$, we get
\[
\begin{aligned}
 &\lim_{n\to\infty}\left( e^{(\tilde\lambda_n 
\K+\rho(\cO-\N))t}\bff_n,\phi_k\right)_{k,n} 
=\lim_{n\to\infty}\sum_{p\geq 
0}\frac{t^p\lambda^{p}}{p!}
e^{-\rho(k+p)t}\\
&\qquad\qquad\left(\frac{\rho}{\mu_n}\right)^{k+p}\sum_{M\geq 
k+p}\frac{M!}{(M-k-p)!}
\int f_{N+q}(\uv_{M})(G_k^{*|p|}\phi_k)(\uv_{M})d\uv_{N+q}\, .
\end{aligned}
\]
and the derivation of \eqref{densiO} is complete.

Turning to \eqref{proptot} we set
\begin{align*}
\mathcal A_q(\phi_k)=&\sum_{p_0, p_1 , \ldots,p_{q} \geq 0}
\int_{0\leq t_q\leq\cdots\leq t_1\leq t}\prod_{i=0}^q 
e^{-\rho(t_{i}-t_{i+1})(|p|_i-i)}\frac{(t_{i}-t_{i+1})^{p_i}}{p_i!}\, 
dt_1\cdots dt_q\cdot\\
&\qquad G_{k+|p|_q-q}^{*p_{q}}I_{k+|p|_q-q+1}\cdots 
G_{k+p_0-1}^{*p_1}I_{k+p_0}G_k^{*p_0}\phi_k\nonumber
\end{align*} 
For the rest of this section we will 
neglect the number of variables subscript in order to make expressions more 
readable. To understand the structure of $\mathcal 
A_q(\phi\otimes\psi)$, first look at $\mathcal 
A_2(\phi\otimes\psi)$.
Combining \eqref{fact} and \eqref{deriI} we can write
\begin{subequations}
\begin{align}
&\sum_{p_0,p_1,p_2}\prod_{i=0}^2
\frac{(e^{-(t_{i}-t_{i+1})}(t_{i}-t_{i+1}))^{p_i}}{p_i!}
G^{*p_2}IG^{*p_1}IG^{*p_0}(\phi\otimes\psi)=\nonumber\\
&\quad\sum_{p^1_0,p^1_1,p^1_2}\sum_{p^2_0,p^2_1,p^2_2}
\prod_{i=0}^2
\frac{(e^{-(t_{i}-t_{i+1})}(t_{i}-t_{i+1}))^{p^1_i}}{p^1_i!}
\prod_{i=0}^2
\frac{(e^{-(t_{i}-t_{i+1})}(t_{i}-t_{i+1}))^{p^2_i}}{p^2_i!}
\nonumber\\
&\qquad\qquad\left( 
G^{*p^1_2}IG^{*p^1_1}IG^ {*p^1_0}\phi\otimes
\label{primot}
G^{*p^2_2}G^{*p^2_1}G^{*p^2_0}\psi\right.\\
&\qquad\qquad+ 
G^{*p^1_2}IG^{*p^1_1}G^{*p^1_0}\phi\otimes\label{secondot}
G^{*p^2_2}G^{*p^2_1}IG^{*p^2_0}\psi\\
&\qquad\qquad+
G^{*p^1_2}G^{*p^1_1}IG^{*p^1_0}\phi\otimes\label{terzot}
G^{*p^2_2}IG^{*p^2_1}G^{*p^2_0}\psi\\
&\qquad\qquad\left.+ 
G^{*p^1_2}G^{*p^1_1}G^ {* p^1_0}\phi\otimes
G^{*p^2_2}IG^{*p^2_1}IG^{*p^2_0}\psi\right)\label{quartot}\, .
\end{align}
\end{subequations}

To simplify \eqref{primot}, we can use that
\[
\sum_{p^2_0,p^2_1,p^2_2}\prod_{i=0}^2
\frac{(e^{-(t_{i}-t_{i+1})}(t_{i}-t_{i+1}))^{p^2_i}}{p^2_i!}
(\bff_n,
G^{*p^2_2}G^{*p^2_1}
G^{*p^2_0}\psi)_{n}=\sum_p \frac{t^p}{p!}e^{-t}(\bff_n,G^ 
{*p}\psi)_{n}
\]
and similarly for \eqref{quartot}. On the other hand, for \eqref{secondot} we 
have
\[
 \begin{aligned}
&\sum_{p^1_0,p^1_1,p^1_2}\sum_{p^2_0,p^2_1,p^2_2}
\prod_{i=0}^2
\frac{(e^{-(t_{i}-t_{i+1})}(t_{i}-t_{i+1}))^{p^1_i}}{p^1_i!}
\prod_{i=0}^2
\frac{(e^{-(t_{i}-t_{i+1})}(t_{i}-t_{i+1}))^{p^2_i}}{p^2_i!}
\nonumber\\
&\qquad\qquad\qquad(\bff_n, 
G^{*p^1_2}IG^{*p^1_1}G^{*p^1_0}\phi)_{n}
(\bff_n,
G^{*p^2_2}G^{*p^2_1}IG^{*p^2_0}\psi)_{n}\\
&\qquad=\sum_{p^1_0,p^1_1}\sum_{p^2_0,p^2_1}
\frac{(e^{-(t-t_1)}(t-t_1))^{p^1_0}
(e^{-t_1}t_1)^{p^1_1}}{p^1_0!p^1_1!}\frac{(e^{-(t-t_2)}(t-t_2))^{p^2_0}(e^{-t_2}
t_2)^{ p^2_1}}{p^2_0!p^2_1!}\\
&\qquad\qquad\qquad\qquad(\bff_n, 
G^{*p^1_1}IG^{*p^1_0}\phi)_{n}(\bff_n,G^{*p^2_1}I
G^{*p^2_0}\psi)_{n}\,.
 \end{aligned}
\]
while \eqref{terzot} gives a similar expression but for the roles of $t_1$ and 
$t_2$ that are inverted. Combining this expressions we get
\[
 \mathcal A_2(\phi\otimes\psi)=\mathcal A_0(\phi)\mathcal A_2(\psi)+
 \mathcal A_1(\phi)\mathcal A_1(\psi)+\mathcal A_2(\phi)\mathcal A_0(\psi)
\]
For the general case we can write
\begin{align*}
&\prod_{i=0}^qe^{-\rho 
(t_{i}-t_{i+1})(k+|p|_i-i)}\frac{(t_{i}-t_{i+1})^{p_i}}{p_i!}G^{*p_
{q+1}}I\cdots 
G^{*p_1}IG^{*p_0}(\phi\otimes\psi) =\\
&\qquad\sum_{p_{i}^1+p_{i}^2=p_i}
\sum_{\sigma_1,\ldots\sigma_q\in\{0,1\}}\\
&\qquad\qquad\prod_{i=0}^q e^{-\rho 
(t_{i}-t_{i+1})(k_1+|p^1|_i-i)}\frac{(t_{i}-t_{i+1})^{p^1_i}}{p^1_i!} 
G^{*p^1_{q+1}}I^{\sigma_q}\cdots 
G^{*p^1_1}I^{\sigma_1}G^{*p^1_0}\phi\\
&\qquad\qquad\otimes\prod_{i=0}^q e^{-\rho 
(t_{i}-t_{i+1})(k_2+|p^2|_i-i)}\frac{(t_{i}-t_{i+1})^{p^2_i}}{p^2_i!}G^{*p^2_{q+
1}}I^{1-\sigma_q}\cdots 
G^{*p^2_1}I^{1-\sigma_1}G^{*p^2_0}\psi\, .
\end{align*}
that, after resummation, gives
\begin{align*}
&A_q(\phi\otimes\psi)= 
\sum_{q_1+q_2=q}\sum_{p^1_0, p^1_1 , \ldots,p^1_{q^1+1} 
\geq 0} \sum_{p^2_0, p^2_1 , \ldots,p^2_{q^2+1} 
\geq 0}\lambda^{|p^1|+|p^2|}\\
&\quad\sum_{\pi}^{*}\int_{0\leq t_{\pi(q_1+q_2)}\leq\cdots\leq t_{\pi(1)}\leq 
t}dt_{1,1} \cdots dt_{1,q_1}dt_{2,1}\cdots dt_{2,q_2}\\
&\qquad\prod_{i=0}^{q_1} e^{-\rho(t_{1,i}-t_{1,i+1})(k_1+|p^1|_i-i)}
\frac{(t_{1,i}-t_{1,i+1})^{p^1_i}}{p^1_i!}\quad G^{*p^1_{q_1+1}}I \cdots 
G^{*p^1_1}IG_{k_1}^{*p^1_0}\phi\\
&\qquad\otimes\prod_{j=0}^{q_2} e^{-\rho(t_{2,j}-t_{2,j+1})(k_2+|p^2|_j-j)}
\frac{(t_{2,j}-t_{2,j+1})^{p^2_j}}{p^2_j!}G^{*p^2_{q_2+1}}I \cdots
G^{*p^2_1}IG^{*p^2_0}\psi\\
\end{align*}
where $\sum_{\pi}^{*}$ is the sum over all one-to-one functions $\pi$ from 
$\{1,\ldots,q_1+q_2\}$ to the set 
$\{(1,1),\ldots,(1,q_1),(2,1),\ldots,(2,q_2)\}$ such that if, for $i>j$ and 
$\sigma\in\{1,2\}$, we have $\pi(i)=(\sigma,q)$ and $\pi(j)=(\sigma,q')$ then 
$q>q'$. Observing that
\begin{align*}
\sum_{\pi}^{*}\int_{0\leq t_{\pi(1)}\leq\cdots\leq t_{\pi(q_1+q_2)}\leq 
t}&dt_{1,1} \cdots dt_{1,q_1}dt_{2,1}\cdots dt_{2,q_2}=\\
&\int_{0\leq 
t_{1,1}\leq\cdots\leq t_{1,q_1}\leq 
t}dt_{1,q_1} \cdots dt_{1,1} \int_{0\leq 
t_{2,q_2}\leq\cdots\leq t_{2,1}\leq 
t}dt_{2,1} \cdots dt_{2,q_2}
\end{align*}
we get
\[
 \mathcal A_q(\phi\otimes\psi)=\sum_{q_1+q_2=q}\mathcal A_{q_1}(\phi)\mathcal 
A_{q_2}(\psi)\, .
\]
Propagation of chaos now follows easily.

\medskip

\begin{acknowledgements}  
F.B. gratefully acknowledges National Science Foundation grant DMS-1907643. 
F.B. thanks M. Loss and E. Carlen for many enlightening discussions and 
suggestions.
\end{acknowledgements}

%
%



\end{document}